\documentclass[manuscript]{aastex61}

\usepackage{enumerate}
\usepackage{euscript}
\usepackage{gensymb}
\usepackage{graphicx}
\usepackage{xcolor} 

\newif\ifmod
\modtrue
\newif\ifdel
\deltrue

\newcommand{\rew}[1]{\ifmod\textcolor{blue}{{#1}}\else{#1}\fi}

\newcommand{\kms}{km~s$^{-1}$}

\newcommand\hcop{HCO$^+$}
\newcommand\hcopj{HCO$^+$ (1$-$0)}
\newcommand\hcopjb{HCO$^+$~(3$-$2)}

\newcommand\hcoppj{H$^{13}$CO$^+$~(1$-$0)}
\newcommand\hcnj{HCN~(3$-$2)}

\newcommand\hcoj{H$_2$CO~(2$_{12}$$-$1$_{11}$)}

\newcommand{\ta}{$T_{\rm A}^*$}
\newcommand{\tr}{$T_{\rm R}^*$}

\shorttitle{Inflow study on HMPOs}
\shortauthors{Yoo et al.}

\begin{document}

\title{Inflow Motions associated with High-mass Protostellar Objects}

\author{Hyunju Yoo}
\affiliation{Department of Astronomy and Space Science, Chungnam National University, Daejeon, Korea}
\affiliation{Korea Astronomy and Space Science Institute, 776 Daedeokdae-ro, Yuseong-gu, Daejeon 34055, Korea}

\author{Kee-Tae Kim}
\affiliation{Korea Astronomy and Space Science Institute, 776 Daedeokdae-ro, Yuseong-gu, Daejeon 34055, Korea}

\author{Jungyeon Cho}
\affiliation{Department of Astronomy and Space Science, Chungnam National University, Daejeon, Korea}

\author{Minho Choi}
\affiliation{Korea Astronomy and Space Science Institute, 776 Daedeokdae-ro, Yuseong-gu, Daejeon 34055, Korea}

\author{Jingwen Wu}
\affiliation{National Astronomical Observatories, Chinese Academy of Sciences, 20A Datun Road, Chaoyang District, Beijing, 100012, People's Republic of China}

\author{Neal J. Evans~II}
\affiliation{Department of Astronomy, The University of Texas at Austin, 2515 Speedway, Stop C1400, Austin, TX 78712-1205, USA}
\affiliation{Korea Astronomy and Space Science Institute, 776 Daedeokdae-ro, Yuseong-gu, Daejeon 34055, Korea}

\author{L. M. Ziurys}
\affiliation{Department of Chemistry, Department of Astronomy, Arizona Radio Observatory, and Steward Observatory, University of Arizona, 933 N. Cherry Avenue, Tucson, AZ 85721, USA}




\begin{abstract}

We performed a molecular line survey of 82 high-mass protostellar objects in search for inflow signatures associated with high-mass star formation.
Using the \hcoppj\ line as an optically thin tracer, 
we detected a statistically significant excess of blue asymmetric line profiles in the \hcopj\ transition, but nonsignificant excesses in the \hcopjb\ and \hcoj\ transitions. The negative blue excess for the \hcnj\ transition suggests that the line profiles are affected by dynamics other than inflow motion. The \hcopj\ transition thus seems to be the suitable tracer of inflow motions in high-mass star-forming regions, as previously suggested. 
We found 27 inflow candidates that have at least one blue asymmetric profile and no red asymmetric profile, and derived the inflow velocities to be 0.23$-$2.00~\kms\ for 20 of them using a simple two-layer radiative transfer model. Our sample is divided into two groups 
in different evolutionary stages. The blue excess of the group in relatively earlier evolutionary stages was estimated to be slightly higher than that of the other in the \hcopj\ transition. 

\end{abstract}




\section{Introduction} 
\label{sec:intro}

High-mass stars (Mass $>$ 8 $M_\odot$, Luminosity $>$ 10$^3$ $L_\odot$) have great physical and chemical importance in the interstellar environments. 
They affect energetics of galaxies via jets/outflows, strong radiation, and supernova explosions, and play a role in triggering next-generation star formation with the associated shocks. Furthermore, they enrich the surrounding systems and change chemical abundances. 
In spite of the importance of high-mass stars, there still remain questions on their formation process due to difficulties arising from observational and evolutionary limitations. In order to investigate the formation of high-mass stars, it is necessary to identify sources in different evolutionary phases, especially protostellar stage, and understand their environments.                                                                                                                                                                                                                                                                                                                                                                                                                                                         

Gravitational collapse is the most important and fundamental key process in the early stage of star formation. Unfortunately, collapsing motion is relatively harder to observe than other motions (e.g. rotation, outflow). 
According to the ``inside-out" collapse model and observations of low-mass protostellar cores
\citep{Shu77, Zhou93, Myers96, Evans99, Evans03}, an asymmetric double-peaked velocity profile with higher peak at the blue-shifted part (hereafter, blue profile) can be a tool for detecting inflow signatures. 
The blue profile has dependences on the optical depth of the observed line and the velocity of the inflowing material. Therefore, finding adequate inflow tracers for target sources is needed to enhance our understanding. In this work, we study inflow motions toward high mass star-forming regions using several molecular transitions and try to identify suitable inflow tracers.

Although inflow motion is difficult to observe, several surveys detected statistically significant evidence of inflow motions via observations of blue profiles toward low-mass star-forming regions.
The molecular line surveys with various transitions supported that active inflow motions exist regardless of the evolutionary stages of low-mass star formation \citep{Gregersen97, Gregersen00, Mardones97, Evans03}. Therefore, gravitational collapse is a generally accepted phenomenon for low-mass star formation both theoretically 
and observationally. 

On the other hand, inflow studies of high-mass star-forming regions toward different samples in different inflow-tracing molecular lines show inconsistent fractions of blue asymmetries \citep*[see Section \ref{sec3.1} for detailed results]{WuEvans03, Fuller05, Purcell06, Wyrowski06, Wu07, Reiter11, He15, Jin16}. 
Such studies are expected to give us clues to understand the influence and importance of gravitational collapse on the formation and evolution of high-mass stars. Furthermore, they may enable us to discuss similarity between high-mass star-forming process and the low-mass counterpart. However, it is difficult not only to characterize the evolutionary stages of high-mass star-forming regions but also to find appropriate inflow tracers. Accordingly, there should be more comprehensive researches with advanced facilities toward adequate targets to understand the formation of high-mass stars. 

In this study, we surveyed 82 high mass protostellar object (HMPO) candidates in multiple molecular lines by using single-dish telescopes to investigate the inflow motions via statistical analysis of the observed line spectra. 
We present the source selection criteria in Section \ref{sec2.1} and the observational details in Section \ref{sec2.2}. 
Statistical analysis with line profiles are in Section \ref{sec3.1} and identification of inflow candidates and examination of their characteristics are in Section \ref{sec3.2}. We discuss appropriateness of determination of profiles and statistics depending on sub-groups in Section \ref{sec:dis}. Finally, we conclude with a summary of the main results in Section \ref{sec:con}.

\section{Source selection and Observations}
\label{sec2}

\subsection{Source selection}
\label{sec2.1}

In order to find potential sites of early stage of star formation, \citet{Richards87} proposed selection criteria for identifying bright compact molecular clouds from the IRAS (InfraRed Astronomical Satellite) point source catalogue. Meanwhile, \citet{WoodChurchwell89} suggested IRAS color criteria for ultracompact H\,{\scriptsize II} regions (UCH\,{\scriptsize II}s), $[25 - 12]$ $\geqq$ 0.57 and $[60 - 12]$ $\geqq$ 1.30
		\footnote{Here, [$\lambda_{2} - \lambda_{1}$] is defined as $\log_{10}[F_{\lambda 2}/F_{\lambda 1}]$, where $F_{\lambda i}$ is the flux density in wavelength band $\lambda_{i} $ in the unit of $\mu m$.}. 
\citet{Palla91} combined those two sets of criteria to discover bright infrared sources in the very early  phases of high-mass star formation without positional coincidence with known H\,{\scriptsize II}  regions or extragalactic objects. They found 260 $IRAS$ point sources and distinguished them into two groups. In the IRAS color-color diagram of $[25 - 12]$ and $[60 - 12]$, 125 of the 260 sources are in the region meeting the criteria of \citet{WoodChurchwell89}, while the remaining 135 are outside the region. They classified the 125 sources as the $\it High$ group and the 135 sources as the $\it Low$ group.
\citet{Molinari96} observed 163 among the 260 sources in the NH$_{3}$ (1, 1) and (2, 2) lines, which are believed as tracers of warm dense gas in the vicinity of embedded (proto)stars, in search for HMPO candidates. They detected NH$_{3}$ line emission in 101 sources. The detection rate of NH$_{3}$ emission is considerably higher for the $\it High$ group (80$\%$ of 80 sources) than for the $\it Low$ group (45$\%$ of 83 sources). Together with the finding of \citet{Palla91} that the detection rate of H$_{2}$O maser emission is significantly higher in the $\it High$ group (26\%) than in the $\it Low$ group (9\%), this might suggest that the two groups represent different evolutionary phases of high-mass star formation, i.e., the $\it Low$ group is in an earlier evolutionary stage than the $\it High$ group \citep*[see also][]{Molinari98}. 
We selected 72 of the 101 NH$_3$-detected sources in the catalog of \citet{Molinari96}: 47 $\it High$ and 25 $\it Low$ sources. We also added 10 $\it High$ sources from the HMPO candidate catalogs of \citet{Walsh98, Walsh01}, \citet{Hunter00}, and \citet{Sridharan02}. 
Table~\ref{tab1} presents the information of the 82 sources, such as the IRAS names, equatorial coordinates, kinematic distances ($d_{\rm kin}$), and infrared luminosities ($L_{\rm IR}$). We adopted the information from the original catalogs except the distances and luminosities for
IRAS05391-0152 \citep{Qin08}, IRAS06084-0611 \citep{Gomez02}, and IRAS07427-2400 \citep{Kumar02}.

\subsection{Observations} 
\label{sec2.2}

We made single-point observations toward the 82 sources in up to five transitions, such as \hcopj, \hcopjb, \hcnj, \hcoj\ and \hcoppj\ lines. The used telescopes were the Caltech Submillimeter Observatory (CSO) 10.4~m, the Arizona Radio Observatory (ARO) 12~m, the Submillimeter telescope (SMT) 10~m, and the Korean VLBI Network (KVN) 21~m. 
Table~\ref{tab2} summarizes the observational details, 
including the observed transition, frequency, telescope, main-beam size of the telescope, main-beam efficiency of the telescope, velocity resolution, and number of the observed sources.

\subsubsection{CSO 10.4~m Observations}

We used the CSO telescope to survey 48 sources in the \hcnj\ line and
11 sources in the \hcopjb\ line. The observations were undertaken in 
2005 October. The system temperatures ranged between 350 and 500~K.
We employed the 50~MHz bandwidth acousto-optical spectrometer (AOS) 
with 1024 channels as the backend and obtained typical rms noise levels ($T_{\rm A}^*$) of
0.17~K and 0.27~K at a velocity resolution of 0.22~\kms\ after smoothing for \hcnj\ and \hcopjb, respectively.

\subsubsection{ARO 12~m and SMT Observations}

We surveyed 60, 52, and 45 sources in the \hcopj, \hcoj, and \hcoppj\ lines, respectively, using the previous generation 12~m telescope. The backends were filterbanks with 256 channels and a bandwidth of 25.6~MHz (100~kHz resolution) each. The system temperatures were around 350~K at $\sim$89~GHz and 400~K at 140~GHz. The receivers were dual polarization, single-sideband SIS systems where the image rejection was typically 20~dB, obtained by tuning the mixer backshort. The observed temperatures were obtained on the \tr\ scale \citep{Mangum00}. The rms noise levels were 0.07~K at $\sim$0.34~\kms\ resolution and 0.08~K at 0.21~\kms\ resolution for $\sim$89 and 140~GHz, respectively. 

We also surveyed 32 sources in the \hcopjb\ line using the SMT. The 64~MHz filterbanks with 256 channels were used  with a spectral resolution of 250~kHz. The system temperatures were around 900~K. The single polarization receiver in this case utilized the ALMA Band 6 sideband-separating SIS mixers. The observed temperatures were obtained on the \ta\ scale. The rms noise level was 0.12~K at a resolution of 0.28~\kms. 
These observations were all conducted in 2006 June.

\subsubsection{KVN 21~m Observations}

We observed 14, 19, and 30 sources in the \hcopj, \hcoj, and \hcoppj\ lines, respectively, using the KVN 21~m telescopes at the Yonsei and Tamna stations \citep{Kim11}. The observations were performed in 2014 May and June, 2015 February, and 2016 June. The system temperatures usually ranged between 200 and 300~K. The backends were the digital spectrometers with 4096 channels and a bandwidth of 64~MHz each. The noise levels ($T_{\rm A}^*$) were typically 0.04, 0.06, and 0.02~K 
at velocity resolutions of 0.21, 0.27, and 0.22~\kms\ after smoothing for the \hcopj, \hcoj, and \hcoppj\ transitions, respectively, with two exceptions. The velocity resolution was 0.43~\kms\ for IRAS05137+3919 and IRAS05345+3157 in the \hcoppj.
\

\section{Results}

\subsection{Line Asymmetries} 
\label{sec3.1}

\citet{Mardones97} proposed a dimensionless parameter $\delta v$ to quantify asymmetry of the observed optically thick line profiles : $\delta v = (v_{thick} - v_{thin} ) / \vartriangle \!\!v_{thin}$. Here $v_{thin}$  and $v_{thick}$ are the peak velocities of the optically thin and thick lines, respectively, and  $\vartriangle \!\!v_{thin}$ is the full width at half maximum (FWHM) of the optically thin line. This parameter has been generally used in previous inflow studies of low- and high-mass star-forming regions. 
The $v_{thick}$ is directly obtained from the observed profile 
while $v_{thin}$ and $\vartriangle \!\!v_{thin}$  are measured by Gaussian fitting to the observed profile. 
The optically thin line used in this study is the \hcoppj\ line. 
Table~\ref{tab3} presents the determined parameters for the optically thin and thick lines. 
In order to confirm that \hcoppj\ line emission is optically thin, 
we derived the peak optical depth, $\tau_{thin}$, using the following equations:

\begin{equation}
 \tau_{thin} = - \ln \biggl[1 - \frac{T_{thin}}{(T_{ex} - J_{\nu}(T_{bg}))}\biggr], 
\end{equation} 

\noindent
and
\begin{equation}
 J_{\nu} (T) = \frac{h\nu}{k} \frac{1}{(e^{h\nu/kT} - 1)}.
\end{equation}

\noindent
Here T$_{thin}$ is the brightness temperature of the optically thin line, \hcoppj\  in this work, and T$_{bg}$ is the background brightness temperature, which is assumed as 2.73 K. We assumed that the \hcopj\ line is optically thick (1$-$ e$^{- \tau}$ $\approx$ 1) and that both \hcopj\ and \hcoppj\ lines arise from the same volume with the same excitation temperature, T$_{ex}$. 
The excitation temperature can be calculated from the equation
\begin{equation}
 T_{ex} = \frac{h\nu}{k} \biggl[ \ln \biggl(1+ \frac{(h\nu/k)}{T_{thick}+J_{\nu}(T_{bg})} \biggr) \biggr]^{-1},
 \end{equation}
 
 \noindent
 where T$_{thick}$ is the brightness temperature of the optically thick line, \hcopj. 
Table~\ref{tab3} lists the estimated optical depths of the \hcoppj\ line emission in the second column. The values range from 0.04 to 0.36, with an average value of 0.13. Thus the \hcoppj\ line emission seems to be optically thin enough to measure the systemic velocity and the velocity dispersion of the target clumps.
The above assumption of same emitting volume and excitation temperature for the HCO$^+$ and H$^{13}$CO$^+$ lines may be questionable because of different optical depths. That makes the two lines trace different depths of the clump and hence different volumes. However, the H$^{13}$CO$^+$ line seems to be practically the best choice because other molecular lines might trace more different volumes.

\begin{figure}
\epsscale{1.20}
\plotone{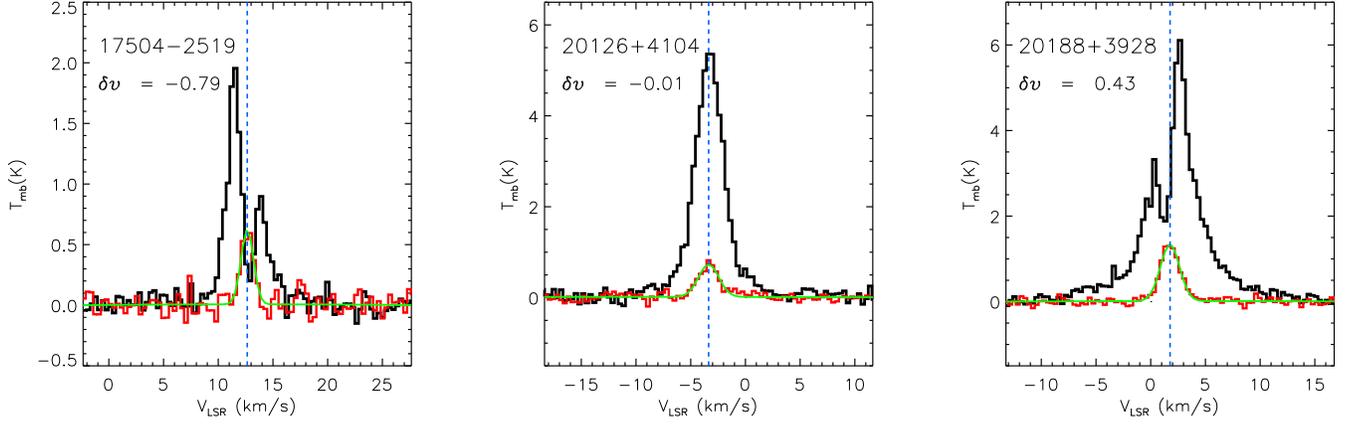}
\caption{Spectra for representative B profiles (left), N profiles (middle) and R profiles (right). 
In each panel, the red and black solid lines show the HCO$^+$ (1--0) and H$^{13}$CO$^+$ (1--0) line profiles, respectively. The blue-dotted vertical line indicates the peak velocity of the H$^{13}$CO$^+$ (1--0) line from the gaussian fitting result in green. The IRAS name and the estimated $\delta v$ are presented at the top left corner (Table~\ref{tab3}).
} 
\label{f1}
\end{figure}

\begin{figure*}
\begin{tabbing}
 \includegraphics[width=0.42\textwidth]{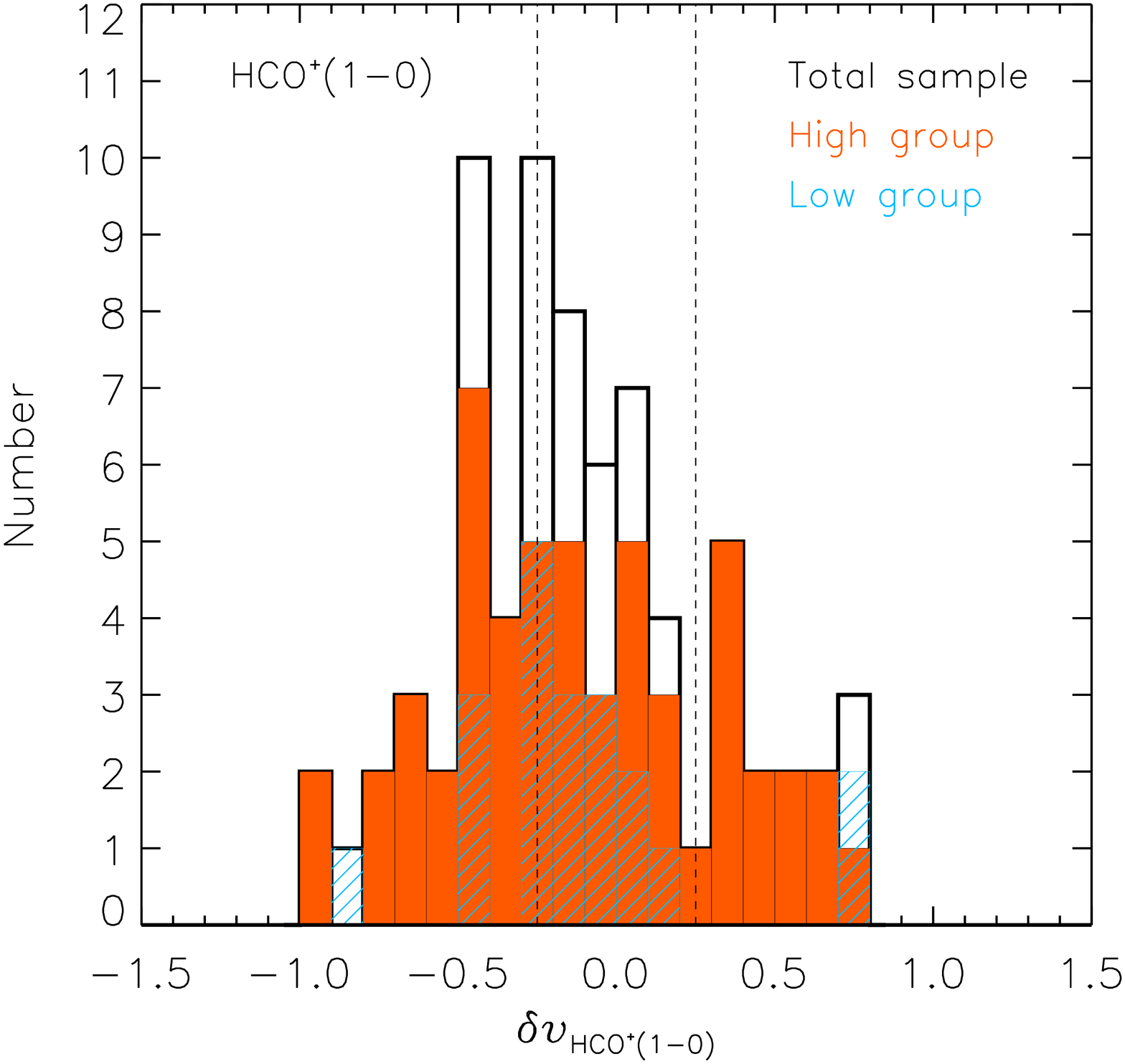}
 \=
  \includegraphics[width=0.42\textwidth]{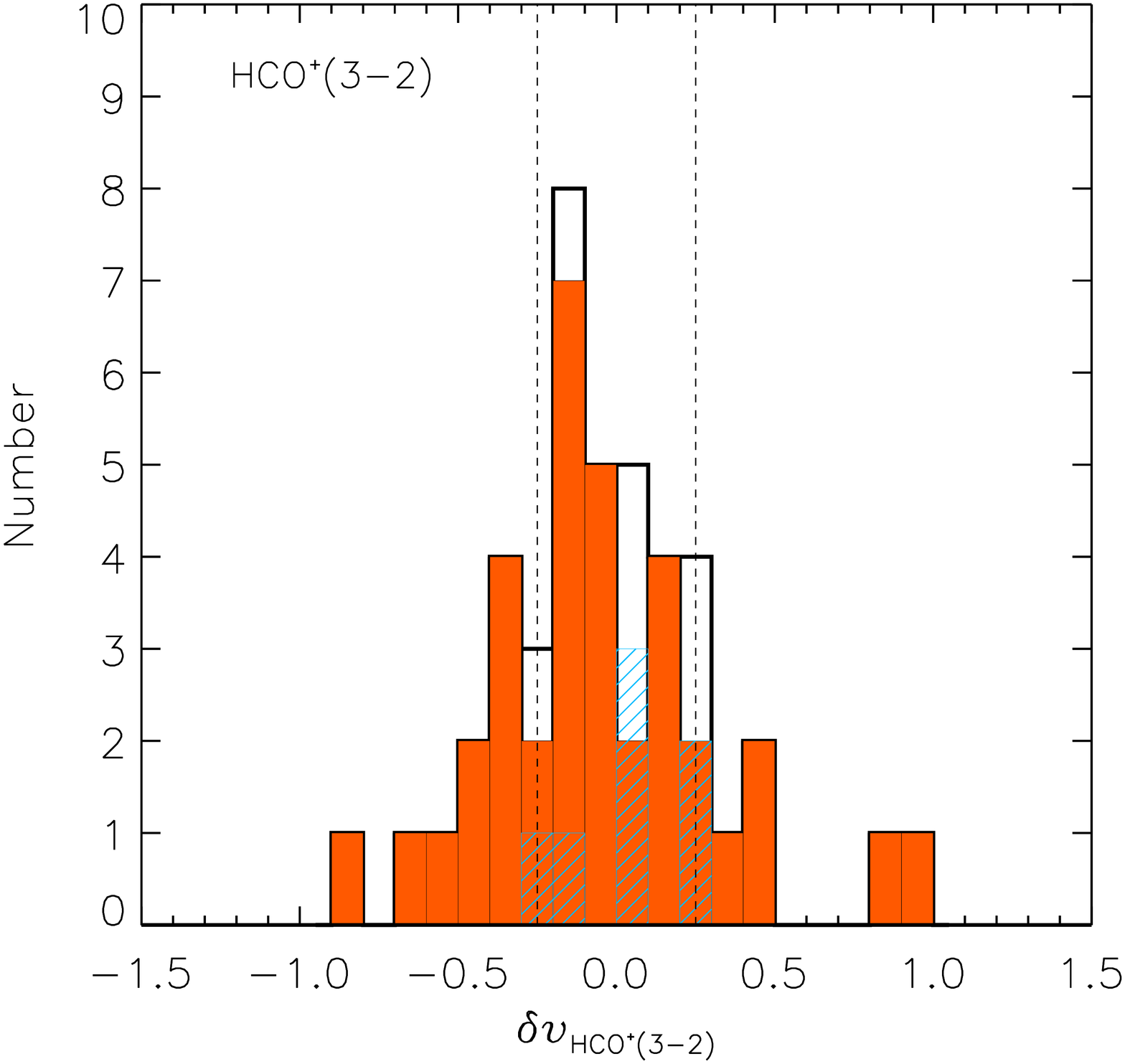} \\
  \includegraphics[width=0.42\textwidth]{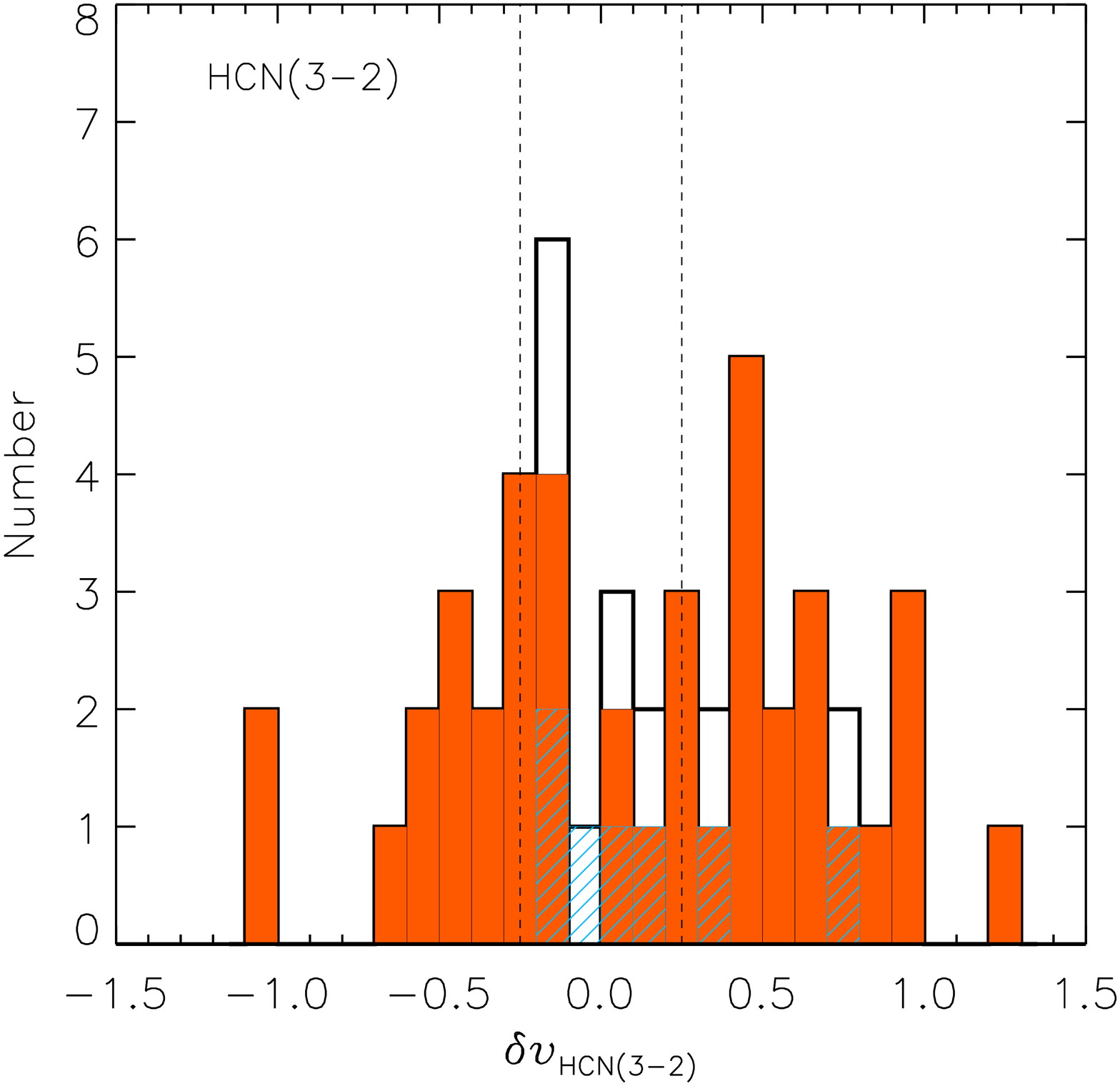}
  \>
  \includegraphics[width=0.42\textwidth]{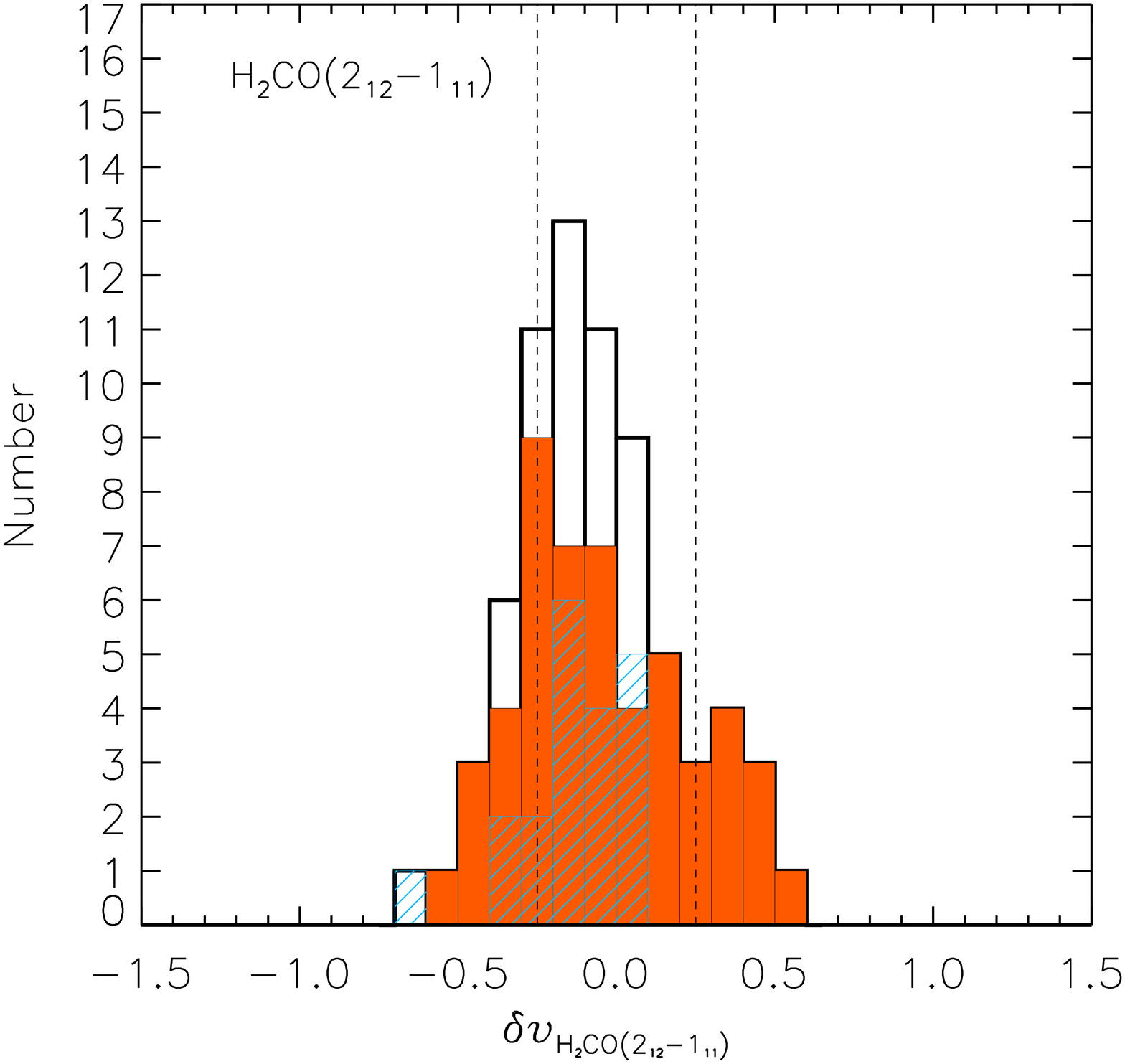} 
  \end{tabbing}
  \caption{Histograms of $\delta v$ for the observed optically thick lines:
HCO$^+$ (1--0), HCO$^+$ (3--2), HCN (3--2), and H$_{2}$CO (2$_{12}$--1$_{11}$). 
The black solid lines are for the entire sample, while the orange-colored bars and the cyan hatched bars are for the $\it High$ and $\it Low$ groups, respectively. The vertical dotted lines indicate the threshold values of  $\delta v$, $\pm$0.25.}
\label{f2}
\end{figure*}

As in most previous inflow studies \citep*[e.g.][]{Mardones97, Gregersen97, Fuller05}, we categorized the optically thick line profiles into three types on the basis of the measured value of $\delta v$: blue (B) for $\delta v < -0.25$, red (R) for $\delta v > 0.25$, neither (N) for $-0.25 \le \delta v \le 0.25$.
Figure~\ref{f1} displays sample \hcopj\ spectra with B, N, and R profiles along with the \hcoppj\ spectra. 
We counted the numbers of B, R, and N profiles ($N_{Blue}$, $N_{Red}$, $N_{Neither}$, and 
N$_{Total} \equiv N_{Blue}$+$N_{Red}$+$N_{Neither}$) for each optically thick line. 
The detection rates of blue profiles D$_{Blue}$ (=N$_{Blue}$/N$_{Total}$)
are 0.39, 0.26, 0.25 and 0.25, and those of red 
profiles D$_{Red}$ (=N$_{Red}$/N$_{Total}$)
are 0.19, 0.16, 0.40 and 0.13 in the \hcopj, \hcopjb, \hcnj,
and \hcoj\ lines, respectively (Table~\ref{tab4}).
%
Table~\ref{tab3} presents the estimated values of $\delta v$ and identified profile types for each source in the individual optically thick lines, and Figure~\ref{f2} exhibits histograms of $\delta v$ for each transition.  

We are able to quantify the dominance of blue profile with respect to red profile by the non-dimensional parameter, so-called blue excess (E), introduced by \citet{Mardones97}~:
\begin{equation}
 E = (N_{Blue} - N_{Red}) / N_{Total}.
\end{equation}

\noindent
Table~\ref{tab4} summarizes our statistical analysis results in four optically thick molecular lines. As mentioned above, the \hcoppj\ line is used as an optically thin tracer to determine $v_{thin}$ and $\vartriangle \!\!v_{thin}$. We found 29 blue profiles and 14 red profiles in the \hcopj\ line for a sample of 74 sources.  
The blue excess E for the \hcopj\ line is 0.20. This value is in good agreement with the estimate (E=0.22) obtained by \citet{He15} for 201 HMPOs in the same transition, and is slightly higher than the estimates of E=0.15 acquired by \rew{\citet{Fuller05} } for 68 HMPOs and E=0.17 by \citet{Wu07} for 29 HMPOs. We found less significant blue excesses of E=0.09 and E=0.13 in the \hcopjb\ and \hcoj\ lines, respectively. For comparison, \citet{Fuller05} obtained E=0.04 in the \hcopjb\ for 24 sources and E=0.19 in the \hcoj\ for 64 sources.
On the other hand, we found 12 blue and 19 red profiles in the \hcnj\ line for 48 sources. Thus, our sample shows a quite significant red excess of E=--0.15, suggesting that the \hcnj\ line traces other dynamics rather than inflows in HMPOs, such as outflow, rotation, and turbulent motions. In contrast, \citet{WuEvans03} measured a blue excess of 0.21 in the \hcnj\ for 28 UCHIIs and compact HII regions, which are mostly more evolved and luminous than the sources in our sample (see Section~\ref{sec:dis}).
The molecular tracer appears to have suitable optical depth and critical density for their sample but this is not the case for most sources in our sample.

In order to get rid of the statistical uncertainty coming from the use of different beam sizes of the telescopes, we also obtained statistics of line parameters separately for the ARO 12~m and the KVN 21~m. That is, we calculated blue excesses of the \hcopj\ line using the \hcoppj\ line as an optically thin line observed by the same telescope. In case of the ARO 12~m, the numbers of blue and red profiles are 19 and 9, respectively, out of total 45 sources, which leads to blue excess of E=0.22. On the other hand, in case of the KVN 21~m, the numbers of blue and red profiles are 5 and 2, respectively, for total 14 sources, which leads to blue excess of E=0.21. However, the blue excess of the latter case has statistical weakness arising from the small number of sources. If we combine two observations, the numbers of blue and red profiles are 24 and 11, respectively, for total 59 sources 
and hence the blue excess is E=0.22. 
Thus our statistics for the entire 74 sample (E=0.20) obtained by the combination of the \hcopj\ and \hcoppj\ lines detected by different telescopes seems to be acceptable. 
%

To evaluate the probability that the measured blue excess is produced by coincidence from a random distribution with the same numbers of blue and red profiles, we performed a binomial test \citep{Fuller05, Wu07, He15, Jin16} defined as 
\begin{equation}
P(X \geq V) = \sum_{x=V}^{N} \frac{N!}{x!(N-x)!} p^{x}(1-p)^{(N-x)}.
\end{equation}

\noindent
Here, $N$ is the number of performances, $V$ is the the number of successes, and $p$ is the probability of occurrence in a single independent performance. 
In our case, $N$ is the total number of blue and red profiles and $V$ is the number of blue profiles. Table~\ref{tab4} also lists the derived values of $P$. The $P$ value for the HCO$^+$(1--0) line is 0.016, which implies that the estimated blue excess (E=0.20) is very unlikely to be generated by chance from an even distribution.

\subsection{Inflow candidates}
\label{sec3.2}

Since the sources in our sample were observed in multiple optically thick lines, one source can be classified as different profile types in different transitions. We thus identified strong inflow candidates as in some previous studies using the two criteria:
1) at least one blue (B) profile and 2) no red (R) profile \citep[e.g.][]{Mardones97, Fuller05}.
There are 27 sources satisfying the criteria. 
The inflow candidates are marked with asterisks in the first column of Table~\ref{tab3} (see also Figure~\ref{fA1}).
Three of them (IRAS05490+2658, IRAS19282+1814, IRAS23545+6508) had been observed by Fuller et al. (2005) as well, but only IRAS05490+2658 was identified as an inflow candidate while the other two did not show any line asymmetry (see their Table~9).
In reality, their observed positions are offset from ours by 12$''$, 76$''$, and 23$''$ for IRAS05490+2658, IRAS19282+1814, and IRAS23545+6508, respectively. The large offsets can cause the non-detection of blue asymmetry in their line profiles of \hcop~(1$-$0) and \hcoj\ for IRAS19282+1814 and IRAS23545+6508, taking into account their beam sizes of 29$''$ at 89 GHz and 17$''$ at 140 GHz.
Therefore, all the 27 sources but IRAS05490+2658 are newly identified inflow candidates. 
It should be noted that the \hcnj\ spectrum of IRAS04579+4703 and the \hcoj\ spectrum of IRAS05345+3157 are quite noisy (Fig. A1) and so their classifications as blue profiles need to be confirmed by more sensitive observations.

We derived the inflow velocities, $v_{in}$, for 20 inflow candidates with blue \hcop~(1$-$0) line profiles using the two-layer radiative transfer model of \citet{Myers96}. The model assumes two uniform layers approaching toward each other with different excitation temperatures. The inflow velocity is determined by both optical depth and excitation temperature.
The estimated inflow velocities range from 0.23 to 2.00~\kms\ with a median value of 0.49~\kms\ (Table~\ref{tab3}). These values are comparable to the estimates (0.1$-$1.8~\kms) of \citet{KlassenWilson07} for 8 high-mass star-forming clumps.  
For further detailed studies, including measurements of the inflow region size and the inflow mass rate, the mapping observations of the blue profiles are required.

On the other hand, we also found 18 sources that have at least one red profile and no blue profile. It is likely that the profiles might be affected by dynamics other than inflow motion. There are 12 sources with both blue and red profiles. They are all classified as $High$ sources, and 7 of them show blue profiles in the \hcopj\ transition but red profiles in the \hcnj\ transition.

\section{Discussion}
\label{sec:dis}

\begin{figure*}
\begin{tabbing}
 \includegraphics[width=0.42\textwidth]{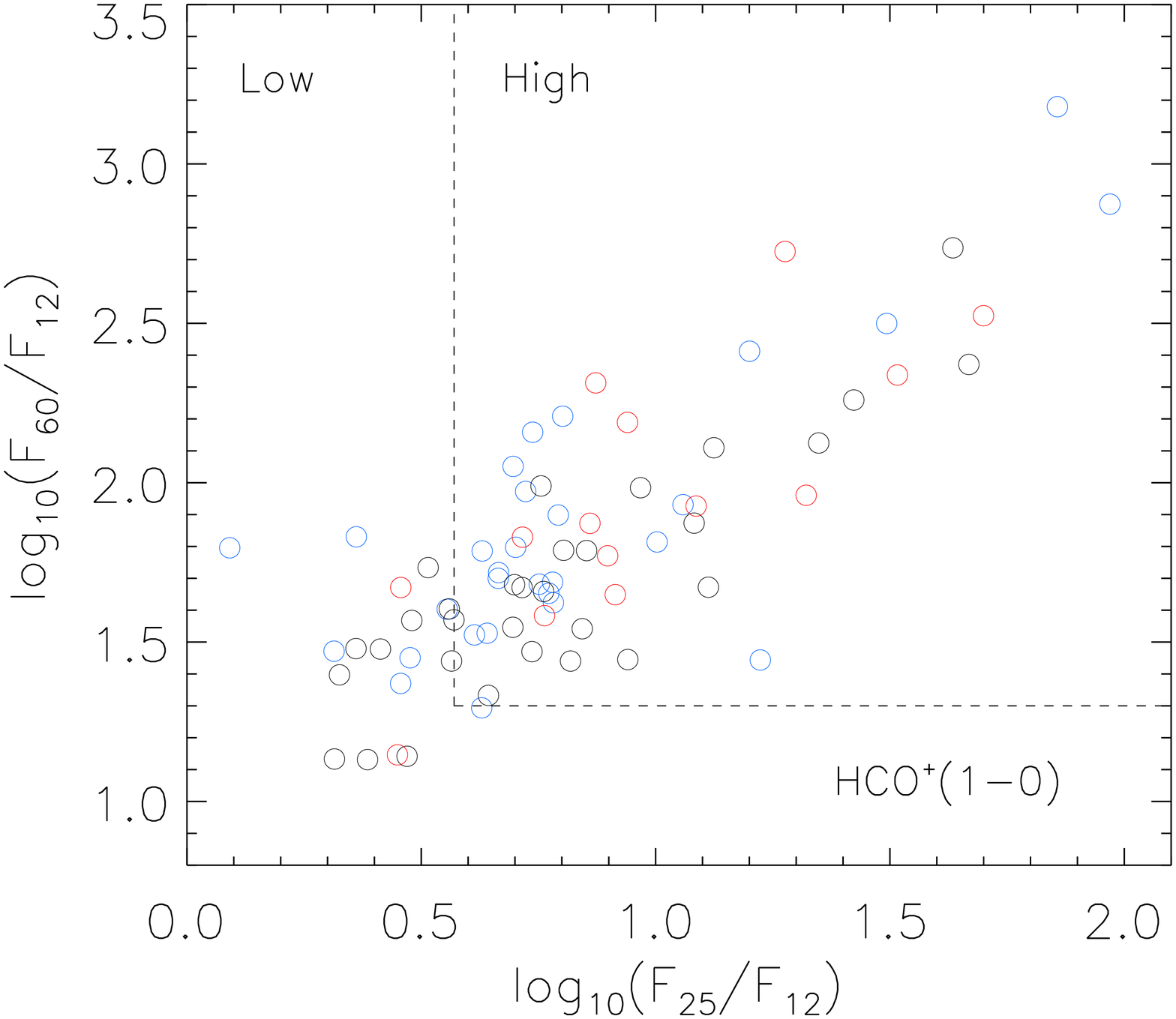}
 \=
  \includegraphics[width=0.42\textwidth]{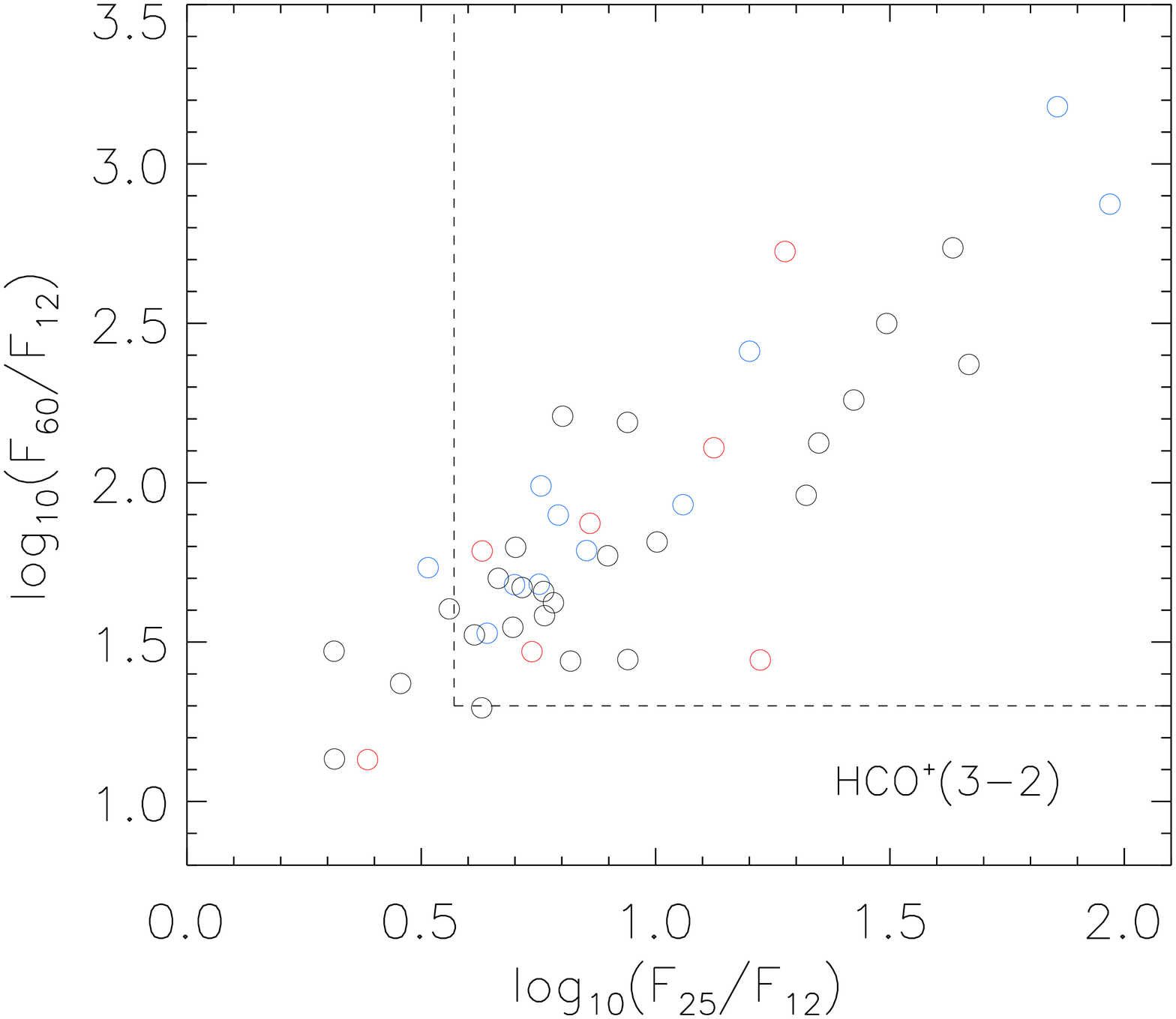} \\
  \includegraphics[width=0.42\textwidth]{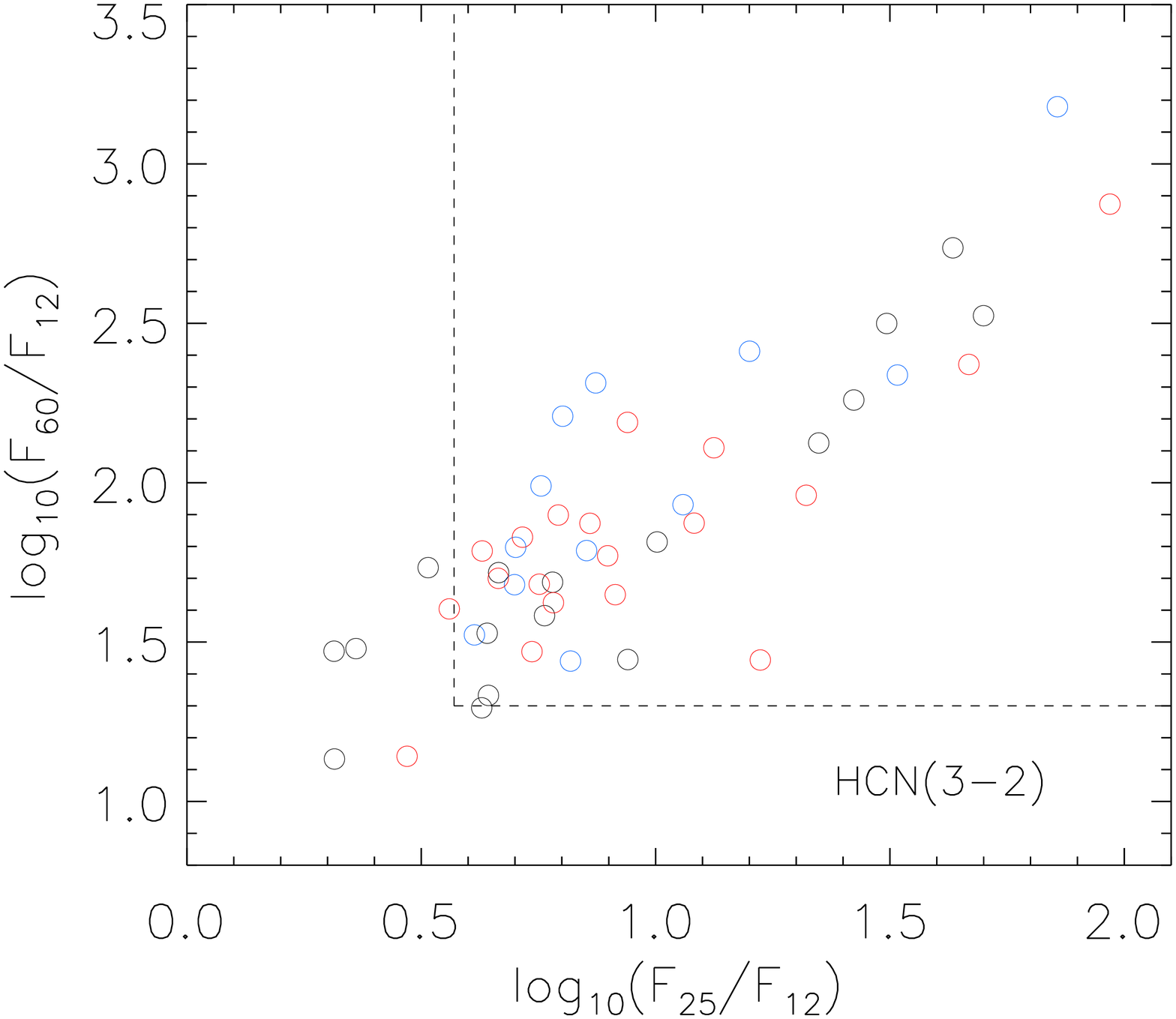}
  \>
  \includegraphics[width=0.42\textwidth]{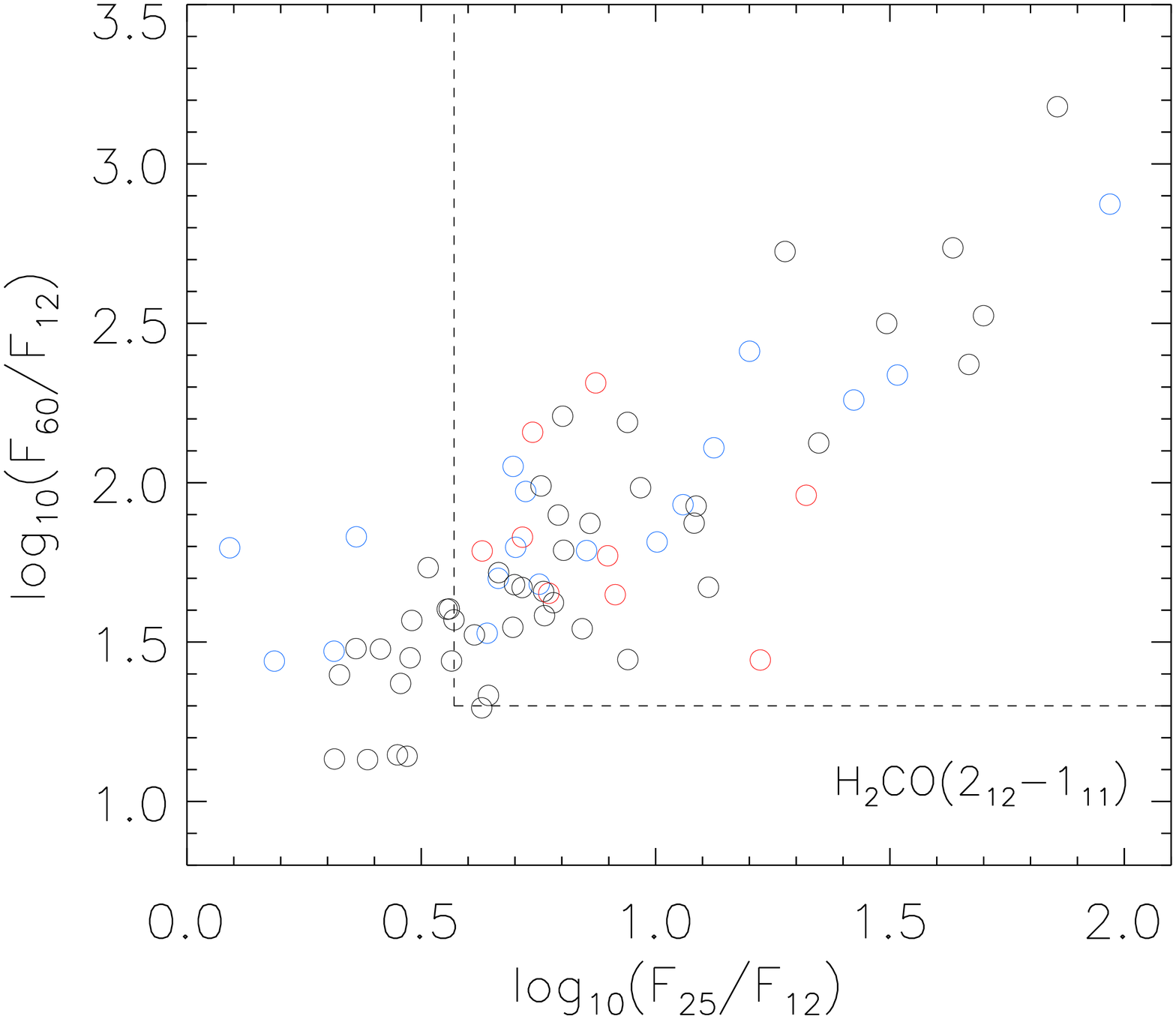} 
  \end{tabbing}
\caption{$IRAS$ Color-color diagrams of [25-12] vs. [60-12] for the optically thick lines. 
The dotted lines are the IRAS color criteria of Wood \& Churchwell (1989) for UCH\,{\scriptsize II} candidates ($High$ sources), $[25 - 12]$ $\geqq$ 0.57 and $[60 - 12]$ $\geqq$ 1.30. The blue, red, and black open circles represent blue, red, and neutral profiles in each line, respectively.}
\label{f3}
\end{figure*}

Our sample is divided into two groups ($\it Low$ and $\it High$) in the IRAS color-color diagram of $[25 - 12]$ and $[60 - 12]$, as discussed in Section~\ref{sec2.1}. The sources in the $High$ group are believed to be in relatively more evolved stage than those of the $Low$ group, although the two groups have similar distributions of infrared luminosities \citep{Palla91, Molinari96, Molinari98}. 
We derive blue excesses separately for the  $Low$ and $High$ groups. 
Table~\ref{tab5} presents the statistical results in all the optically thick transitions (see also Figure~\ref{f2}) and
Figure~\ref{f3} shows the distributions of blue- and red-profile sources in the two groups on the color-color diagram.
In comparison with the $\it High$ group, the $\it Low$ group shows higher blue excesses in the \hcopj\ and \hcoj\ lines but lower excesses in the \hcopjb\ and \hcnj\ lines.
In case of the $\it Low$ group, however, it should be noted that only the \hcopj\ and \hcoj\ lines have statistically sufficient numbers of sources and so the blue excess estimates of the \hcopjb\ and \hcnj\ lines may not be meaningful. 
Thus, further observations for substantially large number of sources are required for the latter transitions to obtain more significant results for evolutionary study of inflow. 

There are very few previous studies investigating the evolutionary effect in the inflow statistics, 
the results of which are inconclusive. \citet{Wu07} reported a higher blue excess for 12 UCHIIs (E=0.58) than 29 HMPOs (E=0.17) in the \hcopj\ transition. \citet{Jin16} also found a similar increasing trend in the HCN (1-0) line for their sample consisting of 26 HMPOs (E=0.15) and 23 UCHIIs (E=0.30), although they obtained the highest blue excess (E=0.42) for 12 high-mass prestellar clumps. On the contrary, \citet{He15} found a decreasing evolutionary trend of blue excess in the \hcopj\ line for a much larger sample: E=0.29 for 84 prestellar clumps, 0.22 for 201 HMPOs, --0.11 for 79 UCHIIs. 
Our result is consistent with that of \citet{He15} although the decreasing trend is less prominent. This difference may be because their sample contains many more evolved UCHIIs than $\it High$ sources in our sample.

\citet{Fuller05} surveyed 77 submillimeter clumps associated with HMPOs in multiple molecular lines for an extensive inflow study toward high-mass star-forming regions.
The blue excesses were derived to be E=0.15 and E=0.19 for the \hcopj\ and \hcoj\ lines,
respectively. When they considered only sources with distances $\le$6~kpc
in order to avoid beam dilution effect due to large distances of the sources, 
the excesses considerably increased and became comparable to the values estimated for low-mass star-forming regions, E $\simeq$ 0.3 \citep{Evans03}. However, we do $not$ find such trend in the same analysis for our sample. For instance, the blue excesses for the \hcopj\ and \hcoj\ lines are E=0.20 and E=0.13 for the entire sample and E=0.12 and E=0.16 for the 65 and 63 sources within 6~kpc, respectively.  This inconsistent result may be caused by large uncertainty in the determination of the distances to the sources. The vast majority of the sources in both samples of ours and \citet{Fuller05} are located in the inner Galaxy, but the distance ambiguity is $not$ resolved for most of them.
           
As noted in Section~\ref{sec3.1}, \citet{WuEvans03} found that the \hcnj\ line shows a significant blue excess (E=0.21) for their sample and thus it is useful for tracing inflow motions toward UCHIIs and compact HII regions. However, we find that the transition is much worse tracer of inflow motions toward HMPOs in our sample than the \hcopj\ line. It is worthwhile to note that their sources are more massive and luminous as well as more evolved than ours, median mass of 8.9$\times$10$^2$~M$_{\odot}$ \& median $L_{\rm IR}$ of 1.06$\times$10$^5$~L$_{\odot}$ \citep{WuEvans03} vs. median $L_{\rm IR}$ of 1.11$\times$10$^4$~L$_{\odot}$. In fact, the critical density of the \hcnj\ transition is two orders of magnitude larger than that of the \hcopj\ transition \citep{Shirley15}. Therefore, we interpret that the \hcopj\ line, which has relatively low critical density, is better suited for investigating inflow motions in our sources than the \hcnj\ line. 
This seems to be consistent with the previous suggestion that the characteristic blue profile will show up only if the critical density of the molecular tracer is suitable for optical depth of the target \citep{Myers96, WuEvans03}.  

Although the blue profile has been widely accepted as an evidence of infall motions, as noted earlier, it is sometimes tricky to interpret the signature due to other kinematics, especially for massive star-forming regions that are much more turbulent and distant than low-mass star-forming regions.
Alternatively, the red-shifted absorption feature can be used to study inflow motions in massive star-forming regions because they are usually bright radio and infrared sources. For example, \cite{Wyrowski12, Wyrowski16}  observed 11 massive molecular clumps in the NH$_{3}$ 3$_{2+}$--2$_{2-}$ line at 1.81~THz using the GREAT instrument onboard SOFIA (beam size = 16$''$). They detected red-shifted NH$_3$ absorption features with respect to the systemic velocities toward 8 sources, and derived the velocity shifts to be 0.3--2.9~\kms. Thier measurements are roughly in agreement with the estimated infall velocities of this study (see Section 3.2). These two kinds of studies utilizing blue profiles and red-shifted absorption features would complement each other.

\section{Conclusions and Summary}
\label{sec:con}

We performed a survey of one optically thin and up to four optically thick molecular lines toward 82 HMPO candidates to understand gravitational collapse in the early stage of high-mass star formation. To quantify asymmetries of the optically thick line profiles, we derived $\delta v$'s of the individual sources in each transition and estimated the blue excess for our sample with $\delta v$=$\pm$0.25 as threshold values. The main results are summarized as follows.

1. We obtained a statistically significant blue excess in the \hcopj\ line (E=0.20), but non-significant excesses in the \hcopjb\ and the \hcoj\ lines (E=0.09 and E=0.13, respectively). The \hcnj\ line shows a negative blue excess of E=$-$0.15. 
The \hcopj\ line thus seems to be the suitable tracer of inflow motions in high-mass star-forming regions as some previous studies proposed \citep*[e.g.][]{Fuller05, Wu07}.
On the contrary, the other lines do not appear to have suitable opacity and critical density for the appearance of blue profile toward most sources in our sample, and may be affected by dynamics other than inflow, such as outflow, rotation, turbulent motions. 

2. We found 27 inflow candidates by adopting the criteria of \citet{Fuller05}, namely, one or more blue profiles and no red profile. All of them are newly identified inflow candidates except one (IRAS05490+2658), which had been classified by \citet{Fuller05} as an inflow candidate.
We derived inflow velocities for 20 out of the 27 candidates using the two-layer radiative transfer model of \citet{Myers96}. The estimated inflow velocities range between 0.23 and 2.00~\kms\ with a median value of 0.49~\kms. On the other hand, there are 18 sources that have red profile(s) but no blue profile. They  might be more affected by dynamics other than inflow motion in the observed optically thick lines.

3. The sources in our sample are all HMPO candidates but they are known to be divided into two different evolutionary stages: $\it Low$ and $\it High$ groups.
The statistical results for the $\it Low$ group show that blue excesses in the \hcopj\ and \hcoj\ lines, of which the number of both groups are statistically meaningful, are slightly higher than those for $\it High$ groups. We also estimated blue excesses for a subsample of sources at relatively small ($\le$6~kpc) distances and obtained less significant excesses than those for the entire sample. This is not consistent with the result of \citet{Fuller05}. This discrepancy may be caused by large uncertainties in determining the distances to the sources in our and their samples.

\acknowledgments

We are grateful to all the staff members at KVN who helped to operate the telescope. The KVN is a facility operated by the Korea Astronomy and Space Science Institute. HY and JC's work is supported by the National R \& D Program through
the National Research Foundation of Korea (NRF), funded by the
Ministry of Education (NRF-2016R1D1A1B02015014).





\startlongtable
\begin{deluxetable}{cccccc}
\tablecaption{Source Information
\label{tab1}}
\tablecolumns{6} 
\tablewidth{0pc} 
\tablehead{
\colhead{IRAS}& \colhead{R.A.} & \colhead{Decl.} & \colhead{$d_{\rm kin}$} & \colhead{$L_{\rm IR}$}& \colhead{} \\
 \colhead{Name} & \colhead{(J2000)} & \colhead{(J2000)} & \colhead{(kpc)} & \colhead{($L_{\odot}$)}& \colhead{Group\tablenotemark{a}} }
\startdata
 $00117+6412$ & 00 14 27.7 & +64 28 46 & 1.80 & 1.38E+03 & H   \\
 $00420+5530$ & 00 44 57.6 & +55 47 18 & 7.72 & 5.15E+04 & L    \\
 $04579+4703$ & 05 01 39.7 & +47 07 23 & 2.47 & 3.91E+03 & H   \\
 $05137+3919$ & 05 17 13.3 & +39 22 23 & 10.8 & 5.61E+04 & L    \\
 $05168+3634$ & 05 20 16.2 & +36 37 21 & 6.08 & 2.40E+04 & H   \\
 $05274+3345$ & 05 30 45.6 & +33 47 52 & 1.55 & 4.53E+03 & H   \\
 $05345+3157$ & 05 37 47.8 & +31 59 24 & 1.80 & 1.38E+03 & L    \\
 $05358+3543$\tablenotemark{b} & 05 39 10.4 & +35 45 19 & 1.80 & 6.31E+03 & H   \\	
 $05373+2349$ & 05 40 24.4 & +23 50 54 & 1.17 & 6.64E+02 & L    \\
 $05391-0152$\tablenotemark{c} & 05 41 38.7  &  -01 51 19 & 0.50 & 1.96E+03 & H   \\  
 $05393-0156$\tablenotemark{c} & 05 41 49.5  &  -01 55 17 & 0.50 & 1.10E+04 & H   \\ 
 $05490+2658$\tablenotemark{b} & 05 52 13.0 & +26 59 34 & 2.10 & 3.16E+03 & H   \\
 $05553+1631$ & 05 58 13.9 & +16 32 00 & 3.04 & 1.17E+04 & H	\\
 $06053-0622$\tablenotemark{c}  & 06 07 46.7 &  -06 23 00  & 0.80 & 2.90E+04 & H    \\	
 $06056+2131$ & 06 08 41.0 & +21 31 01 & 1.50 & 5.83E+03 & H   \\
 $06061+2151$ & 06 09 07.8 & +21 50 39 & 0.10 & 2.78E+01 & H   \\
 $06084-0611$\tablenotemark{c}  &  06 10 51.0 &  -06 11 54 & 1.00 & 9.60E+03 & H   \\	
 $06103+1523$ & 06 13 15.1 & +15 22 36 & 4.63 & 1.91E+04 & H   \\
 $06105+1756$ & 06 13 28.3 & +17 55 33 & 3.38 & 1.60E+04 & H   \\
 $06382+0939$ & 06 41 02.7 & +09 36 10 & 0.76 & 1.63E+02 & L    \\
 $06584-0852$ & 07 00 51.5 & -08 56 29  & 4.48 & 9.08E+03 & L    \\
 $07299-1651$\tablenotemark{c}  & 07 32 10.0 & -16 58 15  & 1.40 & 6.30E+03 & H   \\	
 $07427-2400$\tablenotemark{c}  & 07 44 51.9 & -24 07 41  & 6.40 & 5.00E+04 & H   \\	
 $17417-2851$ & 17 44 53.4 & -28 52 20  & 0.10 & 3.17E+01 & H   \\
 $17450-2742$ & 17 48 09.3 & -27 43 21  & 0.10 & 1.57E+01 & L    \\
 $17504-2519$ & 17 53 35.2 & -25 19 56  & 3.65 & 9.32E+03 & H   \\
 $17527-2439$ & 17 55 49.1 & -24 40 20  & 3.23 & 1.53E+04 & H	\\
 $18014-2428$ & 18 04 29.3 & -24 28 47  & 2.87 & 1.71E+04 & L    \\
 $18018-2426$ & 18 04 53.9 & -24 26 41  & 1.50 & 6.64E+03 & L    \\
 $18024-2119$ & 18 05 25.4 & -21 19 41  & 0.12 & 1.08E+01 & L	\\
 $18089-1732$ & 18 11 51.3 & -17 31 28  & 3.48 & 6.33E+04 & H	\\
 $18134-1942$ & 18 16 22.3 & -19 41 20  & 1.62 & 7.62E+03 & H   \\
 $18144-1723$ & 18 17 24.4 & -17 22 13  & 4.33 & 1.32E+04 & L	\\
 $18151-1208$ & 18 17 57.1 & -12 07 22  & 3.04 & 3.32E+04 & H   \\
 $18159-1550$ & 18 18 47.6 & -15 48 54  & 4.66 & 3.10E+04 & H   \\
 $18159-1648$ & 18 18 53.5 & -16 47 39  & 2.50 & 2.95E+04 & H	\\
 $18162-1612$ & 18 19 07.5 & -16 11 21  & 4.89 & 2.94E+04 & L    \\
 $18236-1205$ & 18 26 24.3 & -12 03 47 & 2.51 & 1.04E+04 & H   \\
 $18256-0742$ & 18 28 20.5 & -07 40 22 & 2.90 & 1.11E+04 & L    \\
 $18258-0737$ & 18 28 34.1 & -07 35 31 & 2.97 & 3.31E+04 & H   \\
 $18316-0602$ & 18 34 19.8 & -05 59 44 & 3.17 & 4.14E+04 & H   \\
 $18317-0513$ & 18 34 25.9 & -05 10 59 & 3.13 & 3.48E+04 & H   \\
 $18360-0537$ & 18 38 40.3 & -05 35 06 & 6.28 & 1.16E+05 & H	\\
 $18372-0541$ & 18 39 56.0 & -05 38 49 & 1.87 & 7.18E+03 & H   \\
 $18396-0431$ & 18 42 18.8 & -04 28 37 & 6.08 & 4.23E+04 & L    \\
 $18488+0000$ & 18 51 24.8 & +00 04 19 & 5.48 & 5.14E+04 & H   \\
 $18507+0121$ & 18 53 17.4 & +01 24 55 & 3.87 & 4.84E+04 & H   \\
 $18511+0146$ & 18 53 38.1 & +01 50 27& 3.86 & 2.01E+04 & L    \\
 $18527+0301$ & 18 55 16.5 & +03 05 07 & 5.26 & 1.63E+04 & L	\\
 $18532+0047$ & 18 55 50.6 & +00 51 22 & 3.96 & 1.27E+04 & H   \\
 $18565+0349$ & 18 59 03.4 & +03 53 22 & 6.77 & 2.62E+04 & L    \\
 $18566+0408$ & 18 59 09.9 & +04 12 14 & 6.76 & 1.02E+05 & H   \\
 $18567+0700$ & 18 59 13.6 & +07 04 47 & 2.16 & 8.39E+03 & L    \\
 $19045+0518$ & 19 06 59.3 & +05 22 55 & 3.80 & 1.09E+04 & H   \\
 $19088+0902$ & 19 11 15.9 & +09 07 27 & 4.71 & 2.99E+04 & H   \\
 $19282+1814$ & 19 30 28.1 & +18 20 53 & 2.11 & 1.63E+04 & H	\\
 $19368+2239$ & 19 38 58.1 & +22 46 32 & 4.44 & 8.63E+03 & H   \\
 $20050+2720$ & 20 07 06.7 & +27 28 53 & 0.73 & 3.88E+02 & H   \\
 $20056+3350$ & 20 07 31.5 & +33 59 39 & 1.67 & 4.00E+03 & H   \\
 $20062+3550$ & 20 08 09.8 & +35 59 20 & 0.08 & 5.40E+00 & H   \\
 $20106+3545$ & 20 12 31.3 & +35 54 46 & 1.64 & 1.79E+03 & L   \\
 $20126+4104$ & 20 14 26.0 & +41 13 32 & 4.18 & 7.12E+04 & H	\\
 $20188+3928$ & 20 20 39.3 & +39 37 52 & 0.31 & 3.43E+02 & H   \\
 $20220+3728$ & 20 23 55.7 & +37 38 10 & 4.49 & 8.09E+04 & H   \\
 $20227+4154$ & 20 24 31.4 & +42 04 17 & 0.10 & 9.14E+00 & H   \\
 $20278+3521$ & 20 29 46.9 & +35 31 39 & 5.02 & 1.08E+04 & L    \\
 $20286+4105$ & 20 30 27.9 & +41 15 48 & 3.72 & 3.90E+04 & H   \\
 $20333+4102$ & 20 35 09.5 & +41 13 18 & 0.10 & 4.57E+01 & L    \\
 $21078+5211$ & 21 09 25.2 & +52 23 44 & 1.49 & 1.34E+04 & H   \\
 $21391+5802$ & 21 40 42.4 & +58 16 10 & 0.75 & 9.39E+01 & H   \\
 $21519+5613$ & 21 53 39.2 & +56 27 46 & 7.30 & 1.91E+04 & H   \\
 $22172+5549$ & 22 19 09.0 & +56 04 45 & 2.87 & 4.78E+03 & L    \\
 $22198+6336$ & 22 21 27.6 & +63 51 42 & 1.28 & 1.24E+03 & H	\\
 $22267+6244$ & 22 28 29.3 & +62 59 44 & 0.45 & 1.10E+02 & H   \\
 $22272+6358$ & 22 28 52.2 & +64 13 44 & 1.23 & 1.97E+03 & H	\\
 $22308+5812$\tablenotemark{d}  & 22 32 46.1 & +58 28 22 & 3.70 & 2.09E+04 & H   \\		
 $22506+5944$ & 22 52 38.6 & +60 00 56 & 5.70 & 2.22E+04 & H   \\
 $23026+5948$ & 23 04 45.7 & +60 04 35 & 5.76 & 1.76E+04 & L    \\
 $23133+6050$\tablenotemark{d}  & 23 15 31.5 & +61 07 08  & 5.20 & 1.20E+05 & H  \\		
 $23140+6121$ & 23 16 11.7 & +61 37 45 & 6.44 & 4.35E+04 & L    \\	
 $23314+6033$ & 23 33 44.4 & +60 50 30 & 2.78 & 1.09E+04 & L    \\		
 $23545+6508$ & 23 57 05.2 & +65 25 11 & 1.27 & 3.89E+03 & H   \\	
\enddata
\tablenotetext{a}{H and L are abbreviations of the $\it High$ and $\it Low$ groups, respectively. }
\tablenotetext{b}{From \citet{Sridharan02}}
\tablenotetext{c}{From \citet{Walsh98}}
\tablenotetext{d}{From \citet{Hunter00}}
\end{deluxetable}

\clearpage


\begin{deluxetable}{lrcclcc}
\tablecaption{Summary of the Observations
\label{tab2}}
\tablewidth{0pt}
\tablehead{
\colhead{Observed} & \colhead{Frequency} & & \colhead{$\theta_{\rm mb}$} &  
&\colhead{$\vartriangle \!\!v$} & \colhead{Number} \\
\colhead{Transition} & \colhead{(GHz)} &  \colhead{Telescope}  & 
\colhead{($''$)} &  \colhead{$\eta_{mb}$} &\colhead{(km s$^{-1}$)} & \colhead{of Sources}  
 }
\startdata
\hcopj\ & 89.188526 	& ARO 12~m 	& 70 & 0.95$^a$ & 0.34 & 60   \\   
	    &    		& KVN 21~m 	& 32 & 0.38 	& 0.21 & 14   \\  
\hcopjb\ & 267.557633 & CSO 10.4~m	& 27 & 0.61	& 0.22 & 11   \\ 
	      & 		& SMT 10~m       & 28 & 0.78 	& 0.28 & 32  \\ 
\hcnj\  & 265.886431	& CSO 10.4~m    & 27 & 0.61 	& 0.22 & 48  \\
\hcoj\  & 140.839515	& ARO 12~m 	& 44 & 0.80$^a$ & 0.21 & 52  \\
	&  			& KVN 21~m 	& 24 & 0.27 	& 0.27 & 19  \\
\hcoppj\ & 86.754330 & ARO 12~m 	& 72 & 0.95$^a$ & 0.35 & 45   \\
	& 			& KVN 21~m 	& 32 & 0.38 	& 0.22 & 30   \\
\enddata
\tablenotetext{a}{Corrected main-beam efficiency $\eta^{*}_{mb}$ defined as $T_{\rm mb}$=\tr/$\eta^{*}_{mb}$, where $T_{\rm mb}$~is the main-beam brightness temperature and \tr~is the corrected radiation temperature (see \citet{Mangum00}).}

\end{deluxetable}


\startlongtable
\begin{longrotatetable}
\begin{deluxetable}{l ccc c cccc c cccc c cccc c c}
\tablecaption{Line Velocities and Asymmetry Parameters
\label{tab3}}
\tabletypesize{\tiny}
\tablecolumns{21} 
\tablewidth{0pt} 
\tablehead{
\colhead{IRAS}  & 
\multicolumn{3}{c}{H$^{13}$CO$^+$ (1--0)} &    
\colhead{}  & 
\multicolumn{4}{c}{$v_{thick}$$^{a}$}
&  
\colhead{}  & 
\multicolumn{4}{c}{$\delta v$} & 
\colhead{}  & 
\multicolumn{4}{c}{Profile} &
\colhead{} &
\multicolumn{1}{c}{} 
\\ \cline{2-4} \cline{6-9} \cline{11-14} \cline{16-19} 
\colhead{Name} & 
\colhead{$\tau_{\rm thin}$$^{b}$}
&\colhead{$v$  $(v_{\rm err})$$^{a}$}
&\colhead{$\vartriangle \!\!v$ ($\vartriangle \!\!v_{\rm err}$)$^{a}$}
&
\colhead{}  & 
\colhead{L1$^{c}$}
&
\colhead{L2$^{c}$}
&
\colhead{L3$^{c}$}
&
\colhead{L4$^{c}$}
&
\colhead{}  & 
\colhead{L1} &
\colhead{L2} &
\colhead{L3} &
\colhead{L4} &
\colhead{}  & 
\colhead{L1} &
\colhead{L2} &
\colhead{L3} &
\colhead{L4} &
\colhead{} &
\colhead{$v_{\rm in}^{a}$}
}
\startdata
$	00117+6412		$	&	0.09	&	-36.21	(	0.11	)	&	2.32	(	0.30	)	&	&	-35.46	&	-36.44	&	-35.63	&	-36.20	&	&	0.32	&	-0.10	&	0.25	&	0.00	&	&	R	&	N	&	N	&	N	&	&	\nodata	\\
$	00420+5530	^*	$	&	0.05	&	-51.44	(	0.21	)	&	1.41	(	0.55	)	&	&	-52.04	&	-51.34	&	-51.71	&	-51.80	&	&	-0.43	&	0.07	&	-0.20	&	-0.26	&	&	B	&	N	&	N	&	B	&	&	0.50	\\
$	04579+4703	^*	$	&	0.18	&	-16.70	(	0.08	)	&	1.43	(	0.20	)	&	&	-16.67	&	-16.55	&	-17.32	&	\nodata	&	&	0.02	&	0.11	&	-0.44	&	\nodata	&	&	N	&	N	&	B	&	\nodata	&	&	\nodata	\\
$	05137+3919	^*	$	&	0.05	&	-25.32	(	0.16	)	&	1.46	(	0.31	)	&	&	-26.58	&	-25.20	&	-25.28	&	-25.27	&	&	-0.86	&	0.08	&	0.03	&	0.04	&	&	B	&	N	&	N	&	N	&	&	0.35	\\
$	05168+3634		$	&	0.24	&	-15.15	(	0.06	)	&	1.18	(	0.18	)	&	&	-15.68	&	\nodata	&	\nodata	&	-14.73	&	&	-0.45	&	\nodata	&	\nodata	&	0.36	&	&	B	&	\nodata	&	\nodata	&	R	&	&	\nodata	\\
$	05274+3345	^*	$	&	0.21	&	-3.36	(	0.04	)	&	2.49	(	0.08	)	&	&	-4.47	&	-3.77	&	-3.63	&	-4.00	&	&	-0.45	&	-0.16	&	-0.11	&	-0.26	&	&	B	&	N	&	N	&	B	&	&	0.48	\\
$	05345+3157	^*	$	&	\nodata	&	-18.16	(	0.17	)	&	1.99	(	0.33	)	&	&	\nodata	&	\nodata	&	\nodata	&	-18.80	&	&	\nodata	&	\nodata	&	\nodata	&	-0.32	&	&	\nodata	&	\nodata	&	\nodata	&	B	&	&	\nodata	\\
$	05358+3543		$	&	0.07	&	-17.31	(	0.04	)	&	2.38	(	0.10	)	&	&	-17.84	&	-14.96	&	-14.21	&	-18.00	&	&	-0.22	&	0.99	&	1.30	&	-0.29	&	&	N	&	R	&	R	&	B	&	&	\nodata	\\
$	05373+2349		$	&	0.32	&	2.23	(	0.03	)	&	1.66	(	0.08	)	&	&	1.95	&	\nodata	&	2.79	&	2.00	&	&	-0.17	&	\nodata	&	0.34	&	-0.14	&	&	N	&	\nodata	&	R	&	N	&	&	\nodata	\\
$	05391-0152		$	&	0.16	&	9.91	(	0.04	)	&	1.21	(	0.10	)	&	&	9.16	&	10.89	&	11.03	&	10.53	&	&	-0.62	&	0.81	&	0.93	&	0.52	&	&	B	&	R	&	R	&	R	&	&	\nodata	\\
$	05393-0156		$	&	0.09	&	10.13	(	0.07	)	&	2.06	(	0.16	)	&	&	9.16	&	11.10	&	11.03	&	10.96	&	&	-0.47	&	0.47	&	0.44	&	0.40	&	&	B	&	R	&	R	&	R	&	&	\nodata	\\
$	05490+2658	^*	$	&	0.13	&	0.55	(	0.07	)	&	1.24	(	0.19	)	&	&	0.11	&	\nodata	&	0.33	&	0.73	&	&	-0.36	&	\nodata	&	-0.18	&	0.15	&	&	B	&	\nodata	&	N	&	N	&	&	0.50	\\
$	05553+1631		$	&	0.12	&	5.65	(	0.05	)	&	1.90	(	0.12	)	&	&	6.37	&	\nodata	&	5.29	&	5.47	&	&	0.38	&	\nodata	&	-0.19	&	-0.10	&	&	R	&	\nodata	&	N	&	N	&	&	\nodata	\\
$	06053-0622		$	&	0.07	&	10.49	(	0.08	)	&	2.64	(	0.22	)	&	&	9.83	&	10.45	&	10.59	&	9.89	&	&	-0.25	&	-0.01	&	0.04	&	-0.23	&	&	N	&	N	&	N	&	N	&	&	\nodata	\\
$	06056+2131		$	&	0.10	&	2.54	(	0.03	)	&	2.22	(	0.06	)	&	&	2.53	&	2.81	&	\nodata	&	2.73	&	&	0.00	&	0.12	&	\nodata	&	0.09	&	&	N	&	N	&	\nodata	&	N	&	&	\nodata	\\
$	06061+2151		$	&	0.13	&	-0.88	(	0.08	)	&	2.02	(	0.19	)	&	&	-1.05	&	\nodata	&	-0.02	&	-1.00	&	&	-0.08	&	\nodata	&	0.43	&	-0.06	&	&	N	&	\nodata	&	R	&	N	&	&	\nodata	\\
$	06084-0611		$	&	0.13	&	11.59	(	0.06	)	&	1.98	(	0.14	)	&	&	11.85	&	11.10	&	11.25	&	11.60	&	&	0.13	&	-0.25	&	-0.17	&	0.00	&	&	N	&	N	&	N	&	N	&	&	\nodata	\\
$	06103+1523		$	&	0.09	&	15.57	(	0.05	)	&	1.87	(	0.11	)	&	&	16.37	&	\nodata	&	16.76	&	16.27	&	&	0.43	&	\nodata	&	0.64	&	0.37	&	&	R	&	\nodata	&	R	&	R	&	&	\nodata	\\
$	06105+1756	^*	$	&	0.04	&	7.78	(	0.14	)	&	1.32	(	0.36	)	&	&	7.24	&	\nodata	&	7.46	&	\nodata	&	&	-0.41	&	\nodata	&	-0.24	&	\nodata	&	&	B	&	\nodata	&	N	&	\nodata	&	&	0.28	\\
$	06382+0939		$	&	0.33	&	5.22	(	0.02	)	&	2.13	(	0.06	)	&	&	4.70	&	\nodata	&	4.94	&	5.31	&	&	-0.24	&	\nodata	&	-0.13	&	0.04	&	&	N	&	\nodata	&	N	&	N	&	&	\nodata	\\
$	06584-0852		$	&	\nodata	&	\nodata	(\nodata	)	&	\nodata	(\nodata	)	&	&	41.13	&	\nodata	&	\nodata	&	\nodata	&	&	\nodata	&	\nodata	&	\nodata	&	\nodata	&	&	\nodata	&	\nodata	&	\nodata	&	\nodata	&	&	\nodata	\\
$	07299-1651		$	&	0.08	&	17.18	(	0.11	)	&	1.78	(	0.30	)	&	&	17.48	&	17.22	&	17.77	&	17.34	&	&	0.17	&	0.02	&	0.34	&	0.09	&	&	N	&	N	&	R	&	N	&	&	\nodata	\\
$	07427-2400	^*	$	&	0.12	&	67.82	(	0.25	)	&	4.88	(	0.58	)	&	&	66.13	&	67.22	&	69.06	&	68.67	&	&	-0.35	&	-0.12	&	0.25	&	0.17	&	&	B	&	N	&	N	&	N	&	&	0.42	\\
$	17417-2851	^*	$	&	0.08	&	-5.33	(	0.17	)	&	2.77	(	0.42	)	&	&	-6.44	&	-5.46	&	-6.30	&	-6.13	&	&	-0.40	&	-0.05	&	-0.35	&	-0.29	&	&	B	&	N	&	B	&	B	&	&	0.68	\\
$	17450-2742		$	&	\nodata	&	\nodata	(\nodata	)	&	\nodata	(\nodata	)	&	&	-16.06	&	-15.64	&	\nodata	&	-15.73	&	&	\nodata	&	\nodata	&	\nodata	&	\nodata	&	&	\nodata	&	\nodata	&	\nodata	&	\nodata	&	&	\nodata	\\
$	17504-2519	^*	$	&	0.18	&	12.65	(	0.06	)	&	1.38	(	0.13	)	&	&	11.56	&	11.70	&	12.97	&	12.08	&	&	-0.79	&	-0.69	&	0.23	&	-0.42	&	&	B	&	B	&	N	&	B	&	&	1.26	\\
$	17527-2439		$	&	0.17	&	13.52	(	0.13	)	&	2.21	(	0.35	)	&	&	14.04	&	\nodata	&	\nodata	&	13.95	&	&	0.24	&	\nodata	&	\nodata	&	0.19	&	&	N	&	\nodata	&	\nodata	&	N	&	&	\nodata	\\
$	18014-2428		$	&	\nodata	&	\nodata	(\nodata	)	&	\nodata	(\nodata	)	&	&	12.67	&	12.64	&	\nodata	&	12.39	&	&	\nodata	&	\nodata	&	\nodata	&	\nodata	&	&	\nodata	&	\nodata	&	\nodata	&	\nodata	&	&	\nodata	\\
$	18018-2426		$	&	0.07	&	10.92	(	0.08	)	&	1.92	(	0.19	)	&	&	10.67	&	11.48	&	\nodata	&	10.82	&	&	-0.13	&	0.29	&	\nodata	&	-0.05	&	&	N	&	R	&	\nodata	&	N	&	&	\nodata	\\
$	18024-2119		$	&	\nodata	&	0.39	(	0.05	)	&	2.14	(	0.11	)	&	&	\nodata	&	\nodata	&	\nodata	&	\nodata	&	&	\nodata	&	\nodata	&	\nodata	&	\nodata	&	&	\nodata	&	\nodata	&	\nodata	&	\nodata	&	&	\nodata	\\
$	18089-1732		$	&	0.21	&	33.41	(	0.13	)	&	3.40	(	0.30	)	&	&	35.43	&	\nodata	&	29.75	&	34.35	&	&	0.59	&	\nodata	&	-1.08	&	0.28	&	&	R	&	\nodata	&	B	&	R	&	&	\nodata	\\
$	18134-1942		$	&	0.19	&	10.38	(	0.05	)	&	1.70	(	0.11	)	&	&	11.00	&	10.92	&	11.10	&	10.61	&	&	0.37	&	0.32	&	0.43	&	0.14	&	&	R	&	R	&	R	&	N	&	&	\nodata	\\
$	18144-1723		$	&	0.08	&	48.23	(	0.16	)	&	3.00	(	0.37	)	&	&	48.81	&	48.84	&	50.57	&	48.47	&	&	0.19	&	0.20	&	0.78	&	0.08	&	&	N	&	N	&	R	&	N	&	&	\nodata	\\
$	18151-1208		$	&	0.06	&	32.96	(	0.16	)	&	2.96	(	0.42	)	&	&	32.63	&	32.94	&	\nodata	&	32.69	&	&	-0.11	&	-0.01	&	\nodata	&	-0.09	&	&	N	&	N	&		&	N	&	&	\nodata	\\
$	18159-1550		$	&	0.14	&	59.21	(	0.12	)	&	2.86	(	0.31	)	&	&	60.88	&	59.56	&	60.58	&	60.23	&	&	0.58	&	0.12	&	0.48	&	0.36	&	&	R	&	N	&	R	&	R	&	&	\nodata	\\
$	18159-1648		$	&	0.24	&	22.38	(	0.07	)	&	3.43	(	0.15	)	&	&	24.62	&	21.68	&	25.39	&	23.06	&	&	0.65	&	-0.20	&	0.88	&	0.20	&	&	R	&	N	&	R	&	N	&	&	\nodata	\\
$	18162-1612		$	&	0.06	&	61.85	(	0.14	)	&	1.73	(	0.31	)	&	&	61.63	&	\nodata	&	\nodata	&	61.69	&	&	-0.12	&	\nodata	&	\nodata	&	-0.09	&	&	N	&	\nodata	&	\nodata	&	N	&	&	\nodata	\\
$	18236-1205	^*	$	&	0.24	&	26.90	(	0.12	)	&	3.20	(	0.26	)	&	&	25.02	&	\nodata	&	\nodata	&	25.67	&	&	-0.59	&	\nodata	&	\nodata	&	-0.38	&	&	B	&	\nodata	&	\nodata	&	B	&	&	1.65	\\
$	18256-0742	^*	$	&	0.15	&	36.82	(	0.10	)	&	1.27	(	0.21	)	&	&	36.20	&	\nodata	&	\nodata	&	35.96	&	&	-0.49	&	\nodata	&	\nodata	&	-0.68	&	&	B	&	\nodata	&	\nodata	&	B	&	&	0.40	\\
$	18258-0737		$	&	0.05	&	37.27	(	0.40	)	&	2.93	(	1.03	)	&	&	38.40	&	\nodata	&	38.78	&	38.43	&	&	0.39	&	\nodata	&	0.52	&	0.40	&	&	R	&	\nodata	&	R	&	R	&	&	\nodata	\\
$	18316-0602		$	&	0.26	&	42.46	(	0.07	)	&	3.38	(	0.20	)	&	&	41.36	&	42.06	&	44.71	&	41.88	&	&	-0.32	&	-0.12	&	0.67	&	-0.17	&	&	B	&	N	&	R	&	N	&	&	\nodata	\\
$	18317-0513		$	&	0.08	&	42.27	(	0.08	)	&	1.33	(	0.22	)	&	&	42.03	&	42.06	&	\nodata	&	42.09	&	&	-0.18	&	-0.16	&	\nodata	&	-0.13	&	&	N	&	N	&		&	N	&	&	\nodata	\\
$	18360-0537	^*	$	&	0.10	&	101.71	(	0.71	)	&	4.16	(	0.37	)	&	&	100.45	&	101.04	&	99.71	&	101.13	&	&	-0.30	&	-0.16	&	-0.48	&	-0.14	&	&	B	&	N	&	B	&	N	&	&	2.00	\\
$	18372-0541		$	&	\nodata	&	22.93	(	0.11	)	&	2.70	(	0.25	)	&	&	\nodata	&	\nodata	&	\nodata	&	\nodata	&	&	\nodata	&	\nodata	&	\nodata	&	\nodata	&	&	\nodata	&	\nodata	&	\nodata	&	\nodata	&	&	\nodata	\\
$	18396-0431		$	&	0.14	&	97.22	(	0.08	)	&	1.80	(	0.18	)	&	&	98.48	&	\nodata	&	\nodata	&	\nodata	&	&	0.70	&	\nodata	&	\nodata	&	\nodata	&	&	R	&	\nodata	&	\nodata	&	\nodata	&	&	\nodata	\\
$	18488+0000		$	&	0.33	&	83.15	(	0.06	)	&	2.56	(	0.16	)	&	&	81.48	&	82.78	&	84.84	&	81.67	&	&	-0.65	&	-0.15	&	0.66	&	-0.58	&	&	B	&	N	&	R	&	B	&	&	\nodata	\\
$	18507+0121		$	&	0.36	&	57.46	(	0.07	)	&	4.06	(	0.17	)	&	&	59.96	&	58.64	&	\nodata	&	56.99	&	&	0.62	&	0.29	&	\nodata	&	-0.11	&	&	R	&	R	&	\nodata	&	N	&	&	\nodata	\\
$	18511+0146		$	&	0.18	&	57.70	(	0.13	)	&	2.97	(	0.30	)	&	&	59.99	&	\nodata	&	\nodata	&	57.55	&	&	0.77	&	\nodata	&	\nodata	&	-0.05	&	&	R	&	\nodata	&	\nodata	&	N	&	&	\nodata	\\
$	18527+0301	^*	$	&	0.16	&	76.05	(	0.10	)	&	1.88	(	0.26	)	&	&	75.50	&	\nodata	&	\nodata	&	75.47	&	&	-0.30	&	\nodata	&	\nodata	&	-0.31	&	&	B	&	\nodata	&	\nodata	&	B	&	&	0.80	\\
$	18532+0047		$	&	0.24	&	58.77	(	0.07	)	&	2.50	(	0.16	)	&	&	58.95	&	\nodata	&	\nodata	&	59.27	&	&	0.07	&	\nodata	&	\nodata	&	0.20	&	&	N	&	\nodata	&	\nodata	&	N	&	&	\nodata	\\
$	18565+0349	^*	$	&	0.10	&	91.42	(	0.13	)	&	2.26	(	0.29	)	&	&	90.76	&	91.18	&	\nodata	&	91.49	&	&	-0.29	&	-0.11	&	\nodata	&	0.03	&	&	B	&	N	&	\nodata	&	N	&	&	1.40	\\
$	18566+0408		$	&	0.13	&	85.52	(	0.21	)	&	3.72	(	0.52	)	&	&	82.58	&	\nodata	&	\nodata	&	87.34	&	&	-0.79	&	\nodata	&	\nodata	&	0.49	&	&	B	&	\nodata	&	\nodata	&	R	&	&	\nodata	\\
$	18567+0700		$	&	0.20	&	29.52	(	0.05	)	&	1.20	(	0.14	)	&	&	29.57	&	\nodata	&	\nodata	&	29.29	&	&	0.04	&	\nodata	&	\nodata	&	-0.18	&	&	N	&	\nodata	&	\nodata	&	N	&	&	\nodata	\\
$	19045+0518		$	&	0.10	&	53.72	(	0.10	)	&	1.88	(	0.41	)	&	&	53.43	&	\nodata	&	53.26	&	53.49	&	&	-0.16	&	\nodata	&	-0.24	&	-0.12	&	&	N	&	\nodata	&	N	&	N	&	&	\nodata	\\
$	19088+0902		$	&	0.17	&	59.92	(	0.19	)	&	2.00	(	0.46	)	&	&	61.45	&	\nodata	&	57.73	&	59.28	&	&	0.77	&	\nodata	&	-1.10	&	-0.32	&	&	R	&	\nodata	&	B	&	B	&	&	\nodata	\\
$	19282+1814	^*	$	&	0.11	&	24.18	(	0.09	)	&	0.95	(	0.24	)	&	&	23.60	&	\nodata	&	\nodata	&	23.78	&	&	-0.61	&	\nodata	&	\nodata	&	-0.42	&	&	B	&	\nodata	&	\nodata	&	B	&	&	0.23	\\
$	19368+2239	^*	$	&	0.10	&	36.51	(	0.12	)	&	2.29	(	0.28	)	&	&	36.23	&	35.70	&	35.68	&	36.29	&	&	-0.12	&	-0.36	&	-0.36	&	-0.10	&	&	N	&	B	&	B	&	N	&	&	\nodata	\\
$	20050+2720		$	&	0.24	&	6.10	(	0.07	)	&	1.98	(	0.18	)	&	&	4.22	&	5.42	&	7.20	&	5.23	&	&	-0.95	&	-0.34	&	0.56	&	-0.44	&	&	B	&	B	&	R	&	B	&	&	\nodata	\\
$	20056+3350	^*	$	&	0.09	&	9.33	(	0.17	)	&	2.05	(	0.46	)	&	&	9.57	&	8.42	&	8.12	&	8.87	&	&	0.12	&	-0.44	&	-0.59	&	-0.22	&	&	N	&	B	&	B	&	N	&	&	\nodata	\\
$	20062+3550	^*	$	&	0.11	&	0.81	(	0.14	)	&	1.31	(	0.33	)	&	&	0.77	&	1.02	&	0.88	&	0.28	&	&	-0.03	&	0.16	&	0.06	&	-0.40	&	&	N	&	N	&	N	&	B	&	&	\nodata	\\
$	20106+3545	^*	$	&	0.08	&	8.08	(	0.09	)	&	1.64	(	0.23	)	&	&	7.63	&	\nodata	&	\nodata	&	7.91	&	&	-0.27	&	\nodata	&	\nodata	&	-0.11	&	&	B	&	\nodata	&	\nodata	&	N	&	&	0.25	\\
$	20126+4104		$	&	0.10	&	-3.36	(	0.07	)	&	2.29	(	0.18	)	&	&	-3.40	&	-3.20	&	-3.01	&	-3.58	&	&	-0.01	&	0.07	&	0.16	&	-0.09	&	&	N	&	N	&	N	&	N	&	&	\nodata	\\
$	20188+3928		$	&	0.18	&	1.77	(	0.09	)	&	2.08	(	0.19	)	&	&	2.68	&	2.20	&	2.76	&	2.46	&	&	0.43	&	0.21	&	0.48	&	0.33	&	&	R	&	N	&	R	&	R	&	&	\nodata	\\
$	20220+3728	^*	$	&	0.04	&	-1.89	(	0.23	)	&	3.40	(	0.56	)	&	&	-2.87	&	-3.68	&	-4.20	&	-2.81	&	&	-0.29	&	-0.53	&	-0.68	&	-0.27	&	&	B	&	B	&	B	&	B	&	&	0.47	\\
$	20227+4154		$	&	0.21	&	5.84	(	0.07	)	&	1.43	(	0.15	)	&	&	6.40	&	\nodata	&	\nodata	&	5.58	&	&	0.39	&	\nodata	&	\nodata	&	-0.18	&	&	R	&	\nodata	&	\nodata	&	N	&	&	\nodata	\\
$	20278+3521		$	&	0.05	&	-3.99	(	0.18	)	&	2.71	(	0.44	)	&	&	-4.67	&	\nodata	&	\nodata	&	-4.61	&	&	-0.25	&	\nodata	&	\nodata	&	-0.23	&	&	N	&	\nodata	&	\nodata	&	N	&	&	\nodata	\\
$	20286+4105	^*	$	&	0.06	&	-3.70	(	0.12	)	&	3.89	(	0.31	)	&	&	-4.30	&	-4.78	&	-4.73	&	-4.97	&	&	-0.16	&	-0.28	&	-0.27	&	-0.33	&	&	N	&	B	&	B	&	B	&	&	\nodata	\\
$	20333+4102	^*	$	&	0.07	&	8.64	(	0.05	)	&	1.25	(	0.12	)	&	&	8.57	&	8.26	&	8.57	&	8.51	&	&	-0.05	&	-0.30	&	-0.05	&	-0.10	&	&	N	&	B	&	N	&	N	&	&	\nodata	\\
$	21078+5211		$	&	\nodata	&	\nodata	(\nodata	)	&	\nodata	(\nodata	)	&	&	-6.60	&	-6.80	&	-6.88	&	-6.42	&	&	\nodata	&	\nodata	&	\nodata	&	\nodata	&	&	\nodata	&	\nodata	&	\nodata	&	\nodata	&	&	\nodata	\\
$	21391+5802	^*	$	&	0.27	&	0.65	(	0.04	)	&	1.97	(	0.11	)	&	&	-0.44	&	-0.30	&	-0.16	&	0.08	&	&	-0.55	&	-0.48	&	-0.41	&	-0.29	&	&	B	&	B	&	B	&	B	&	&	1.10	\\
$	21519+5613		$	&	\nodata	&	\nodata	(\nodata	)	&	\nodata	(\nodata	)	&	&	-62.70	&	-63.06	&	-62.68	&	-62.88	&	&	\nodata	&	\nodata	&	\nodata	&	\nodata	&	&	\nodata	&	\nodata	&	\nodata	&	\nodata	&	&	\nodata	\\
$	22172+5549		$	&	0.09	&	-43.56	(	0.10	)	&	2.78	(	0.28	)	&	&	-43.63	&	-43.38	&	-43.15	&	-43.48	&	&	-0.03	&	0.06	&	0.15	&	0.03	&	&	N	&	N	&	N	&	N	&	&	\nodata	\\
$	22198+6336		$	&	0.11	&	-11.00	(	0.13	)	&	1.48	(	0.25	)	&	&	-11.60	&	-11.52	&	-9.93	&	-11.42	&	&	-0.41	&	-0.35	&	0.72	&	-0.28	&	&	B	&	B	&	R	&	B	&	&	\nodata	\\
$	22267+6244		$	&	0.16	&	-1.32	(	0.07	)	&	1.51	(	0.15	)	&	&	-2.00	&	-1.92	&	0.07	&	-1.39	&	&	-0.45	&	-0.40	&	0.92	&	-0.05	&	&	B	&	B	&	R	&	N	&	&	\nodata	\\
$	22272+6358	^*	$	&	0.14	&	-9.95	(	0.06	)	&	1.11	(	0.13	)	&	&	-10.40	&	-10.88	&	-10.54	&	-10.01	&	&	-0.41	&	-0.84	&	-0.53	&	-0.05	&	&	B	&	B	&	B	&	N	&	&	0.47	\\
$	22308+5812		$	&	0.04	&	-52.29	(	0.23	)	&	5.08	(	0.66	)	&	&	-53.41	&	\nodata	&	\nodata	&	-52.93	&	&	-0.22	&	\nodata	&	\nodata	&	-0.13	&	&	N	&	\nodata	&	\nodata	&	N	&	&	\nodata	\\
$	22506+5944		$	&	0.19	&	-51.46	(	0.05	)	&	2.08	(	0.14	)	&	&	-51.33	&	-50.52	&	-49.51	&	\nodata	&	&	0.06	&	0.45	&	0.94	&	\nodata	&	&	N	&	R	&	R	&	\nodata	&	&	\nodata	\\
$	23026+5948		$	&	0.04	&	-51.10	(	0.12	)	&	1.77	(	0.29	)	&	&	-51.27	&	\nodata	&	\nodata	&	-51.42	&	&	-0.10	&	\nodata	&	\nodata	&	-0.18	&	&	N	&	\nodata	&	\nodata	&	N	&	&	\nodata	\\
$	23133+6050		$	&	0.04	&	-56.55	(	0.08	)	&	2.61	(	0.23	)	&	&	-56.48	&	\nodata	&	\nodata	&	-56.27	&	&	0.03	&	\nodata	&	\nodata	&	0.11	&	&	N	&	\nodata	&	\nodata	&	N	&	&	\nodata	\\
$	23140+6121	^*	$	&	0.15	&	-51.27	(	0.08	)	&	1.77	(	0.17	)	&	&	-52.00	&	\nodata	&	\nodata	&	-51.61	&	&	-0.41	&	\nodata	&	\nodata	&	-0.19	&	&	B	&	\nodata	&	\nodata	&	N	&	&	1.47	\\
$	23314+6033		$	&	0.06	&	-45.24	(	0.14	)	&	2.41	(	0.40	)	&	&	-45.23	&	\nodata	&	\nodata	&	-45.53	&	&	0.00	&	\nodata	&	\nodata	&	-0.12	&	&	N	&	\nodata	&	\nodata	&	N	&	&	\nodata	\\
$	23545+6508	^*	$	&	0.14	&	-18.47	(	0.06	)	&	1.15	(	0.16	)	&	&	-19.58	&	-18.54	&	-18.76	&	-18.72	&	&	-0.96	&	-0.06	&	-0.26	&	-0.22	&	&	B	&	N	&	B	&	N	&	&	0.23	\\
\enddata
\tabletypesize{\small}
\tablenotetext{a}{In units of km s$^{-1}$}
\tablenotetext{b}{Optical depth for the H$^{13}$CO$^+$ (1-0) line.}
\tablenotetext{c}{L1: HCO$^+$ (1--0), L2: HCO$^+$ (3--2), L3: HCN (3--2), L4: H$_{2}$CO (2$_{12}$--1$_{11}$)}

\end{deluxetable}
\end{longrotatetable}



\begin{deluxetable}{lccccccc}
\tablecaption{Blue Excess Statistics for the Entire Sample
\label{tab4}}
\tablewidth{0pt}
\tablehead{\colhead{Transition} & \colhead{N$_{Blue}$} & \colhead{N$_{Red}$} & \colhead{N$_{Total}$} & \colhead{D$_{Blue}$} & \colhead{D$_{Red}$} & 
		\colhead{E} & \colhead{P} }
\startdata
HCO$^+$ (1--0)					&	29	&	14	&	74	&	0.39	&	0.19	&	0.20		& 0.016	\\
HCO$^+$ (3--2)					&	11	&	7	&	43	&	0.26	&	0.16	&	0.09		& 0.240	\\
HCN (3--2)						&	12	&	19	&	48	&	0.25	&	0.40	&	-0.15		& 0.925	\\
H$_{2}$CO (2$_{12}$--1$_{11}$) 		&	18 	&	9	&	71	&	0.25	&	0.13	&	0.13	 	& 0.061	\\
\hline
\enddata
\end{deluxetable}



\begin{deluxetable}{lcccccccc}
\tablecaption{Blue Excess Statistics for the $\it Low$ and $\it High$ groups
\label{tab5}}
\tablewidth{0pt}
\tablehead{\colhead{Transition} & \colhead{Group} & \colhead{N$_{Blue}$} & \colhead{N$_{Red}$} & \colhead{N$_{Total}$} & \colhead{D$_{Blue}$} & \colhead{D$_{Red}$} & \colhead{E} & \colhead{P} }
\startdata
HCO$^+$ (1--0)				&	$\it Low$	&	7	&	2	&	20	&	0.35	&	0.10	&	0.25	& 0.090	\\
							&	$\it High$	&	22	&	12	&	54	&	0.41	 &	0.22	&	0.19	& 0.029	\\
HCO$^+$ (3--2)				&	$\it Low$	&	1 	&	1	&	7	&	0.14	&	0.14	&	0.00	& 0.750	\\
							&	$\it High$	&	10	&	6	&	36	&	0.28	 &	0.17	&	0.11 	& 0.227	\\
HCN (3--2)					&	$\it Low$	&	0	&	2	&	7	&	0.00	&	0.29	&    -0.29	& 1.000	\\
							&	$\it High$	&	12	&	17	&	41	&	0.29	 &	0.41	&    -0.12	& 0.868	\\
H$_{2}$CO (2$_{21}$-1$_{11}$) 	&	$\it Low$	& 	4	&	0	&	20	&	0.20	&	0.00	&	0.20	& 0.063	\\
							&	$\it High$	&	14 	&	9	&	51	&	0.27	 &	0.18	&	0.10	& 0.202	\\
\hline
\enddata
\end{deluxetable}








\clearpage
\appendix
\section{Observed line profiles for all objects}
In this appendix, we present all the detected molecular line spectra of the individual sources.
Table~\ref{tab3} lists detailed information on the inflow statistics of these spectra.
\renewcommand{\thefigure}{A\arabic{figure}}
\setcounter{figure}{0}

\begin{figure*}
\begin{tabbing}
  \includegraphics[width=0.25\textwidth]{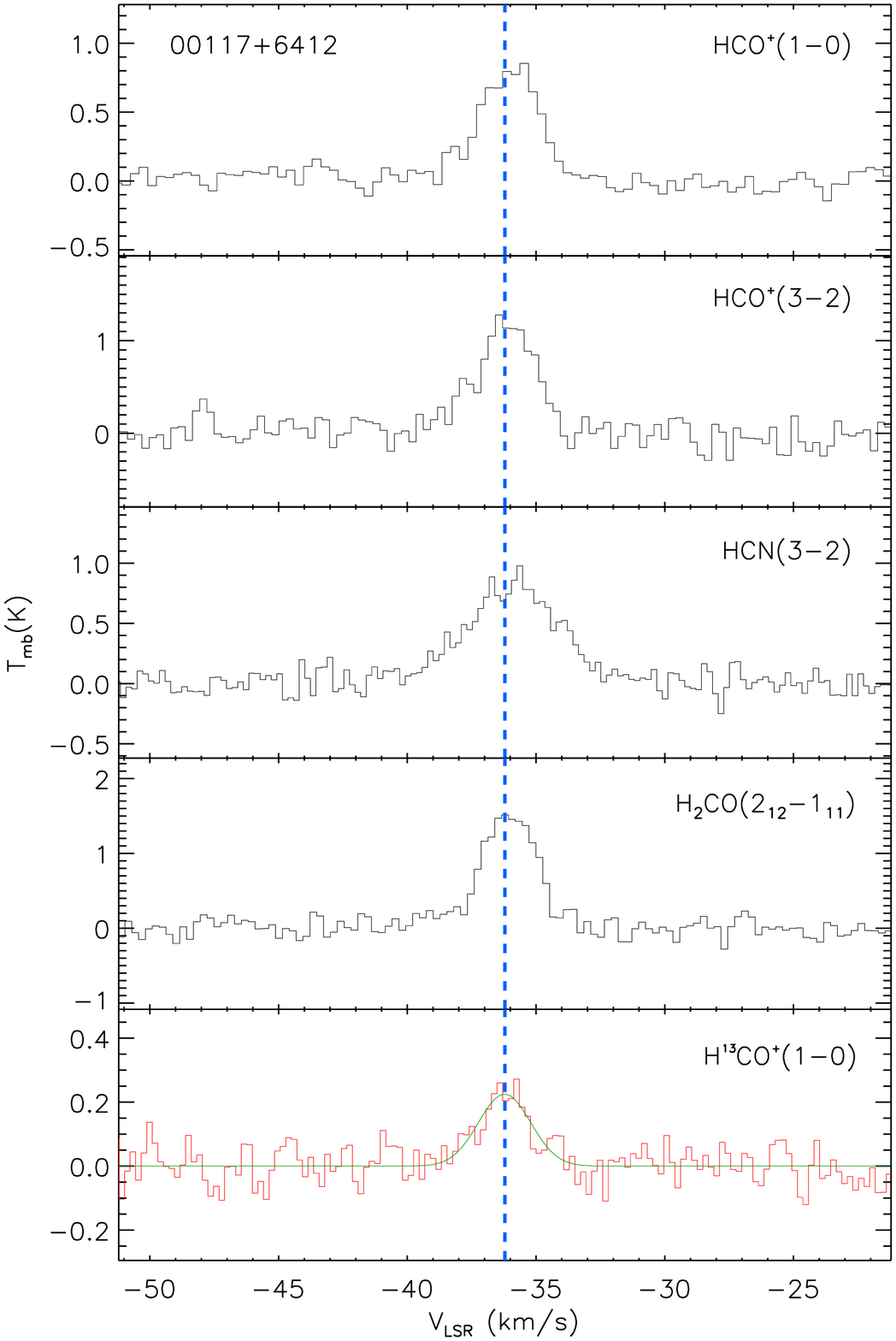}
  \includegraphics[width=0.25\textwidth]{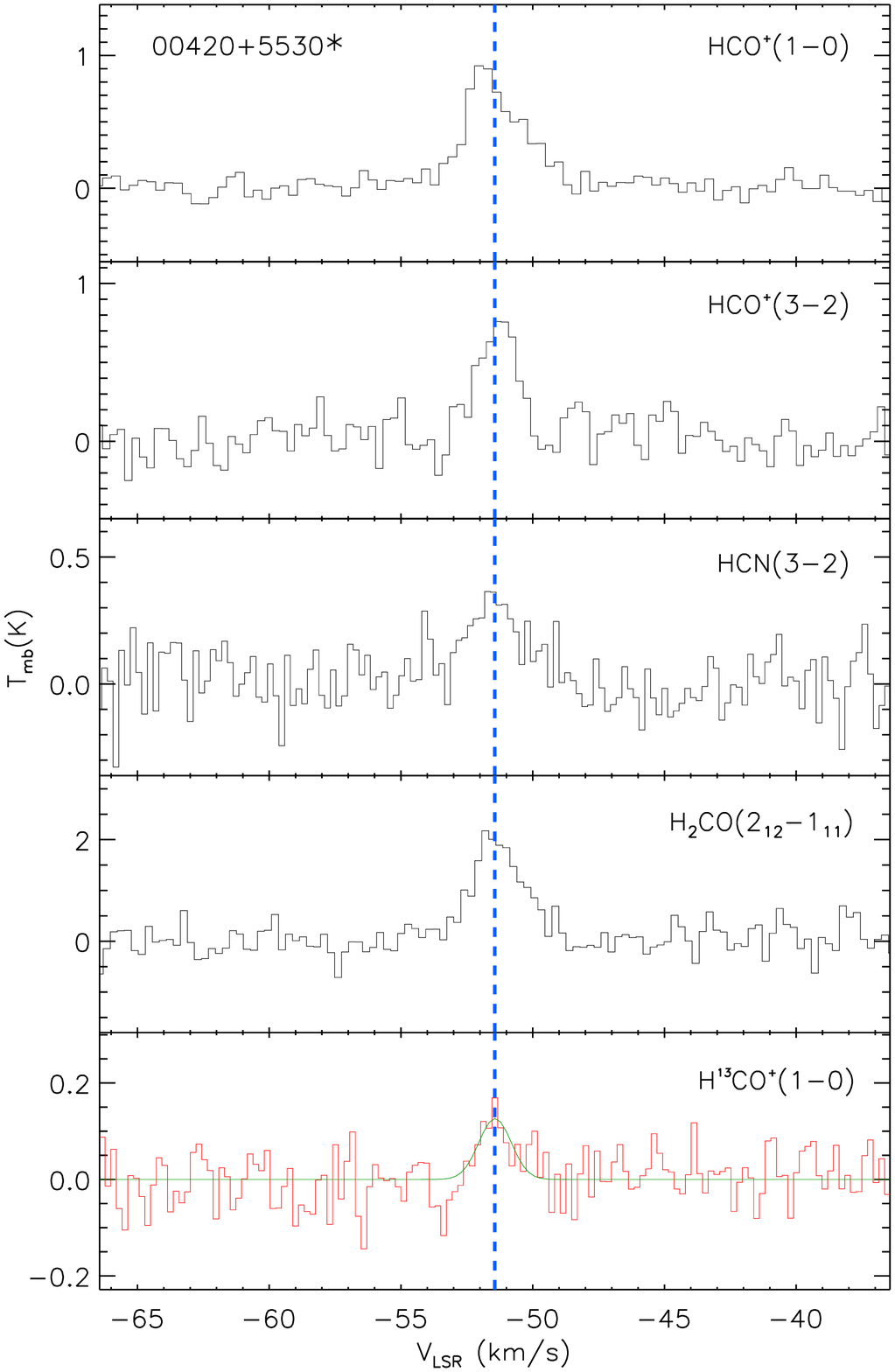} 
  \includegraphics[width=0.25\textwidth]{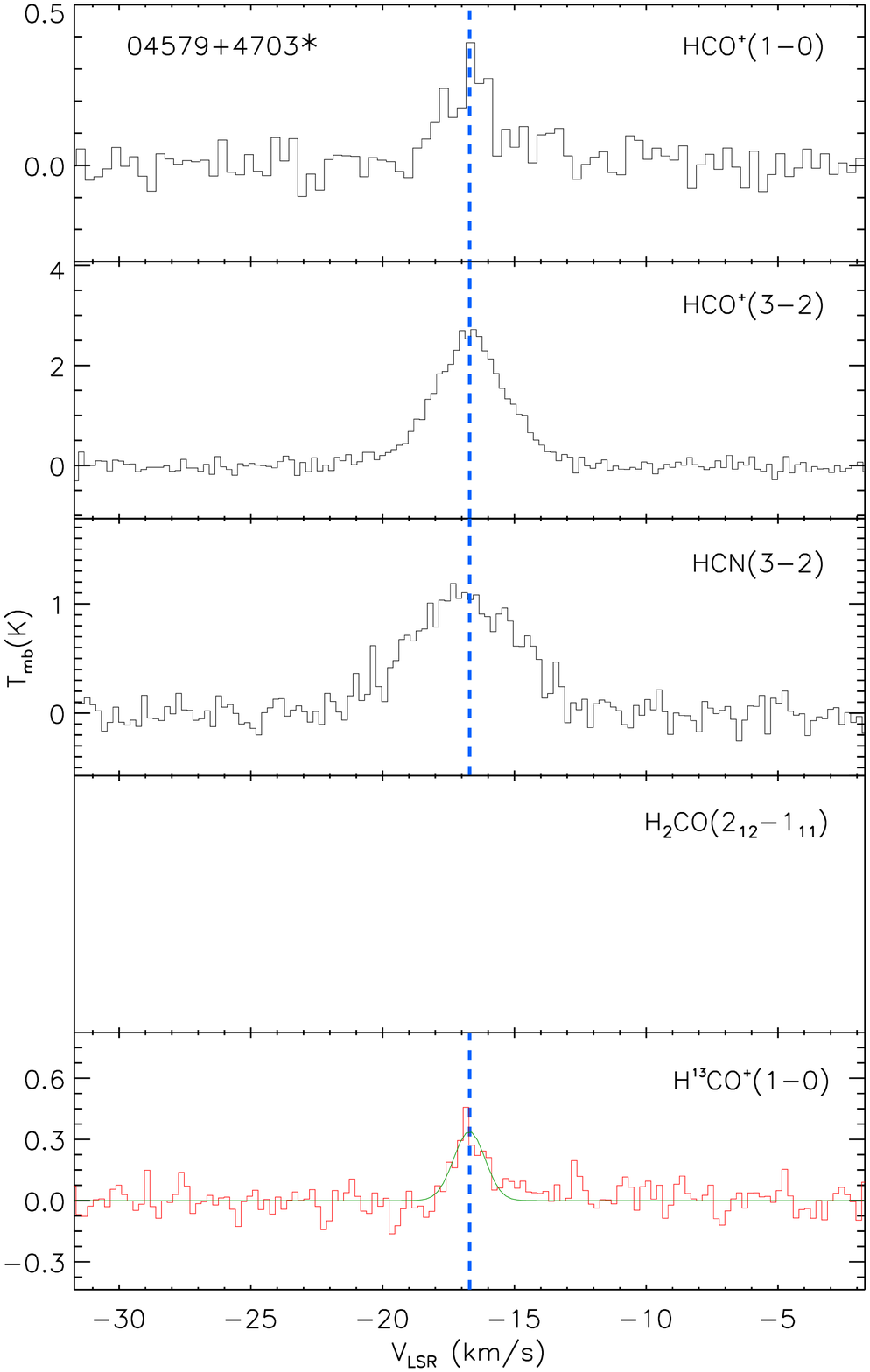}  
  \includegraphics[width=0.25\textwidth]{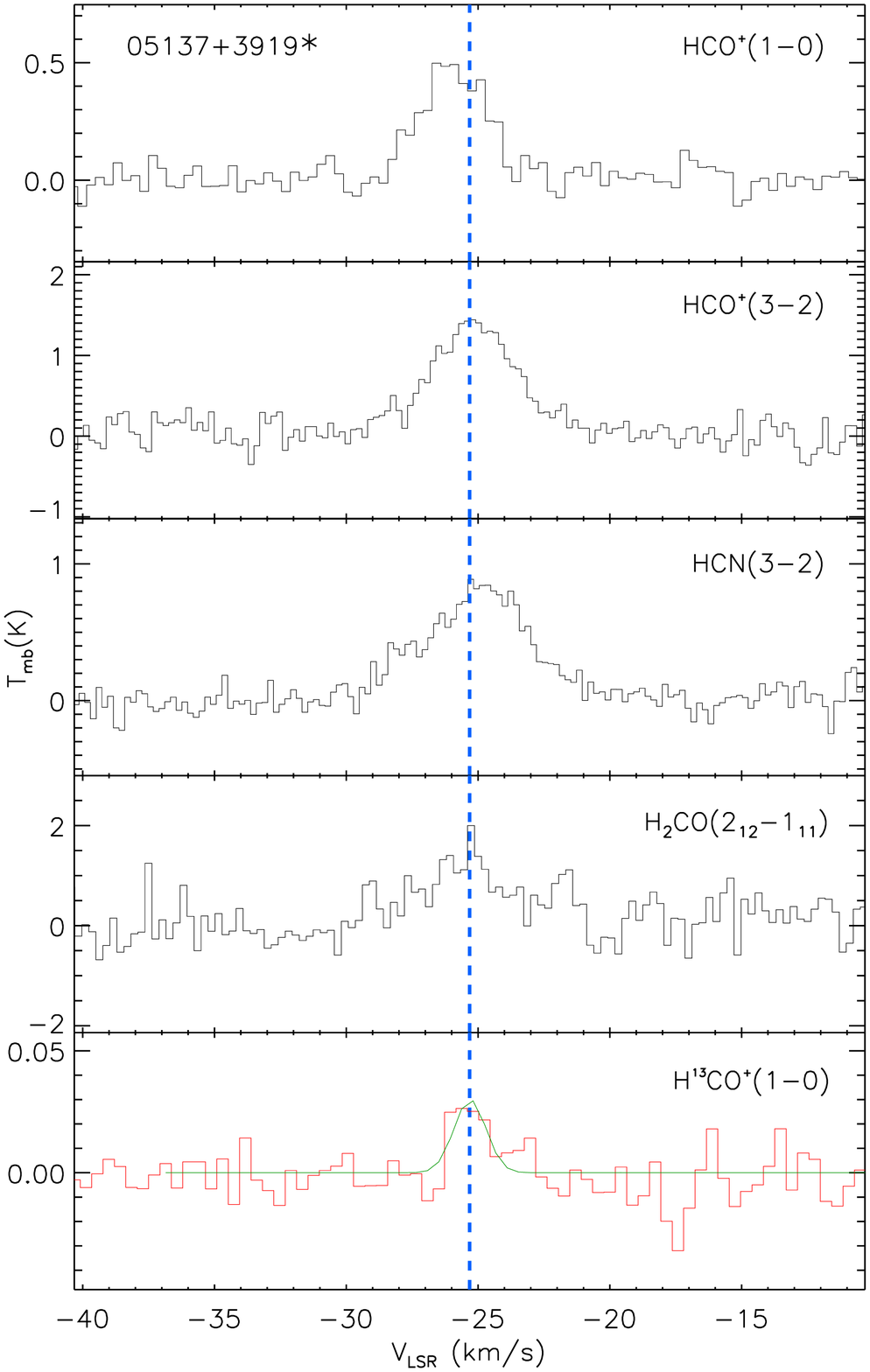} \\
  \includegraphics[width=0.25\textwidth]{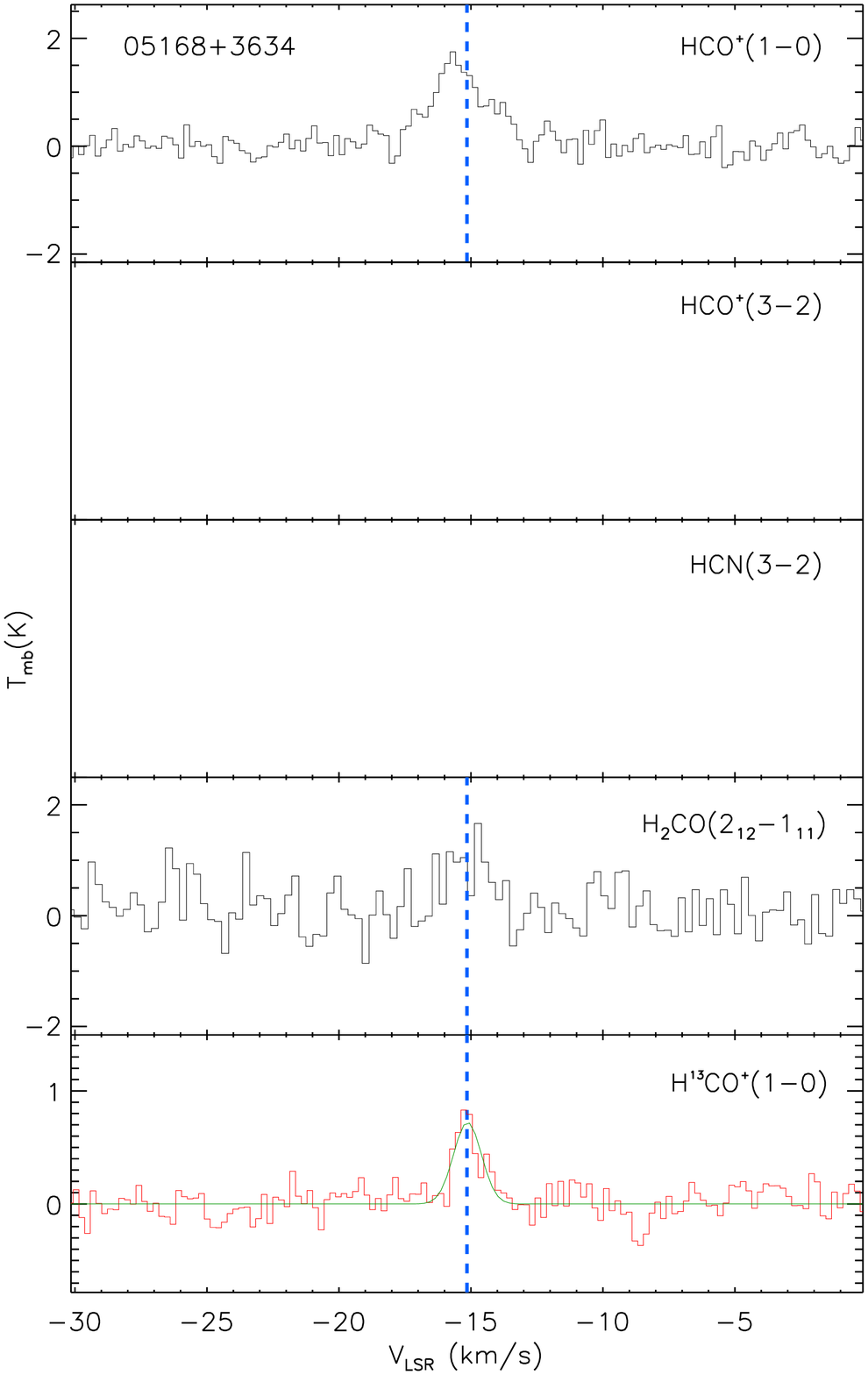} 
  \includegraphics[width=0.25\textwidth]{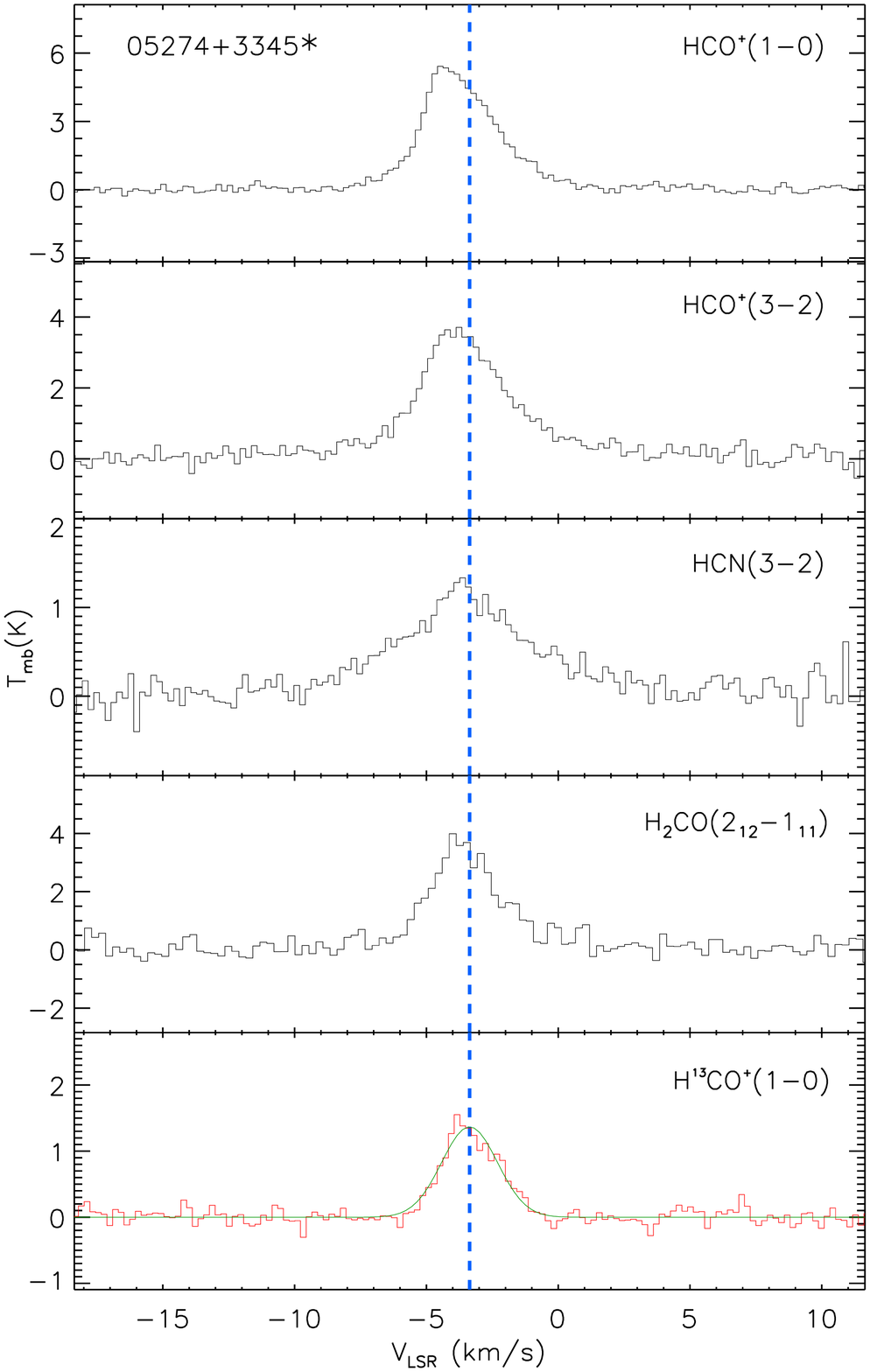} 
  \includegraphics[width=0.25\textwidth]{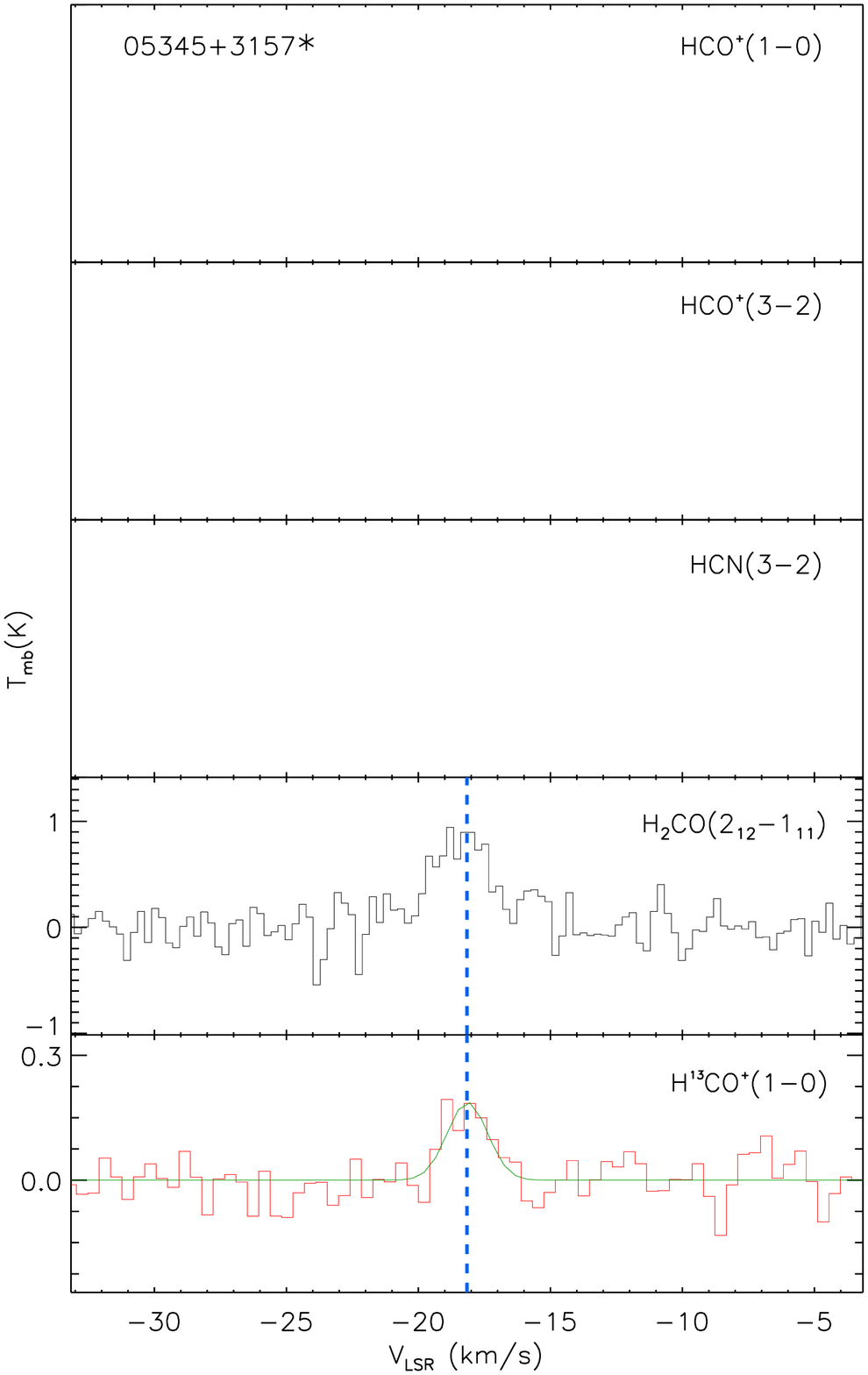} 
  \includegraphics[width=0.25\textwidth]{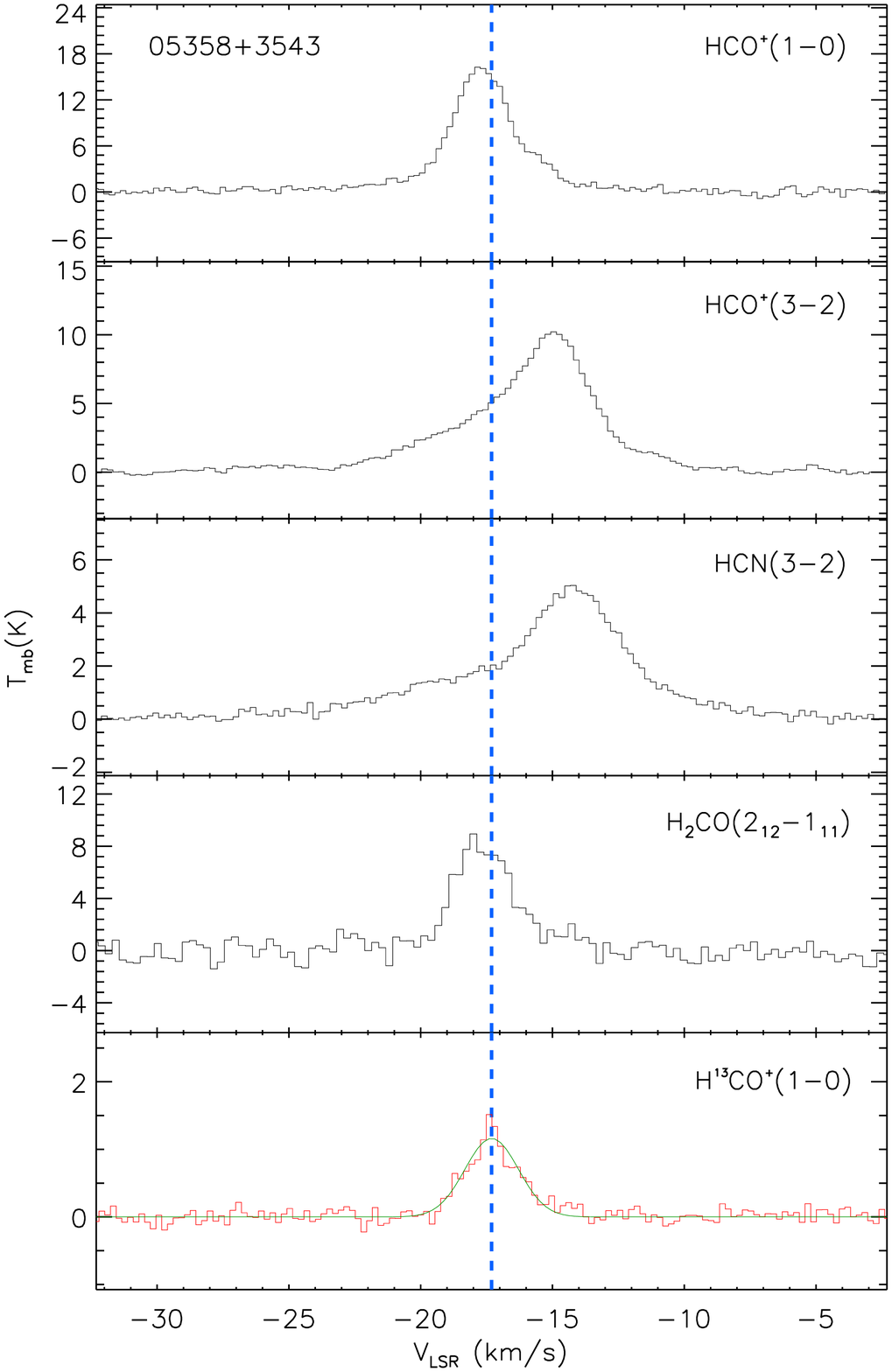} \\
  \includegraphics[width=0.25\textwidth]{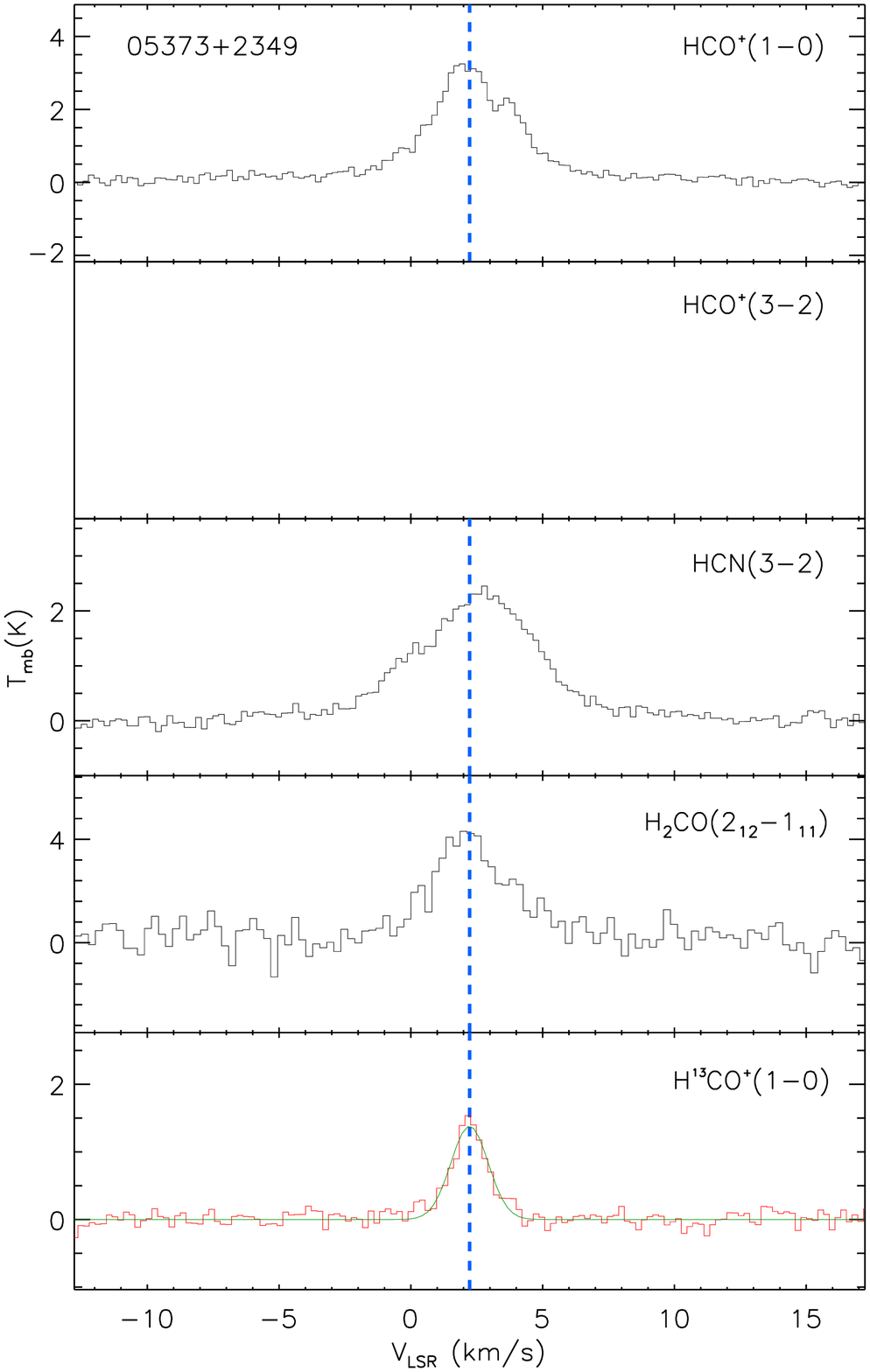} 
  \includegraphics[width=0.25\textwidth]{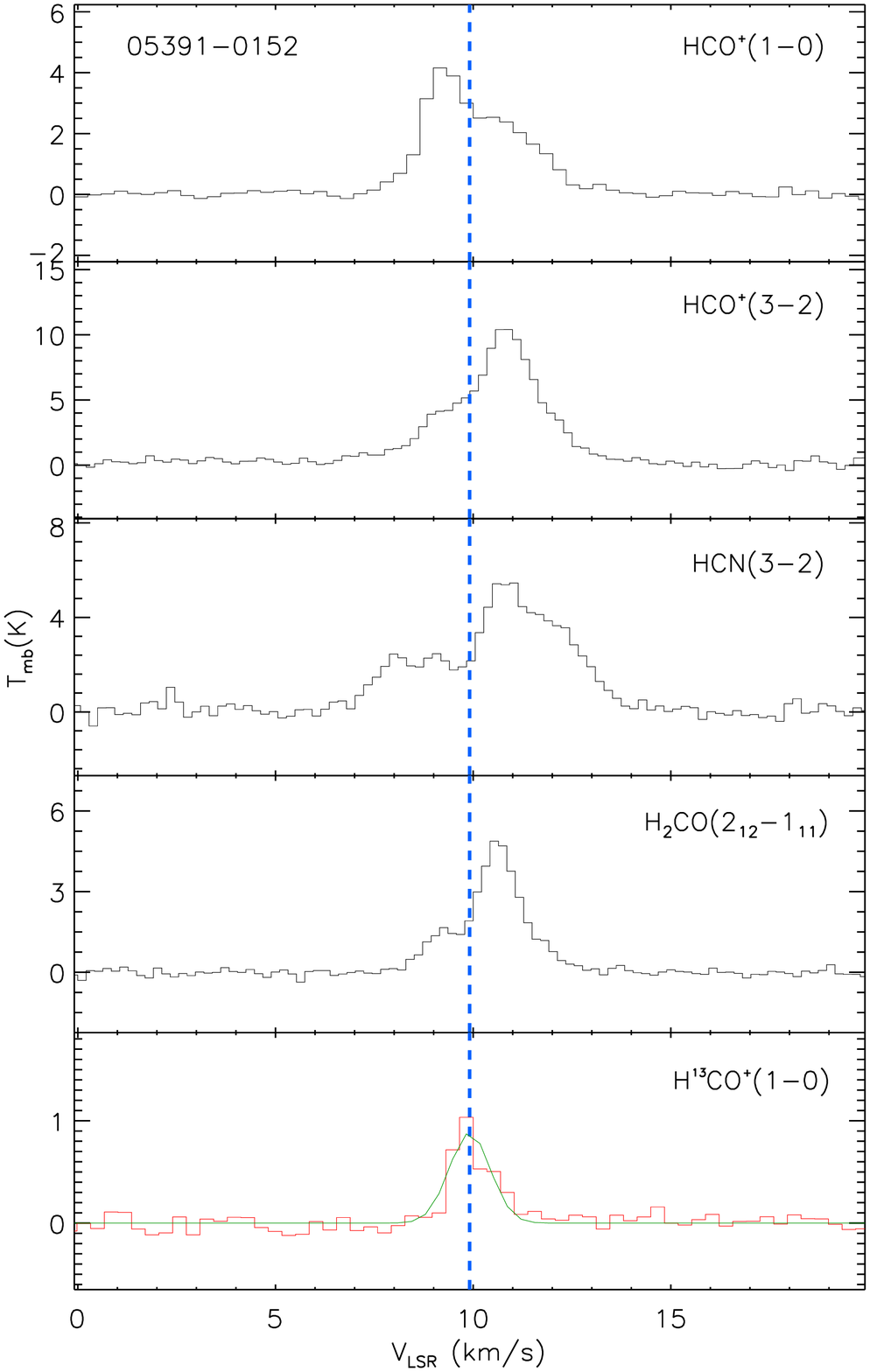} 
  \includegraphics[width=0.25\textwidth]{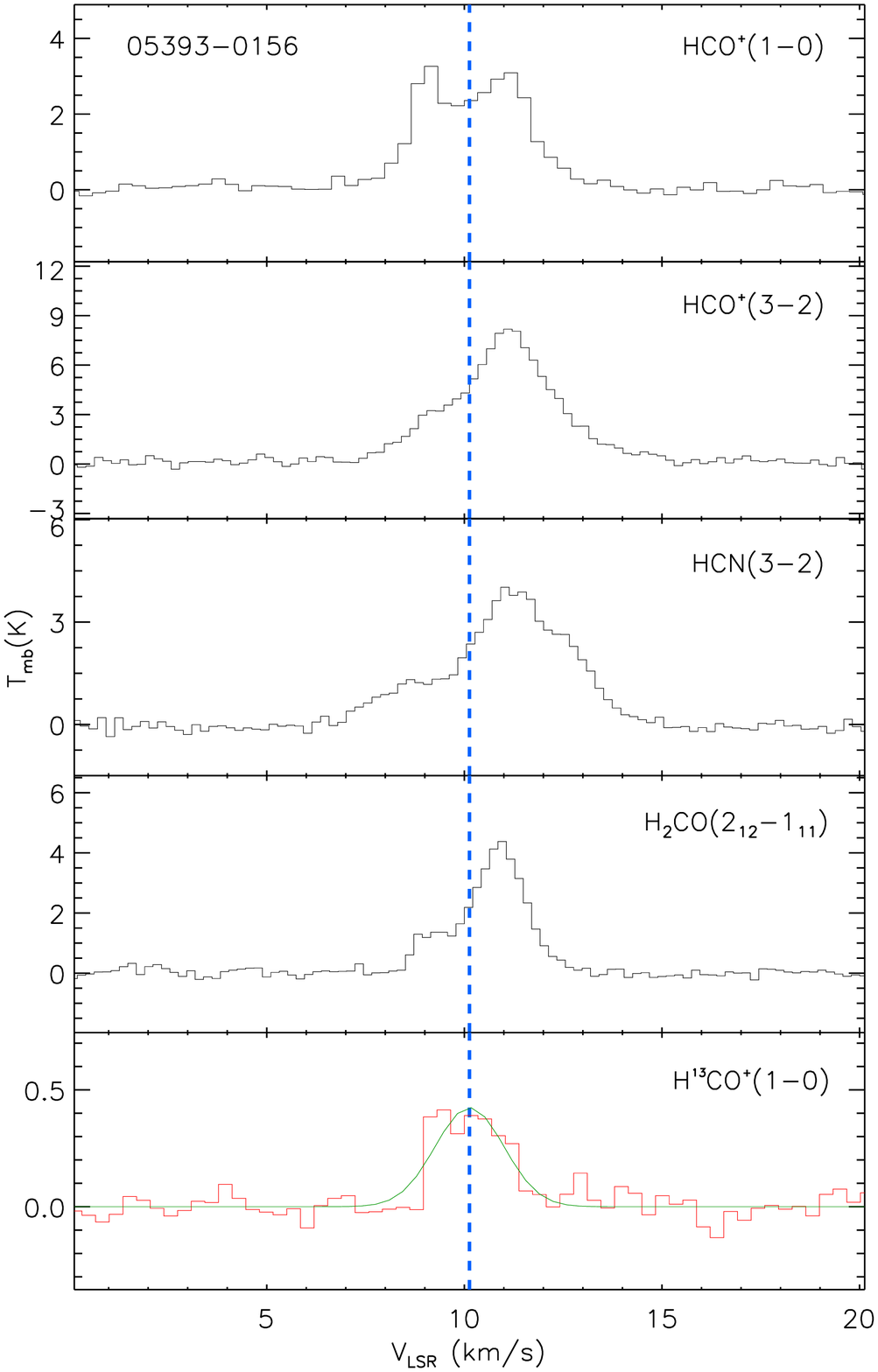} 
  \includegraphics[width=0.25\textwidth]{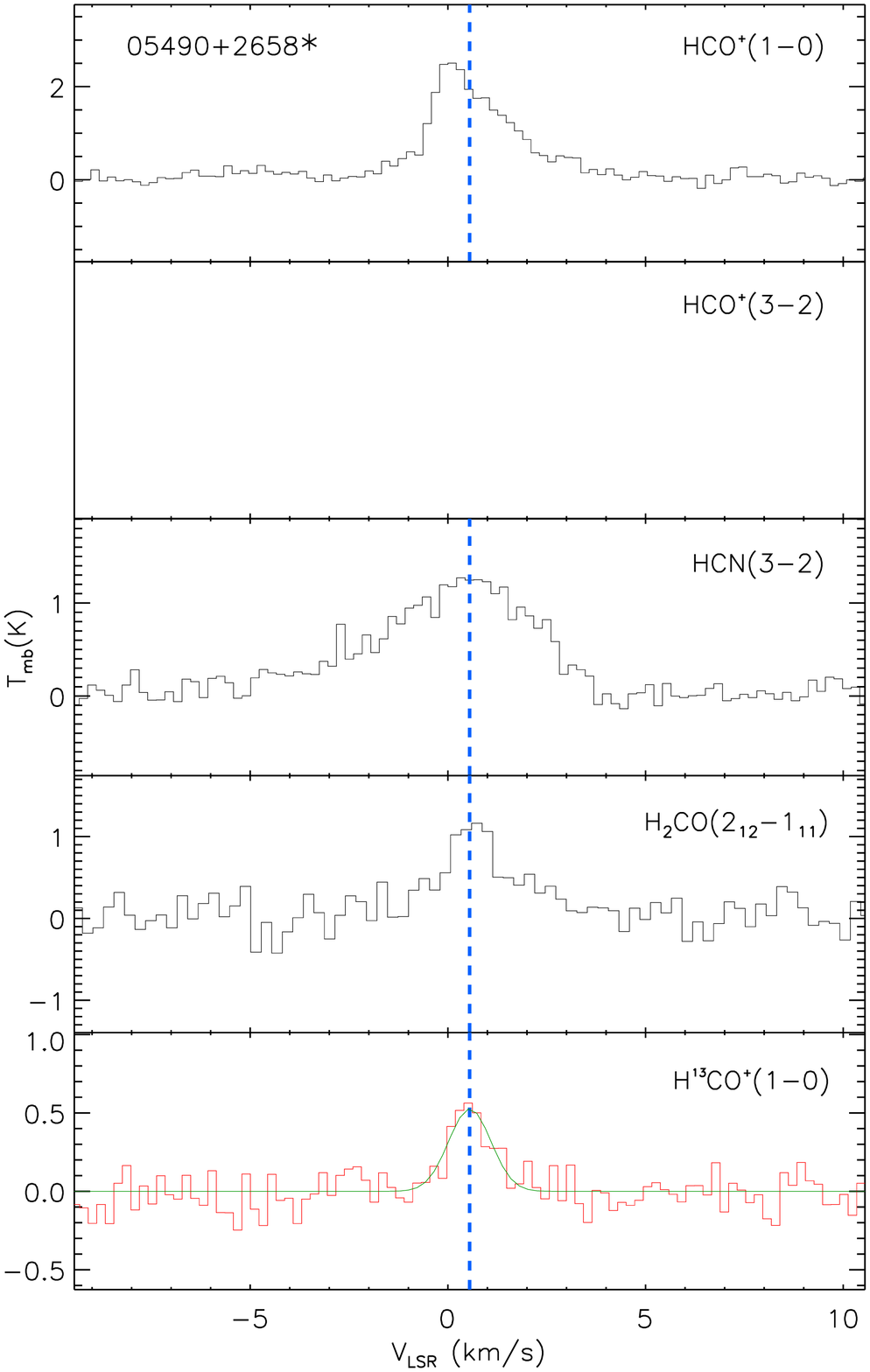} 
\end{tabbing}
 \caption{Detected molecular line spectra of the individual sources. 
For each source, the IRAS name is presented at the upper left corner in the top panel
and the transition is listed at the upper right corner in each panel. The vertical blue dotted line
is the central velocity of the \hcoppj\ line determined by Gaussian fitting. 
The line spectrum is displayed in red with the Gaussian fitting result in green in the bottom panel. 
The inflow candidates are marked with asterisks on the source names. The empty panel means that the transition was not observed for a given source.}
\label{fA1}
\end{figure*}


\begin{figure*}
\begin{tabbing}
  \includegraphics[width=0.25\textwidth]{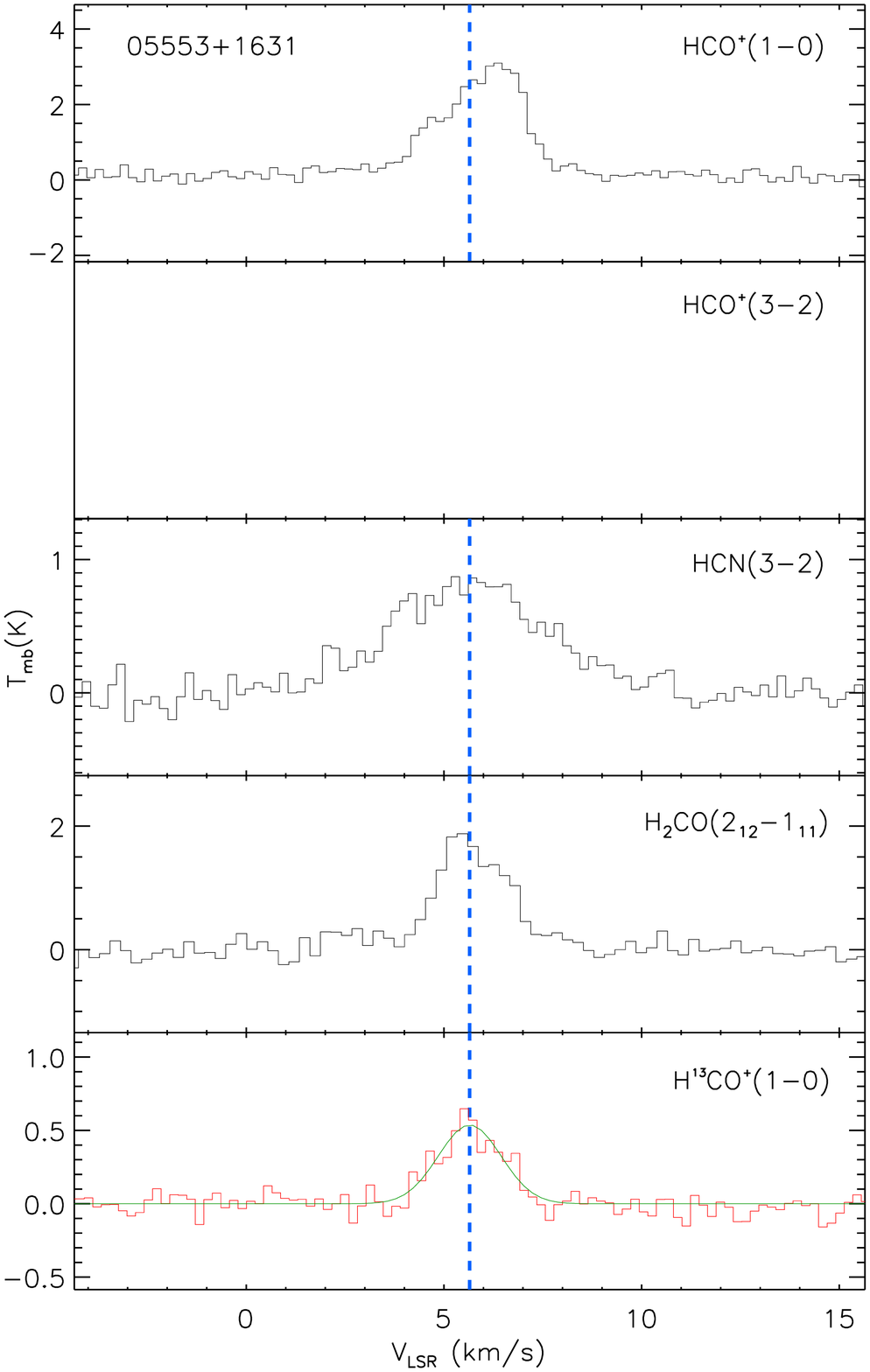} 
  \includegraphics[width=0.25\textwidth]{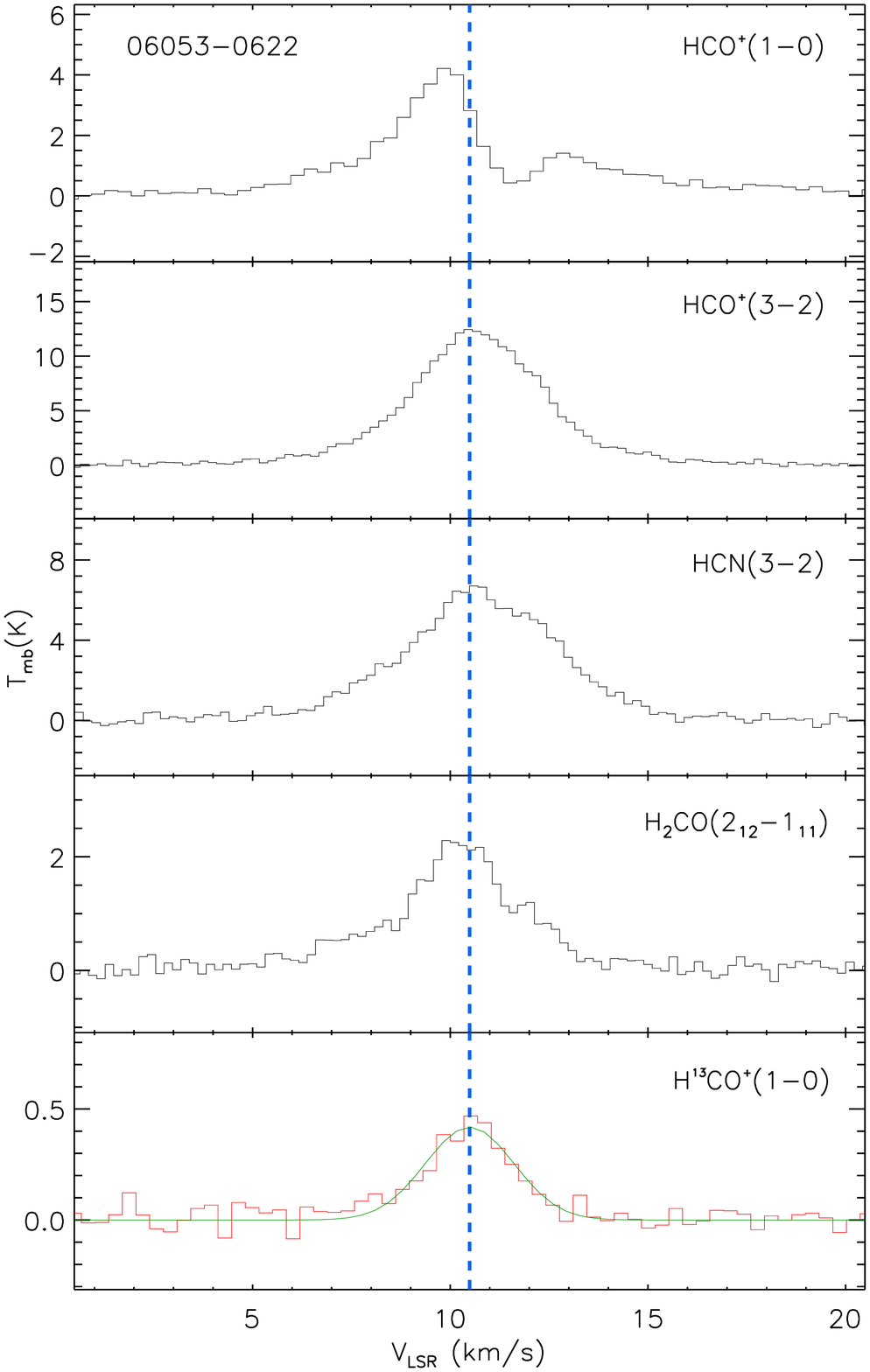}
  \includegraphics[width=0.25\textwidth]{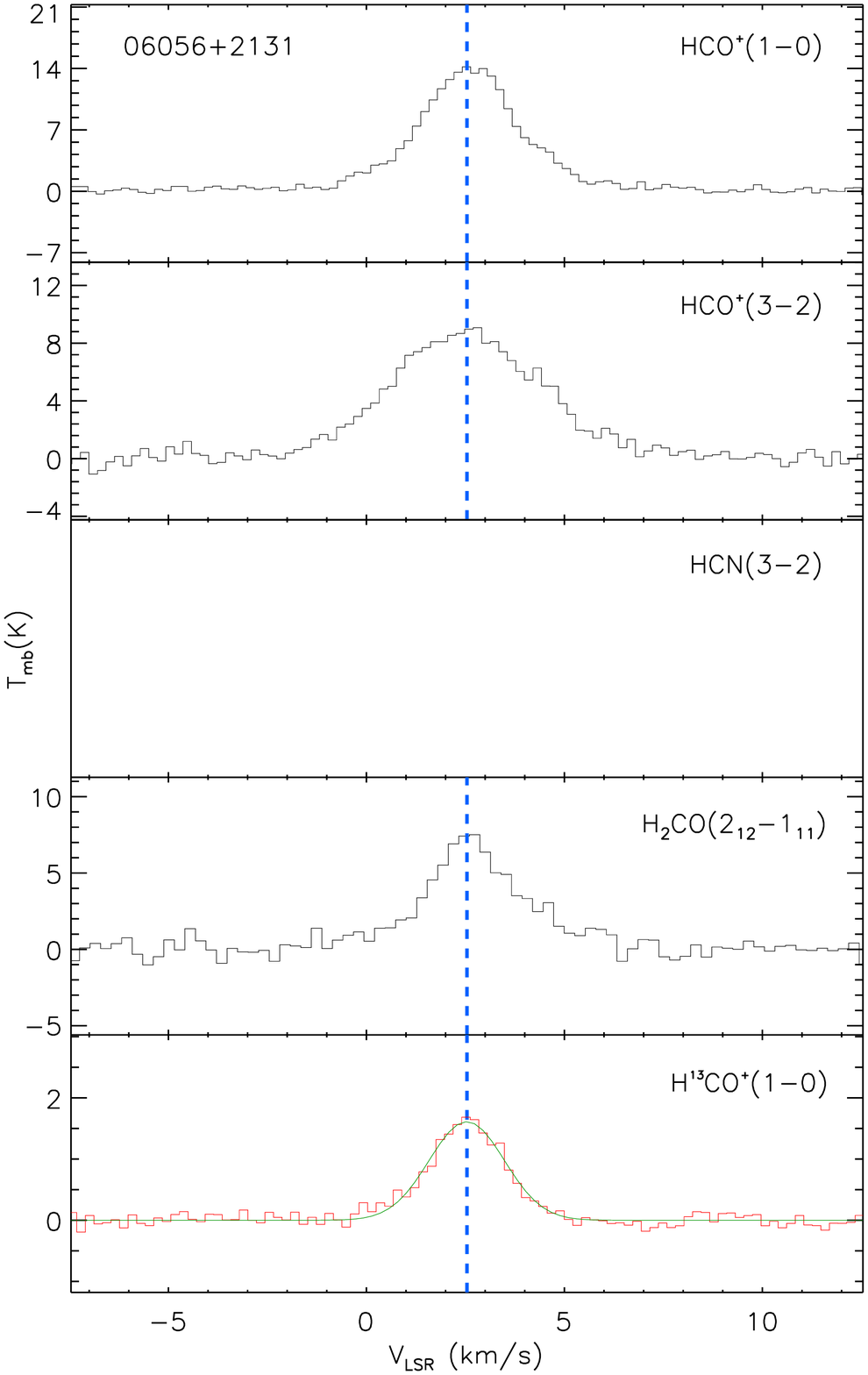} 
  \includegraphics[width=0.25\textwidth]{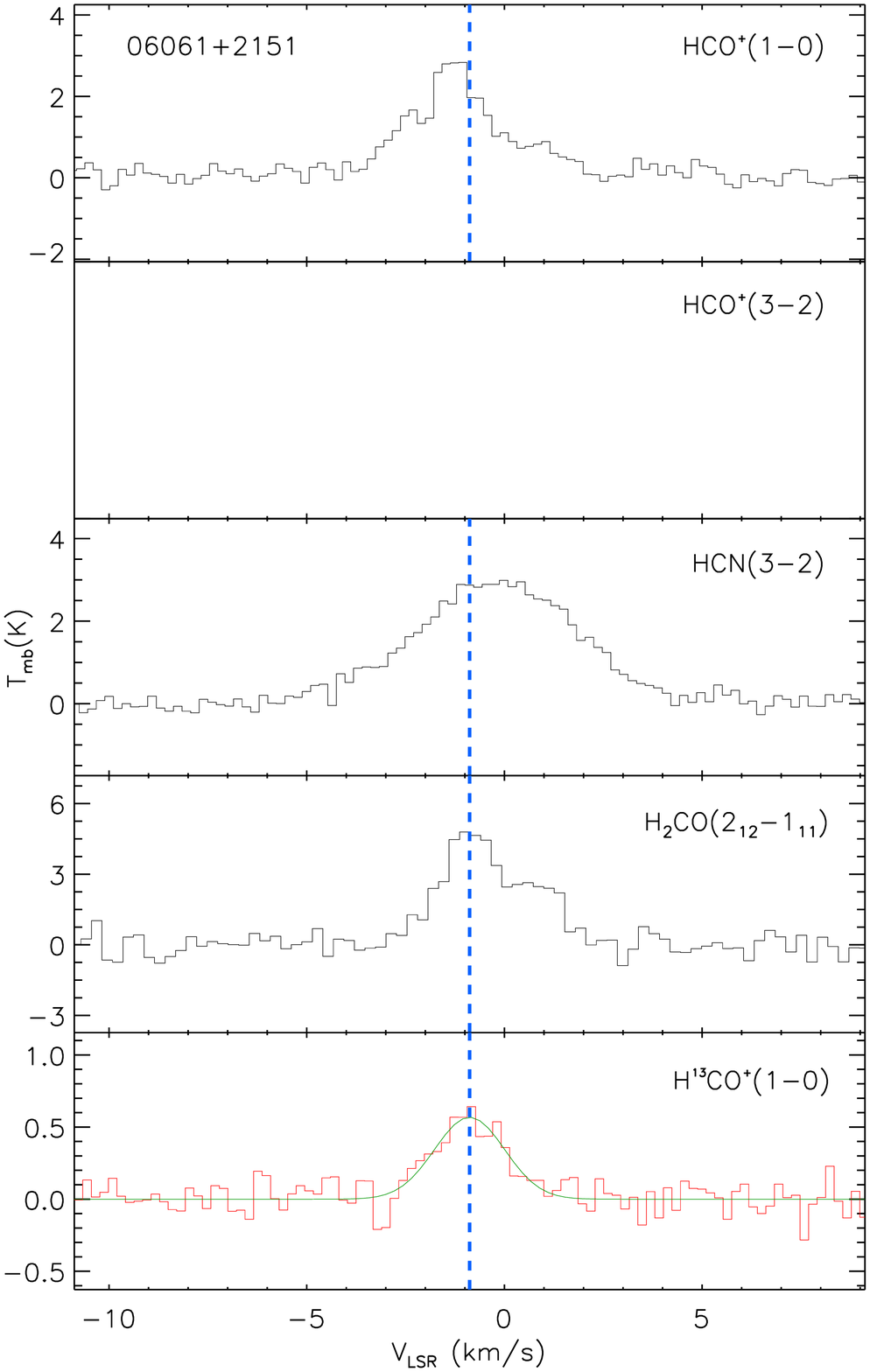} \\  
  \includegraphics[width=0.25\textwidth]{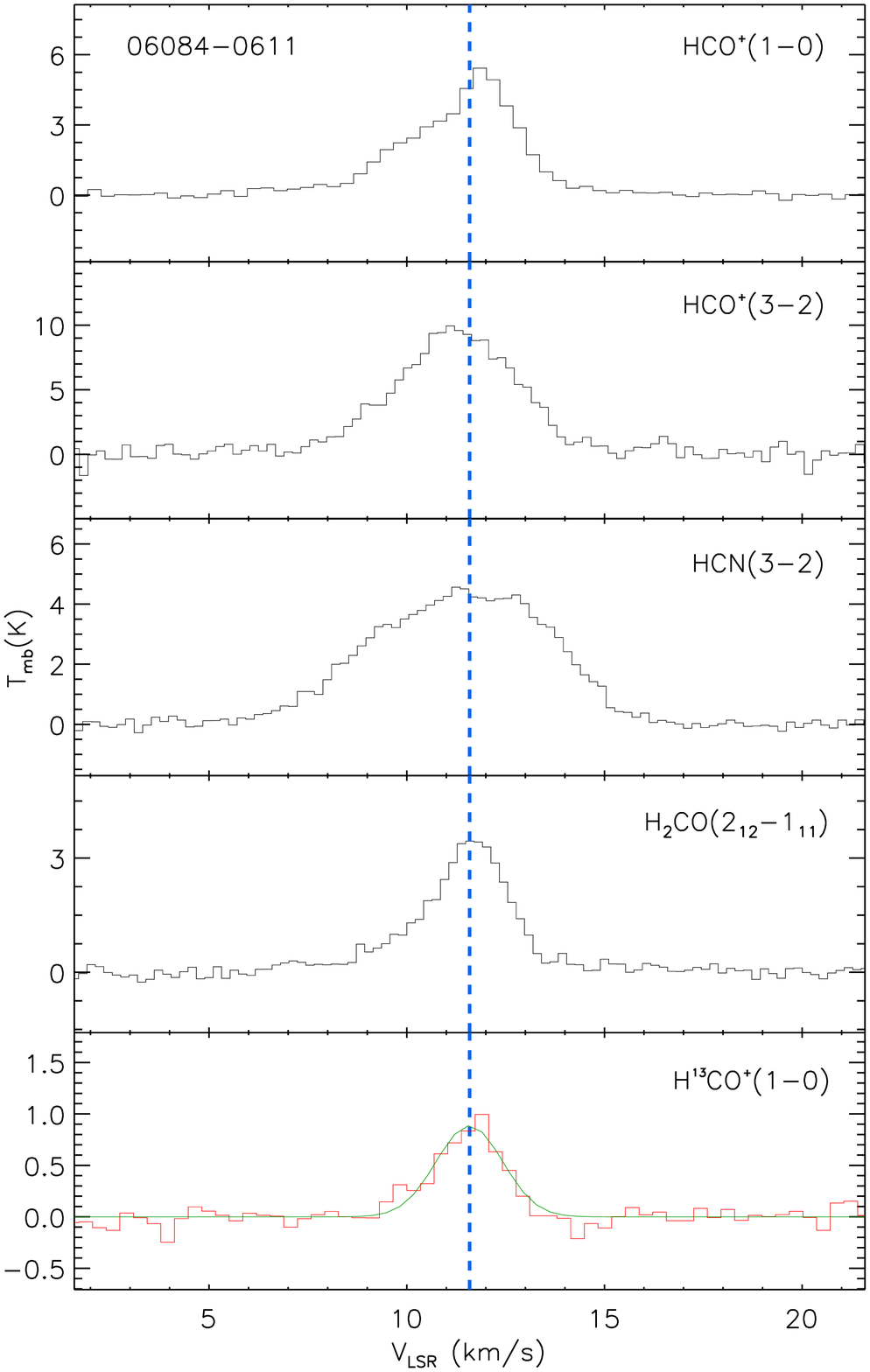} 
  \includegraphics[width=0.25\textwidth]{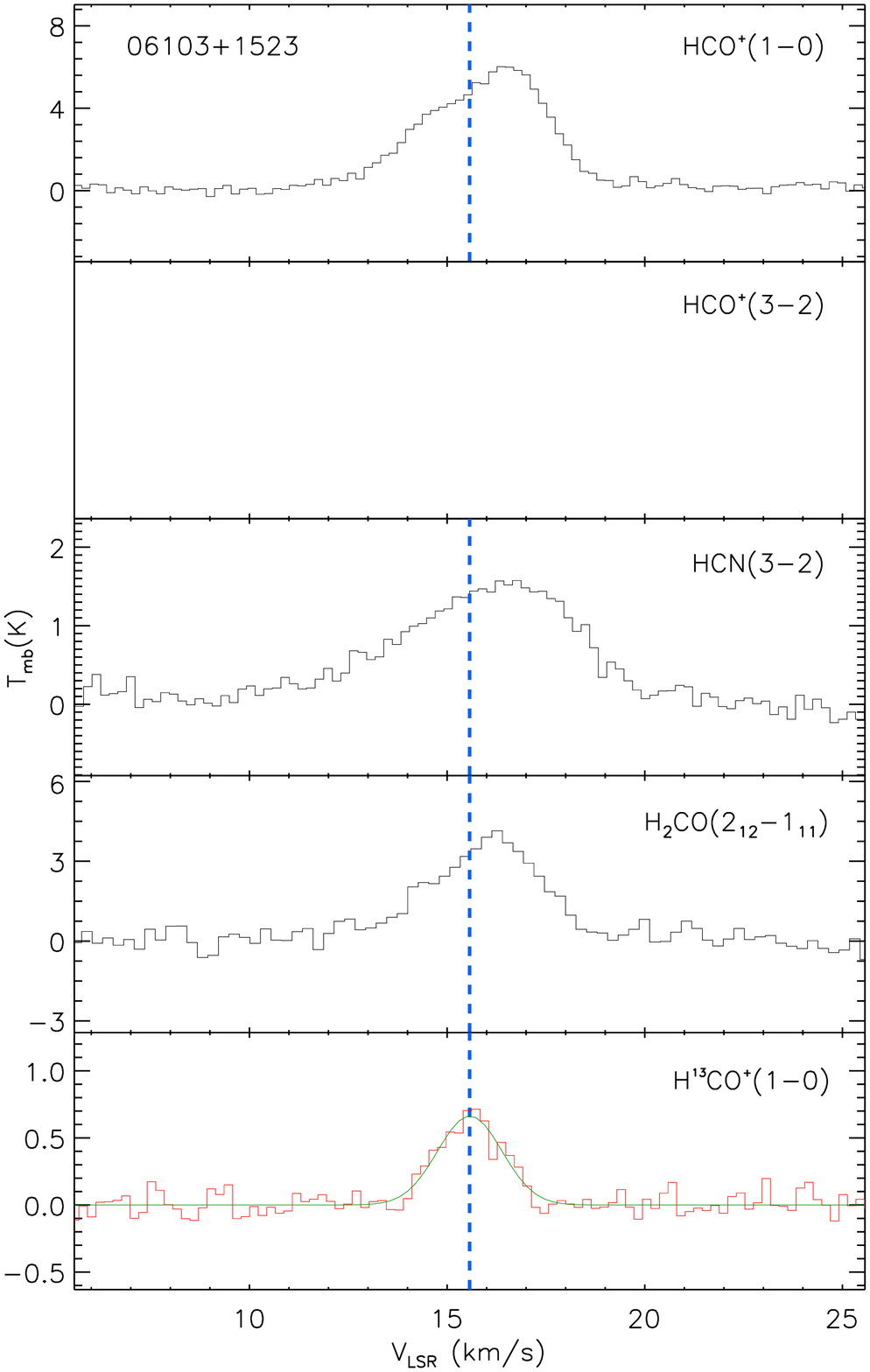} 
  \includegraphics[width=0.25\textwidth]{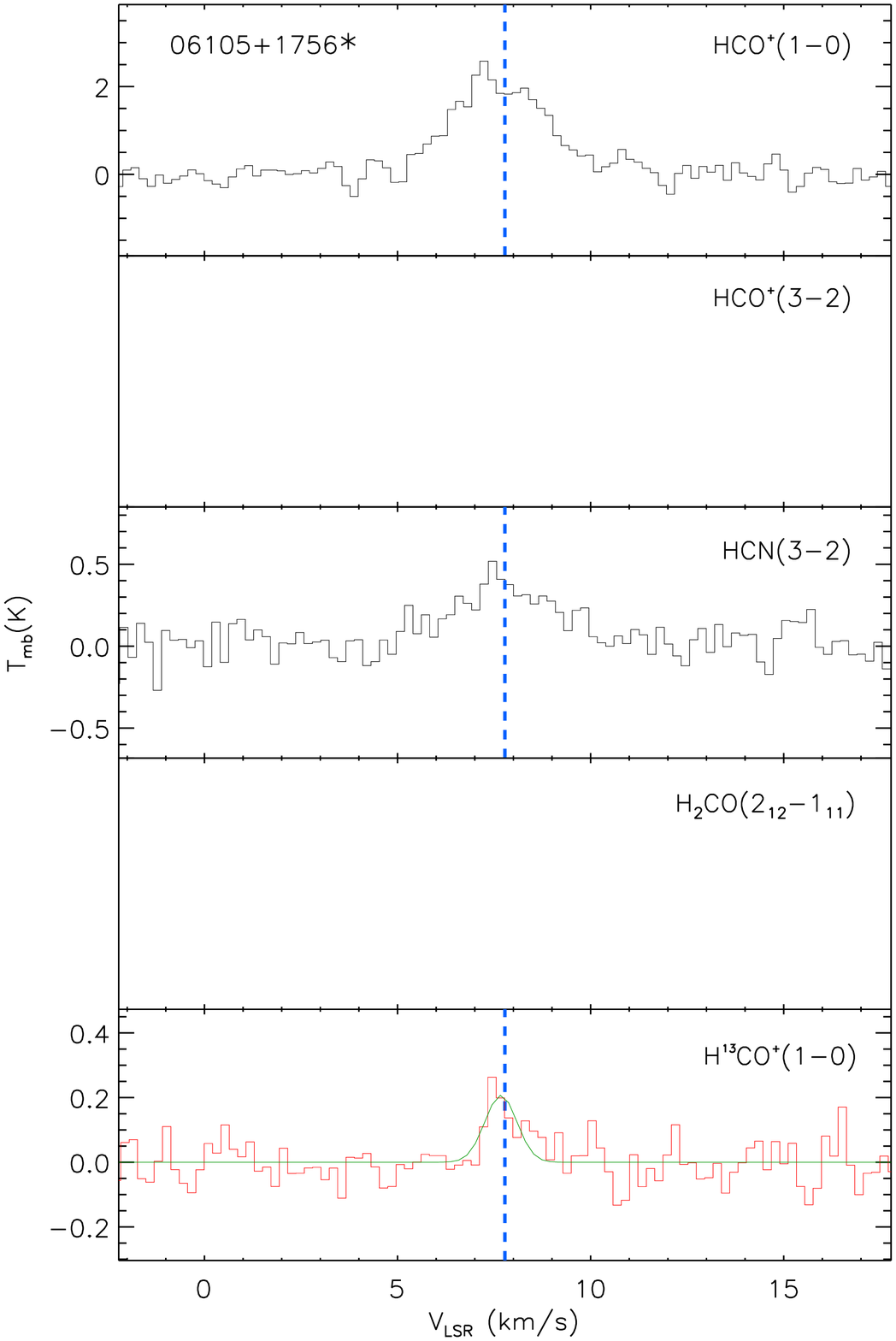} 
  \includegraphics[width=0.25\textwidth]{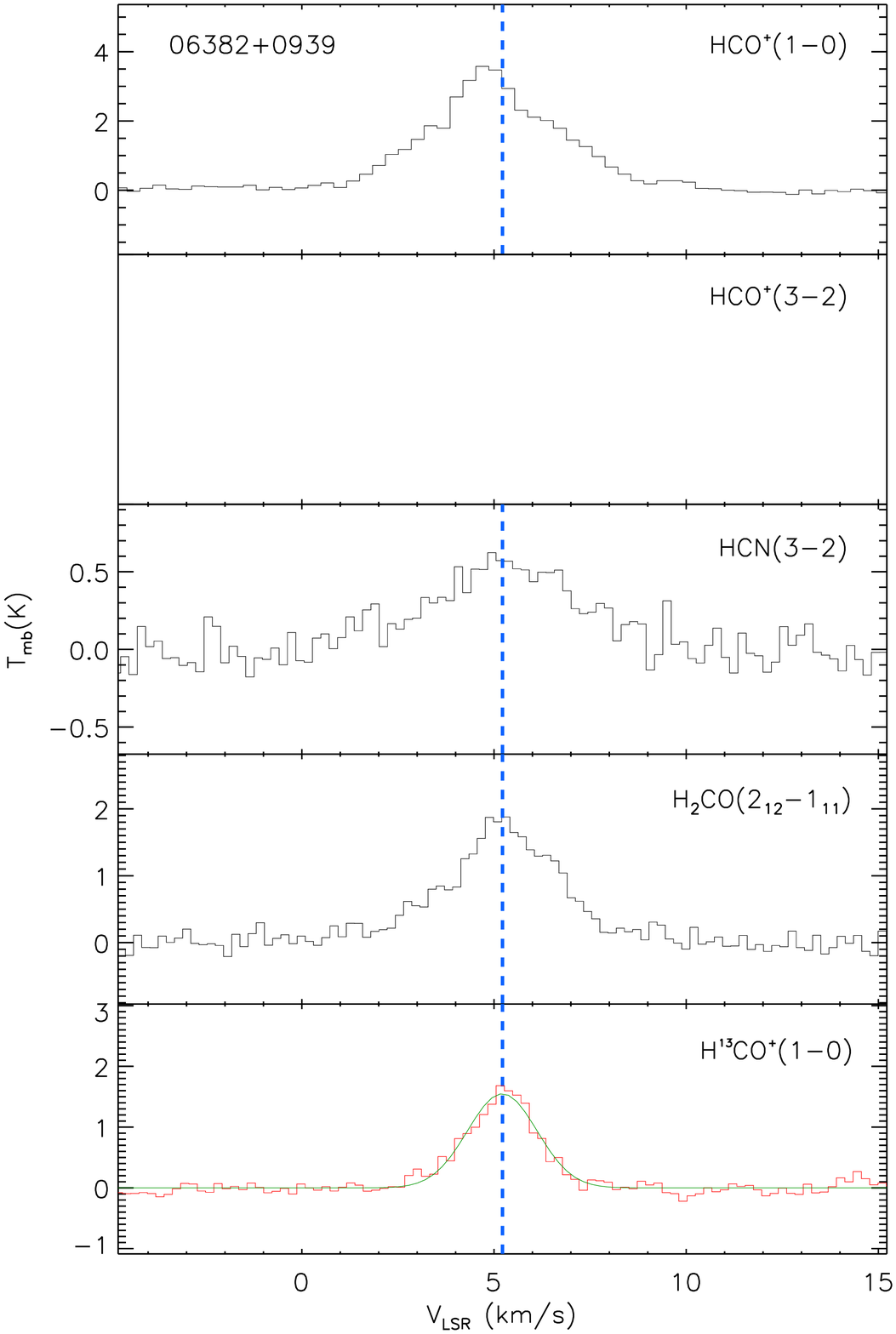} \\
  \includegraphics[width=0.25\textwidth]{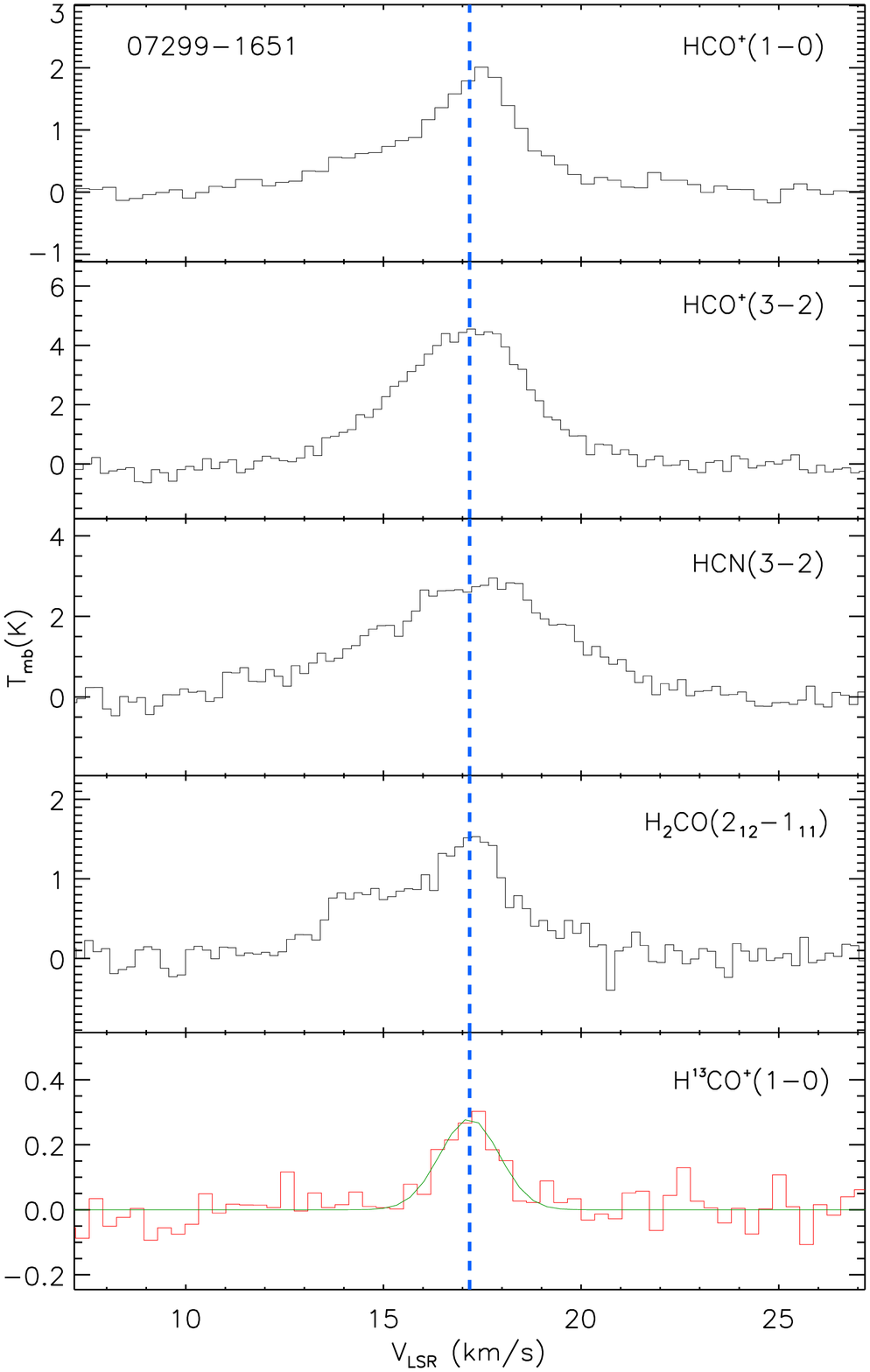} 
  \includegraphics[width=0.25\textwidth]{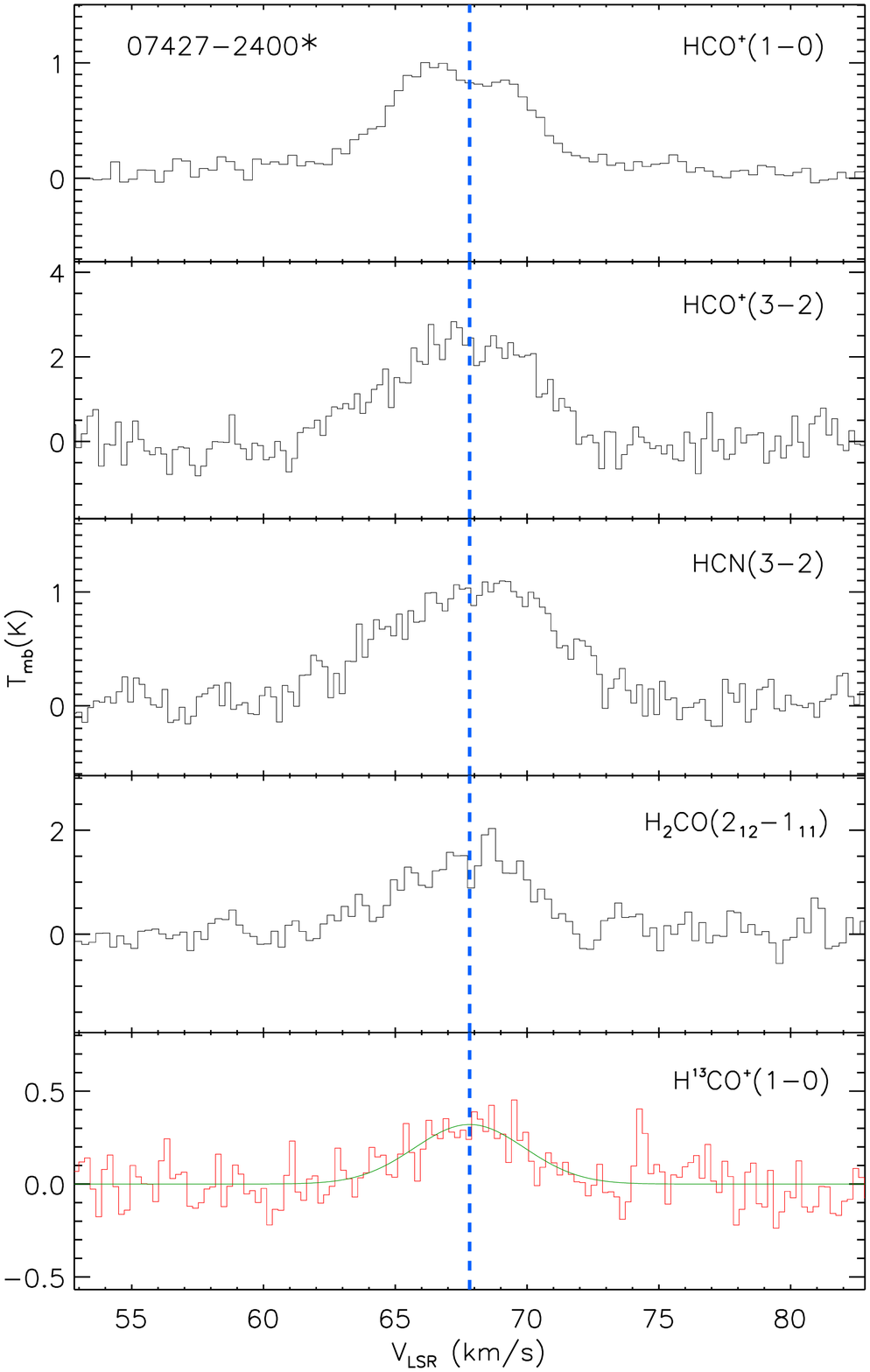} 
  \includegraphics[width=0.25\textwidth]{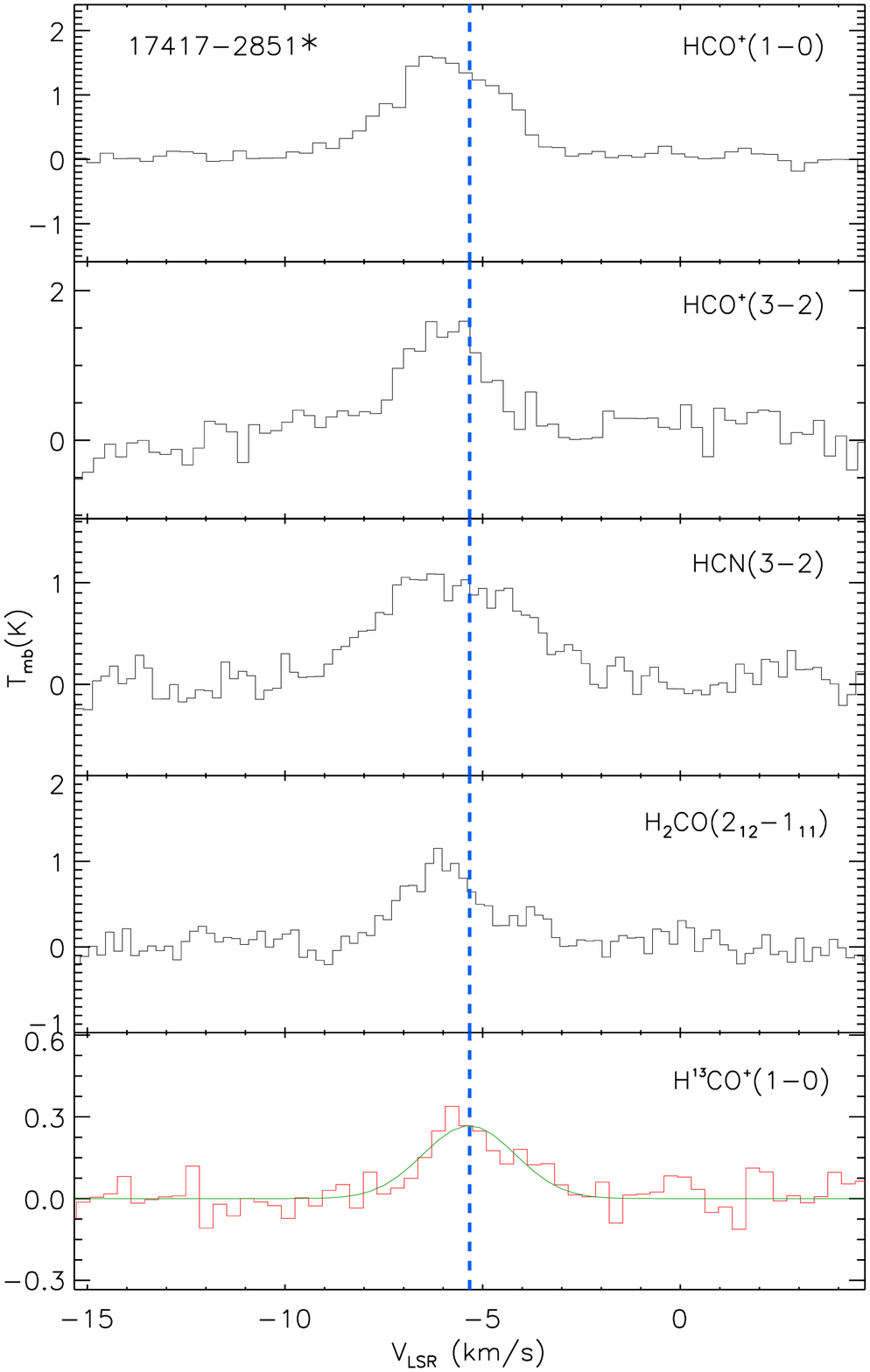} 
  \includegraphics[width=0.25\textwidth]{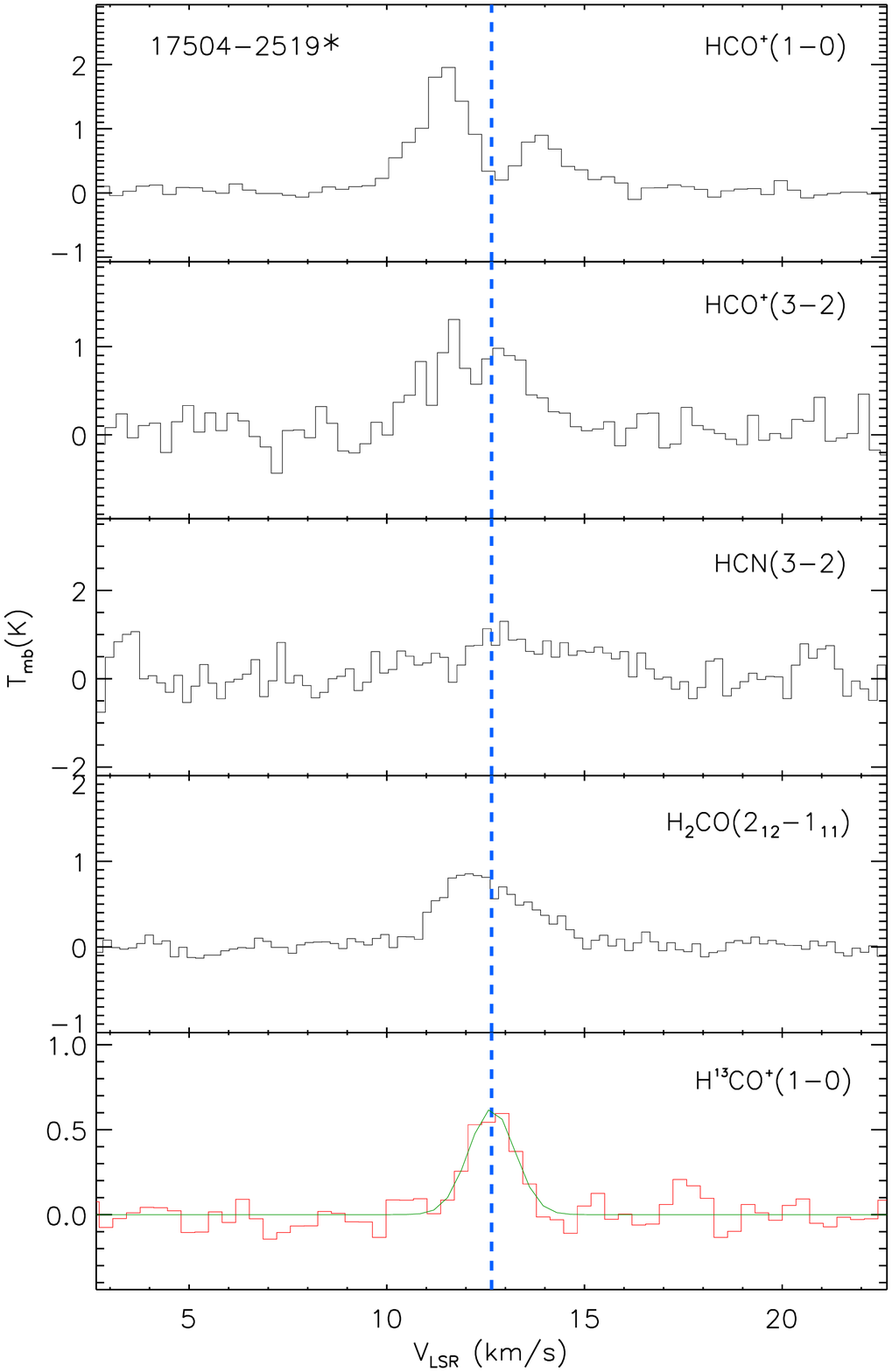} 
\end{tabbing}
\center{\textbf{Figure A1.} continued.}
\label{fA1}
\end{figure*}

\begin{figure*}
\begin{tabbing}
  \includegraphics[width=0.25\textwidth]{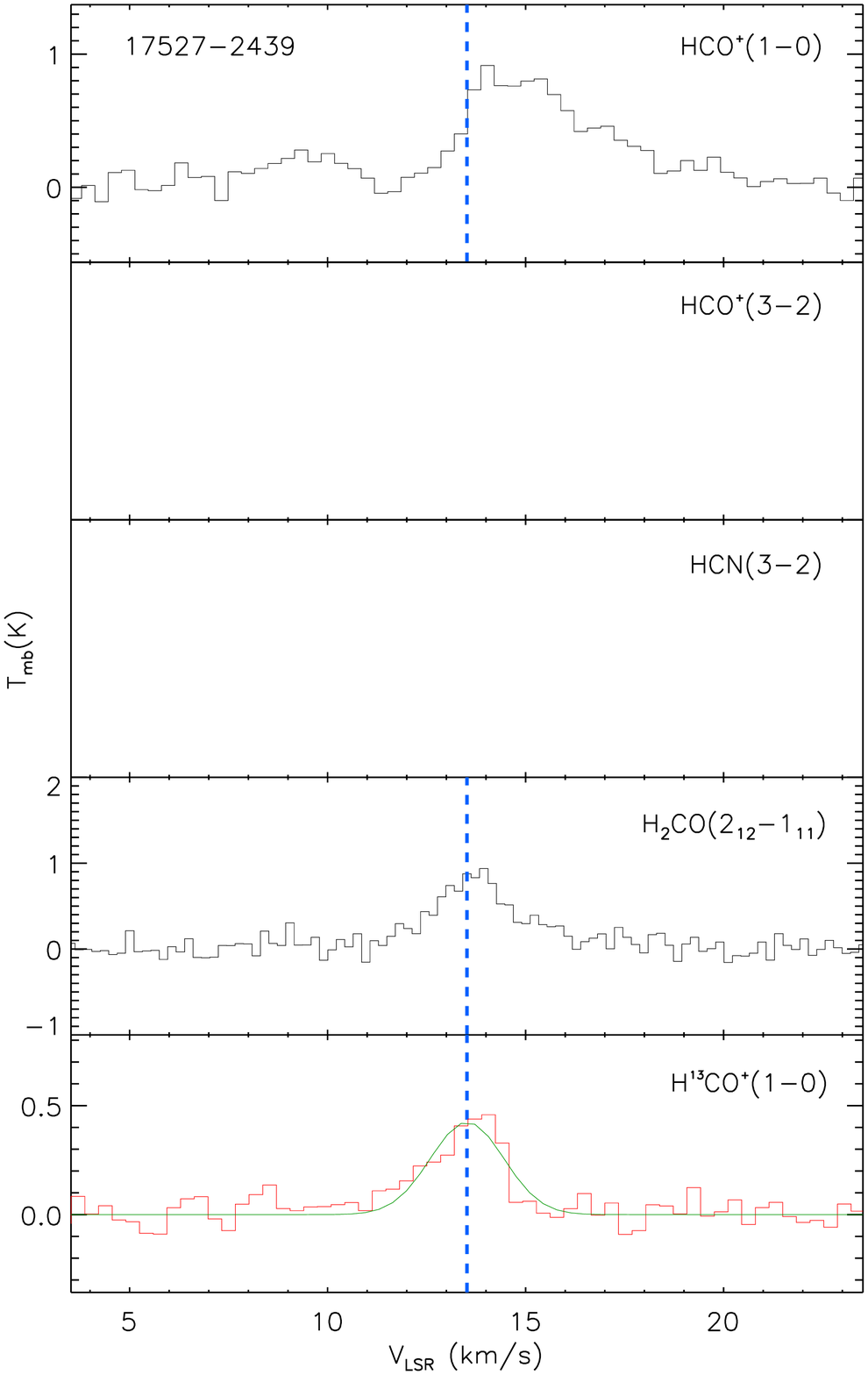} 
  \includegraphics[width=0.25\textwidth]{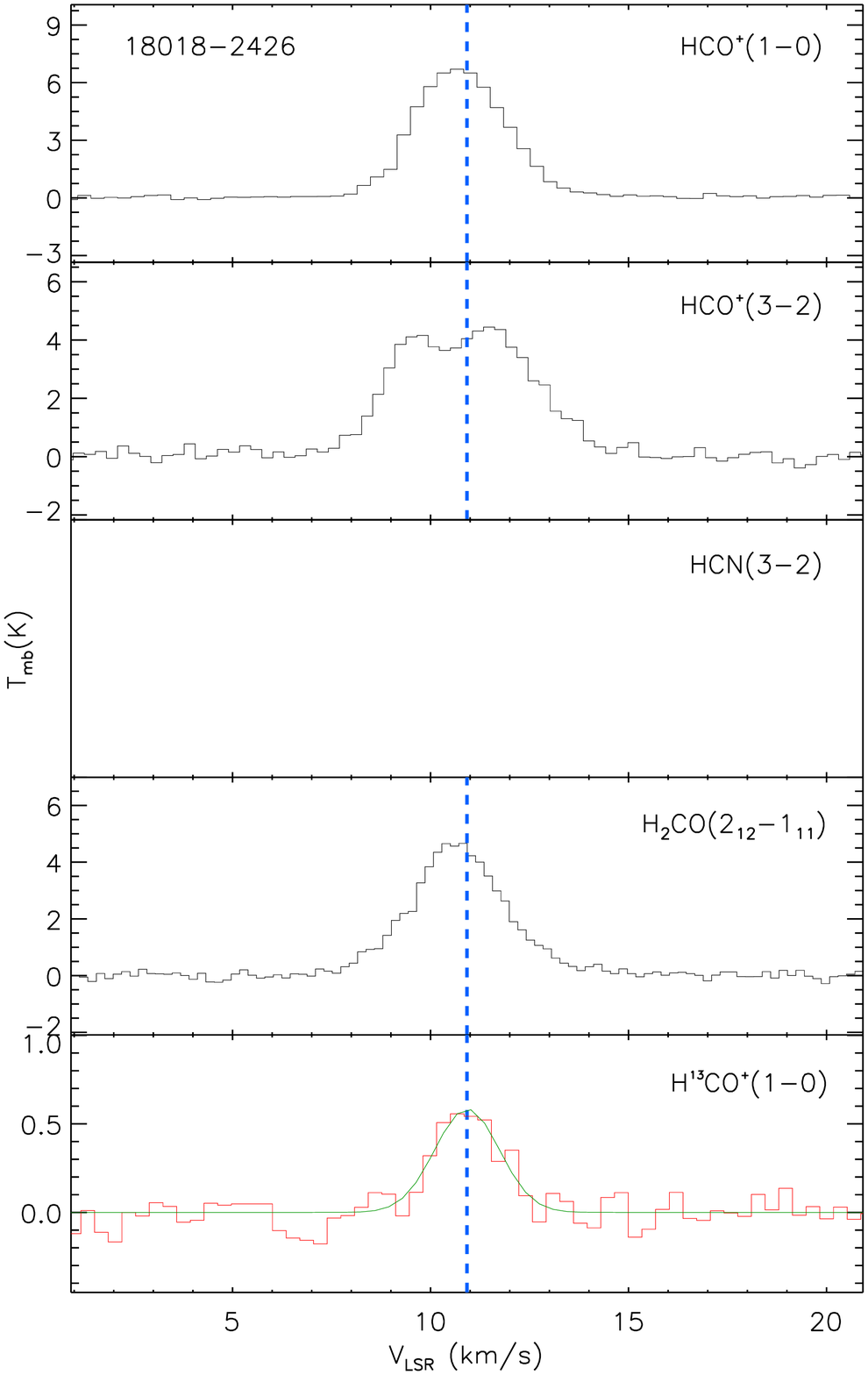}
  \includegraphics[width=0.25\textwidth]{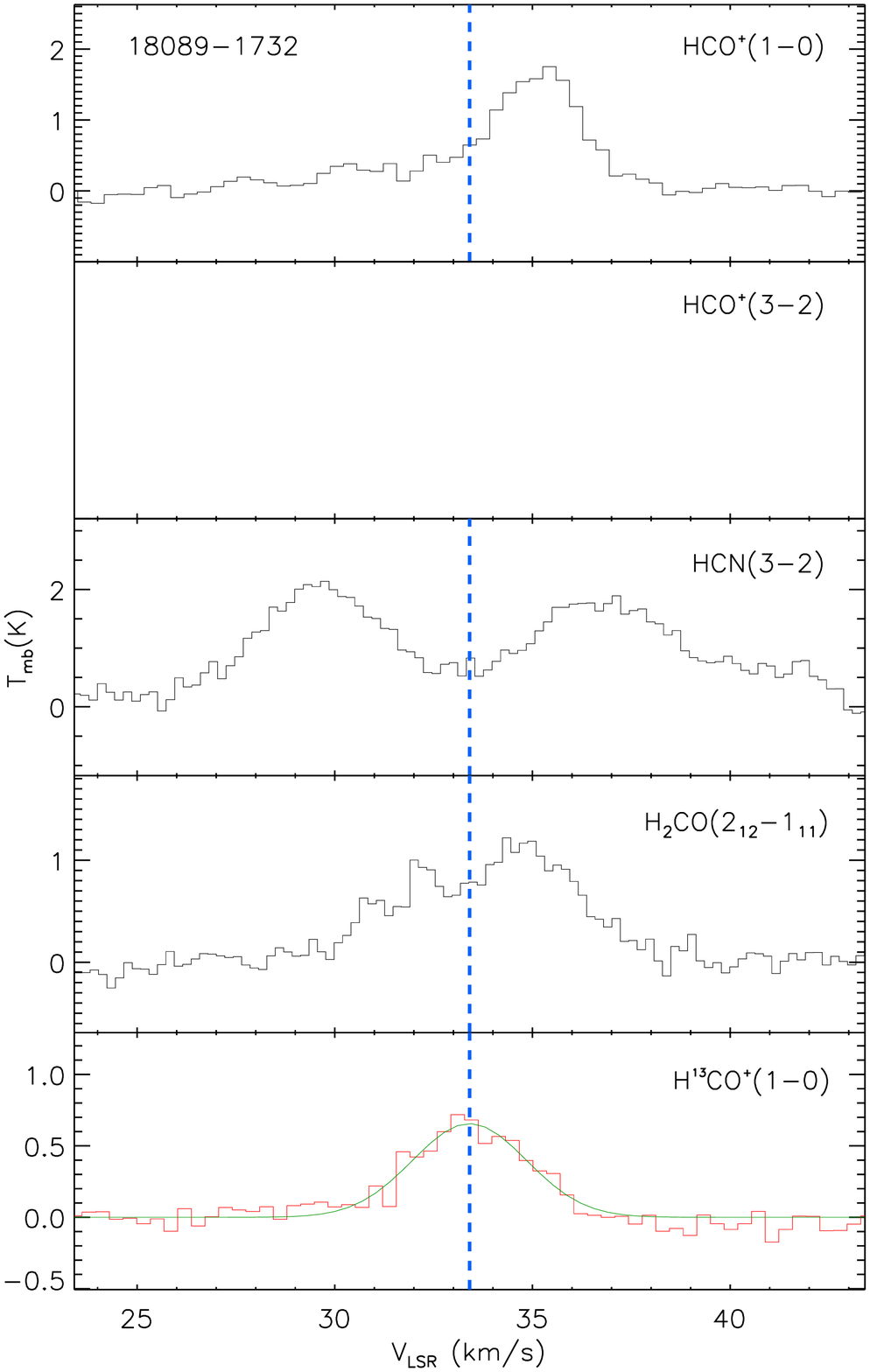} 
  \includegraphics[width=0.25\textwidth]{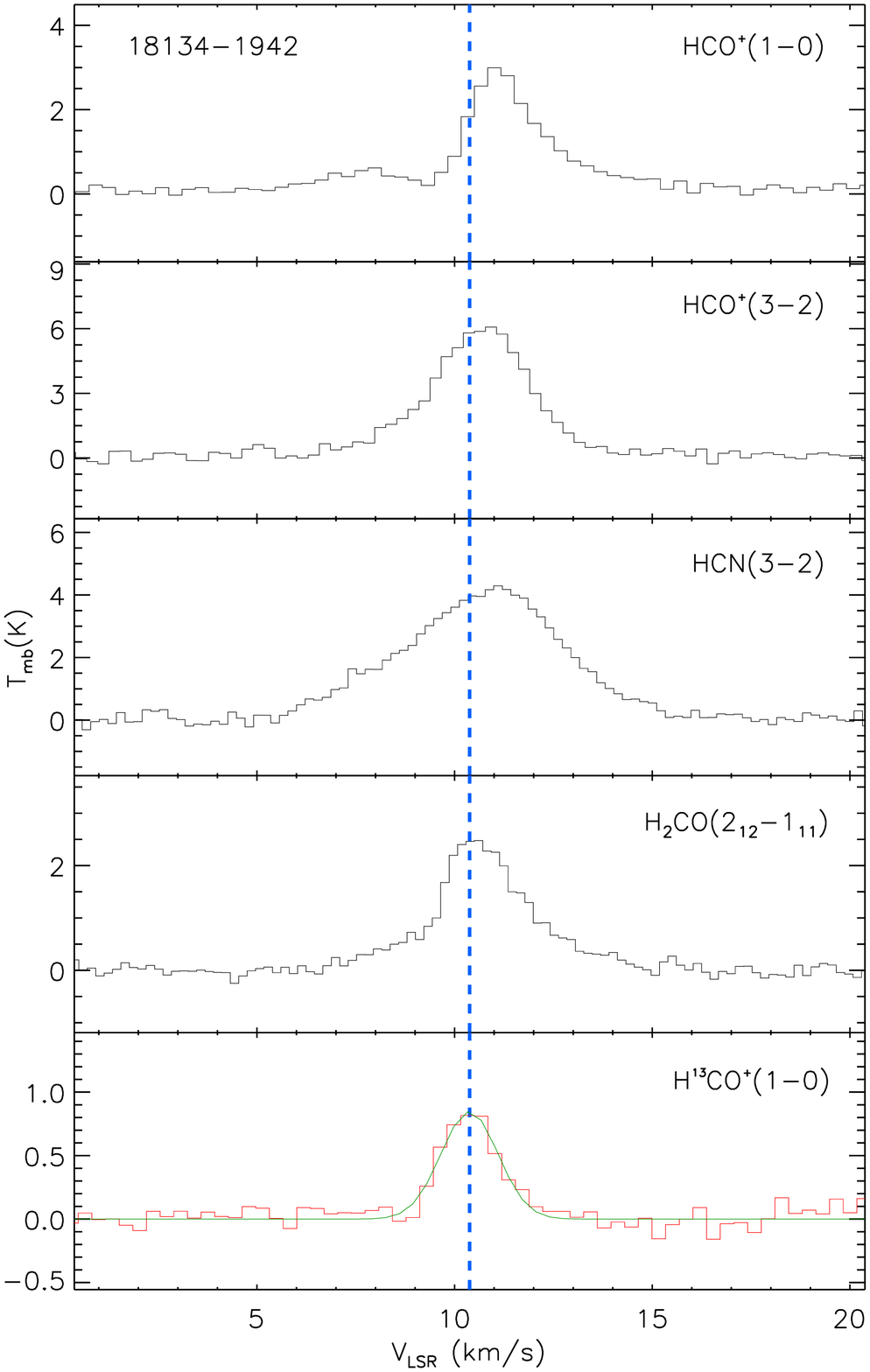} \\
  \includegraphics[width=0.25\textwidth]{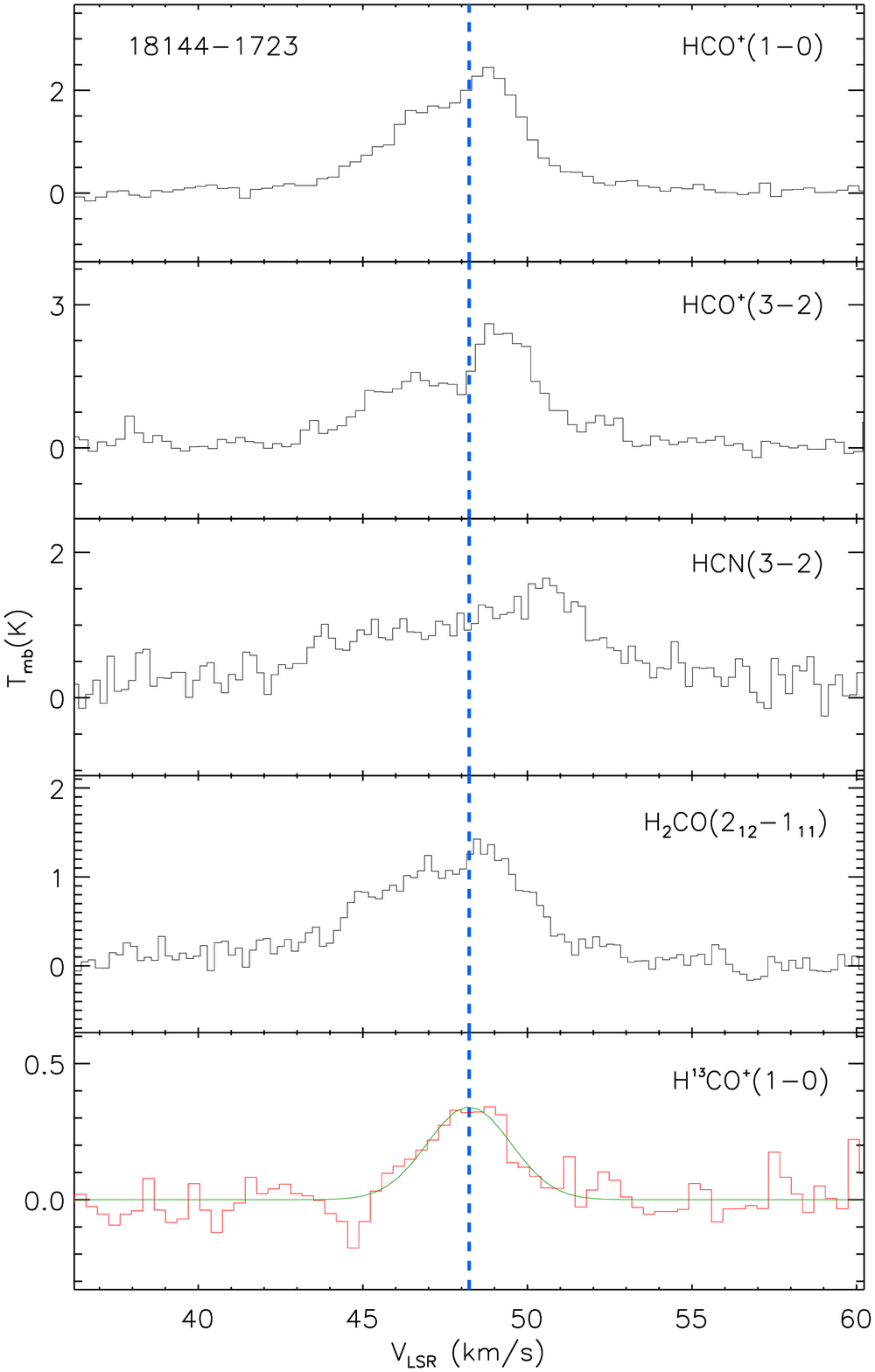} 
  \includegraphics[width=0.25\textwidth]{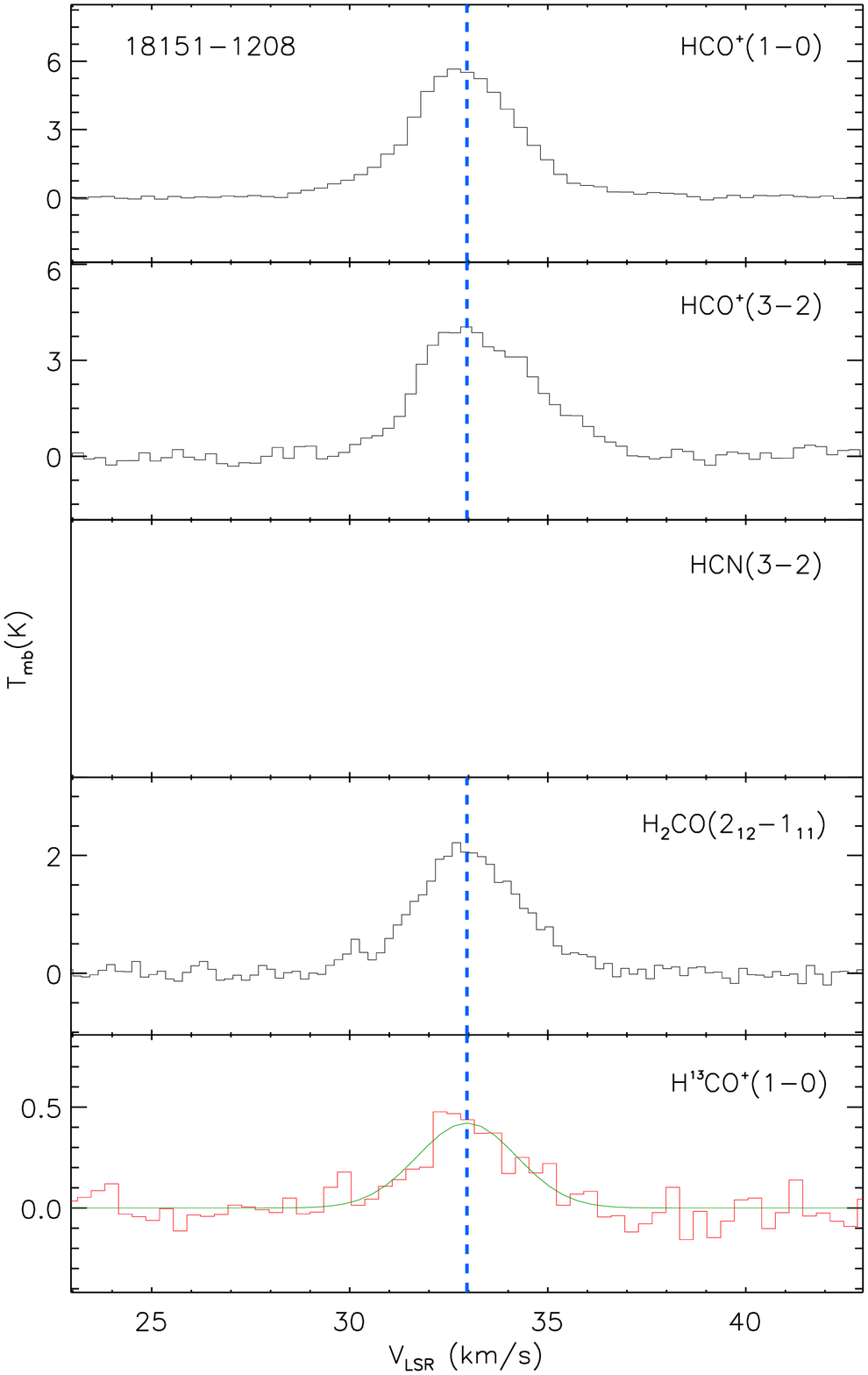} 
  \includegraphics[width=0.25\textwidth]{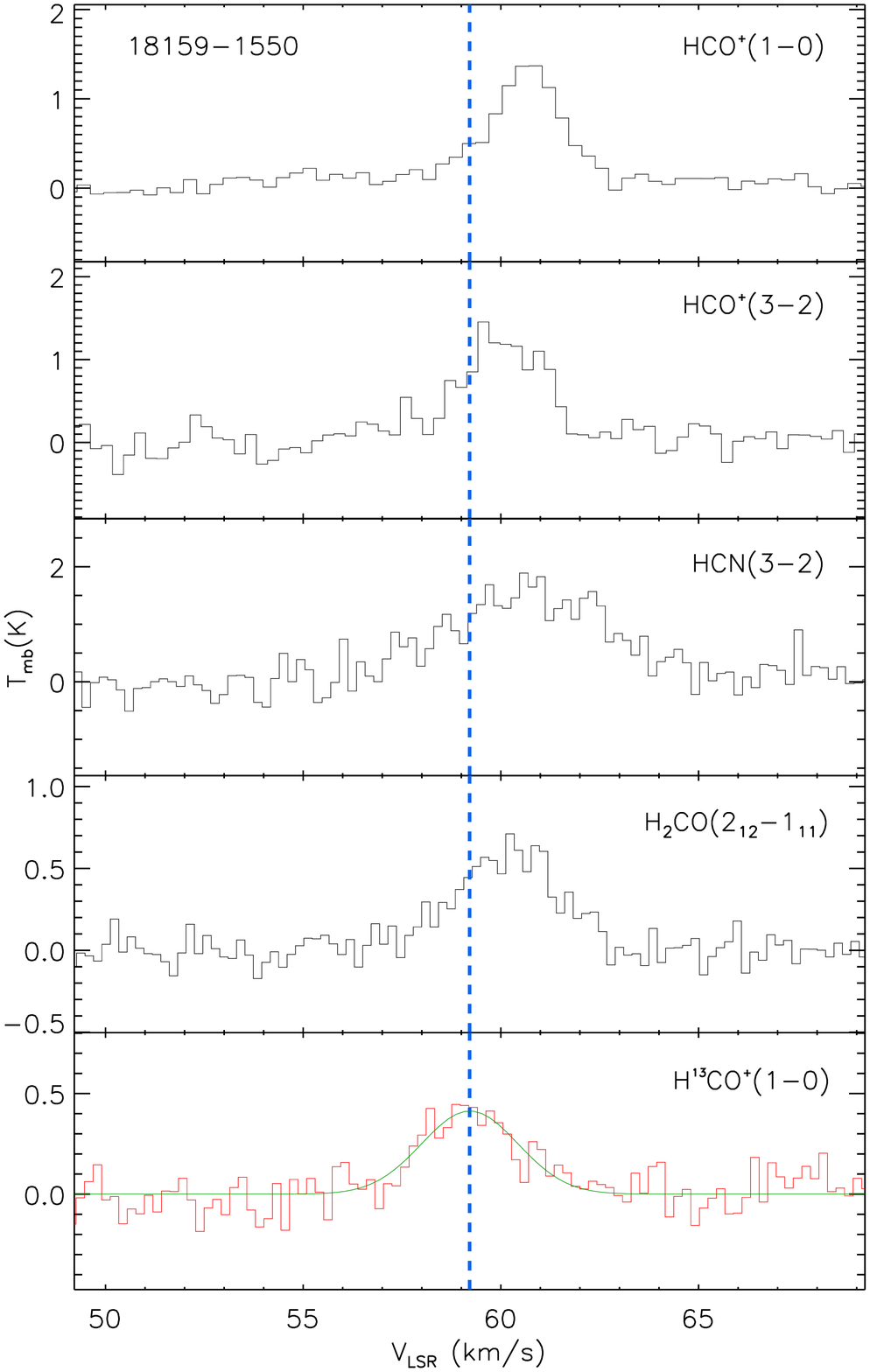} 
  \includegraphics[width=0.25\textwidth]{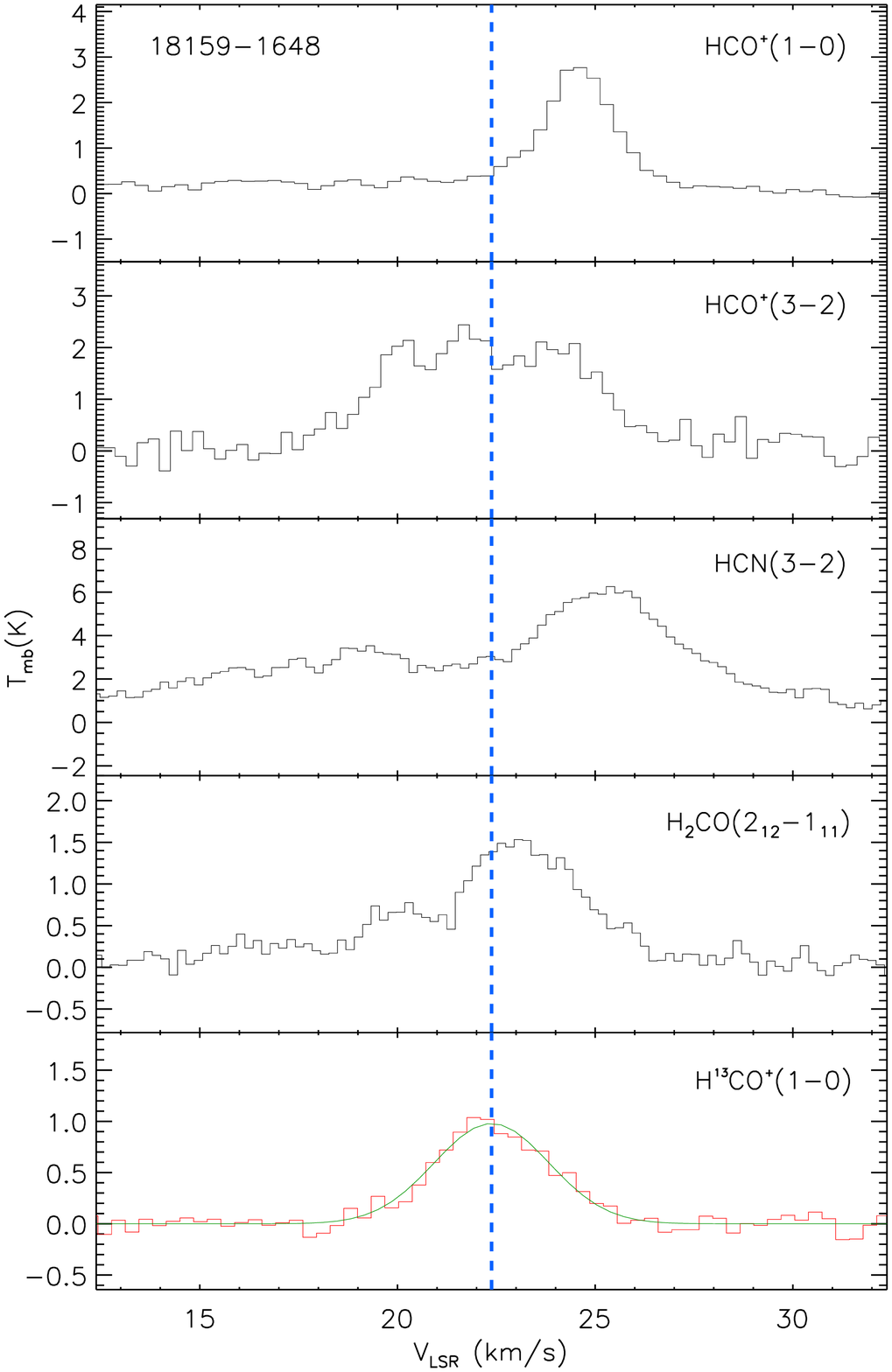} \\
  \includegraphics[width=0.25\textwidth]{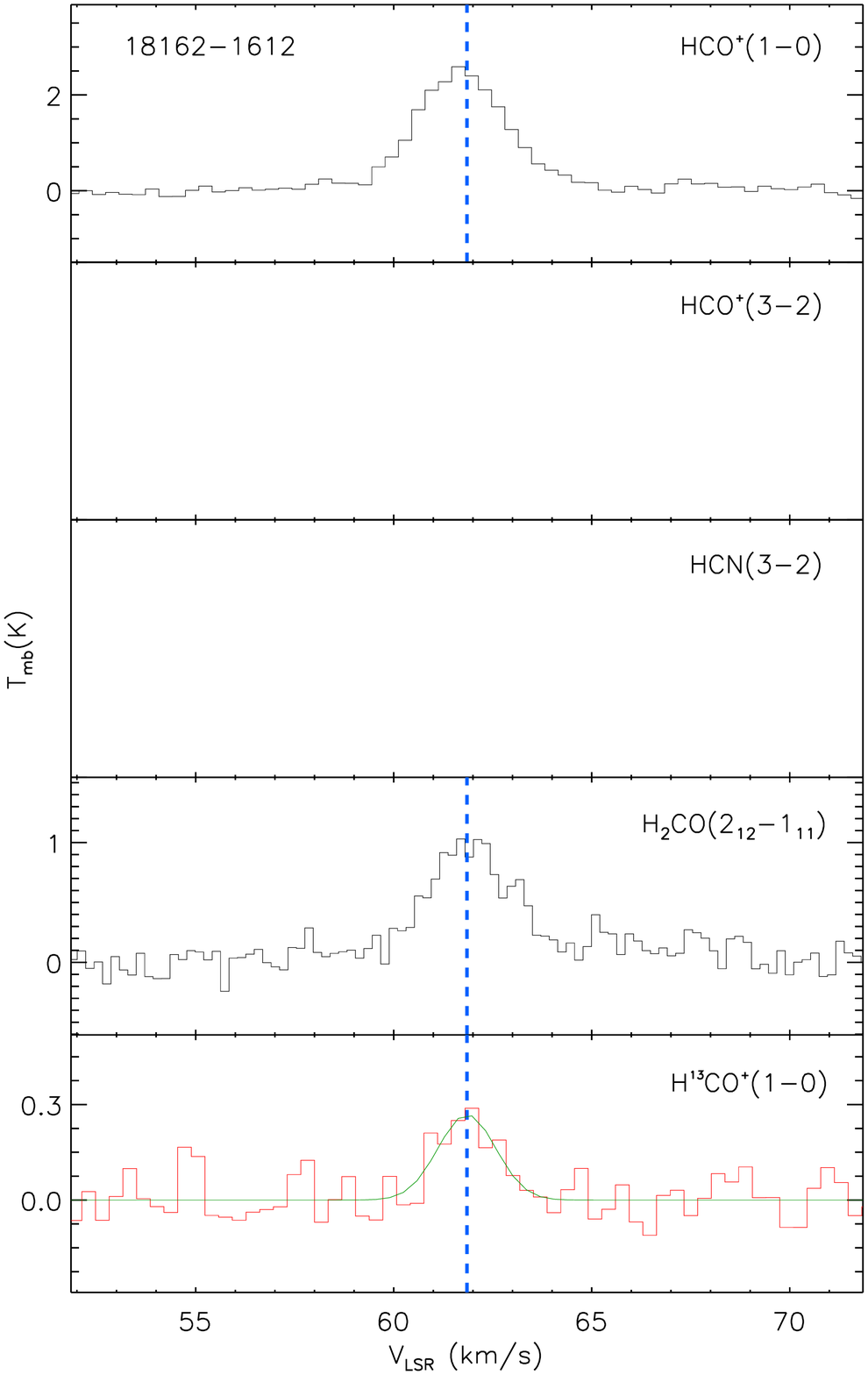} 
  \includegraphics[width=0.25\textwidth]{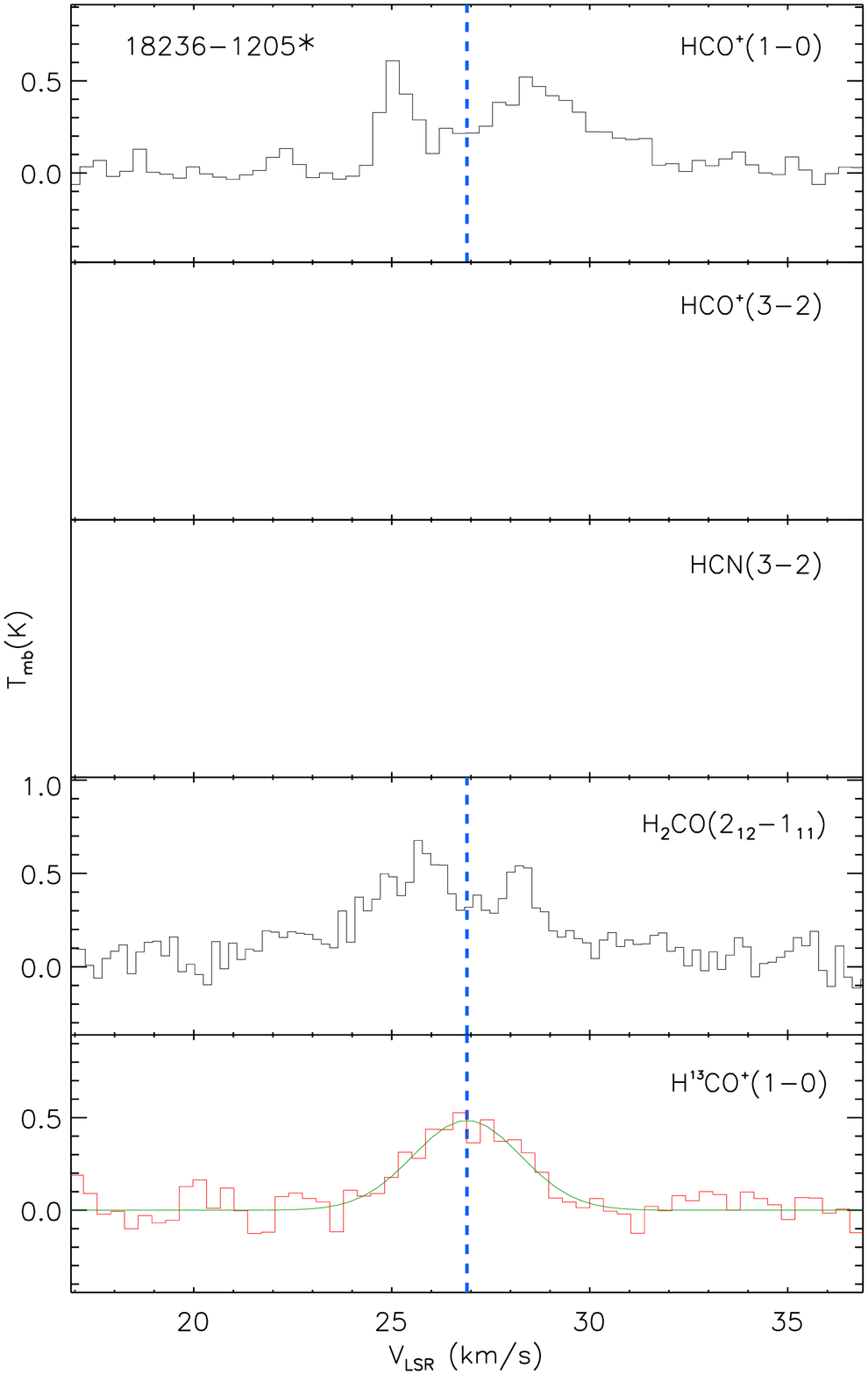} 
  \includegraphics[width=0.25\textwidth]{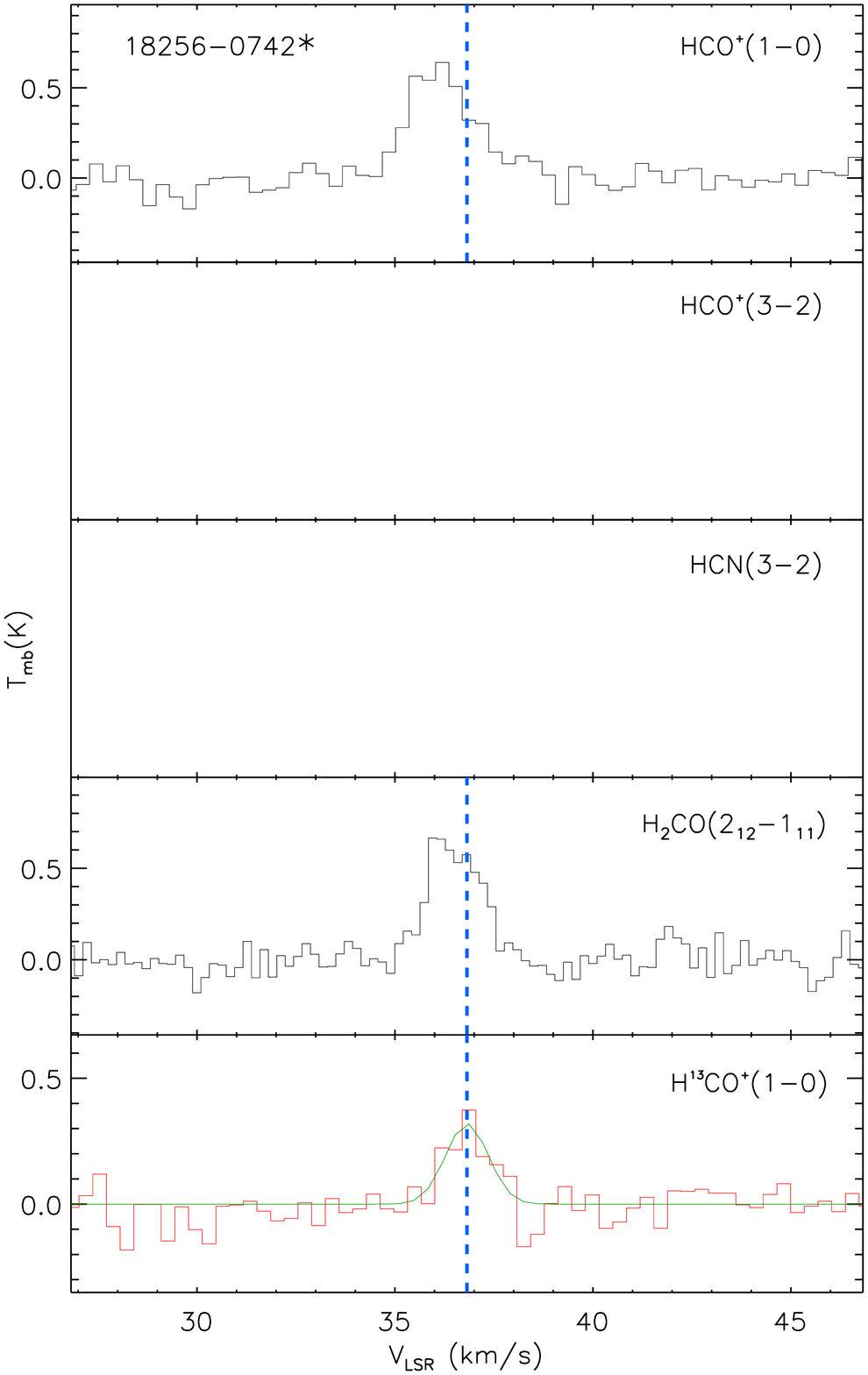} 
  \includegraphics[width=0.25\textwidth]{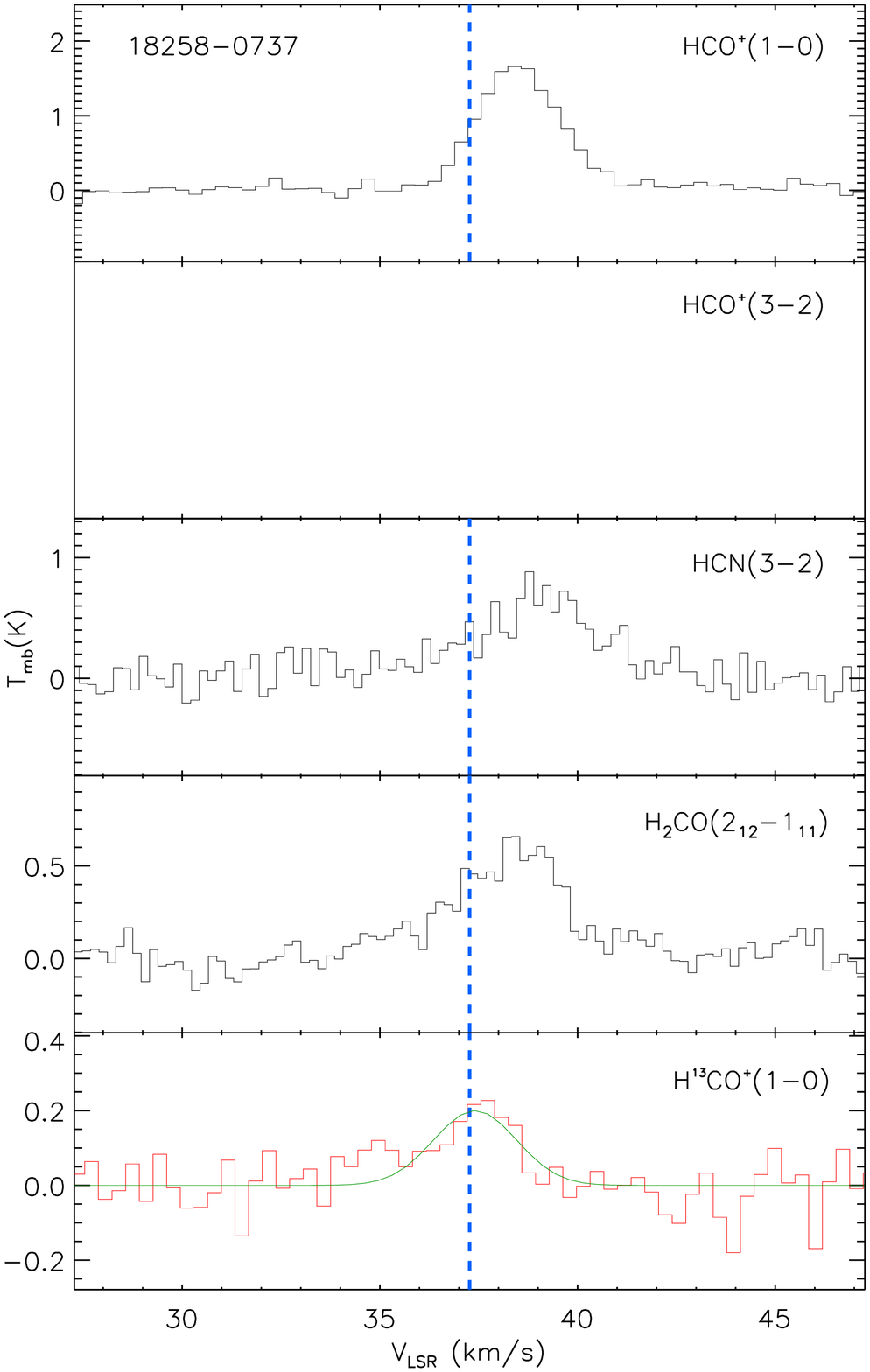} 
\end{tabbing}
\center{\textbf{Figure A1.} continued.}
\label{fA1}
\end{figure*}

\begin{figure*}
\begin{tabbing}
  \includegraphics[width=0.25\textwidth]{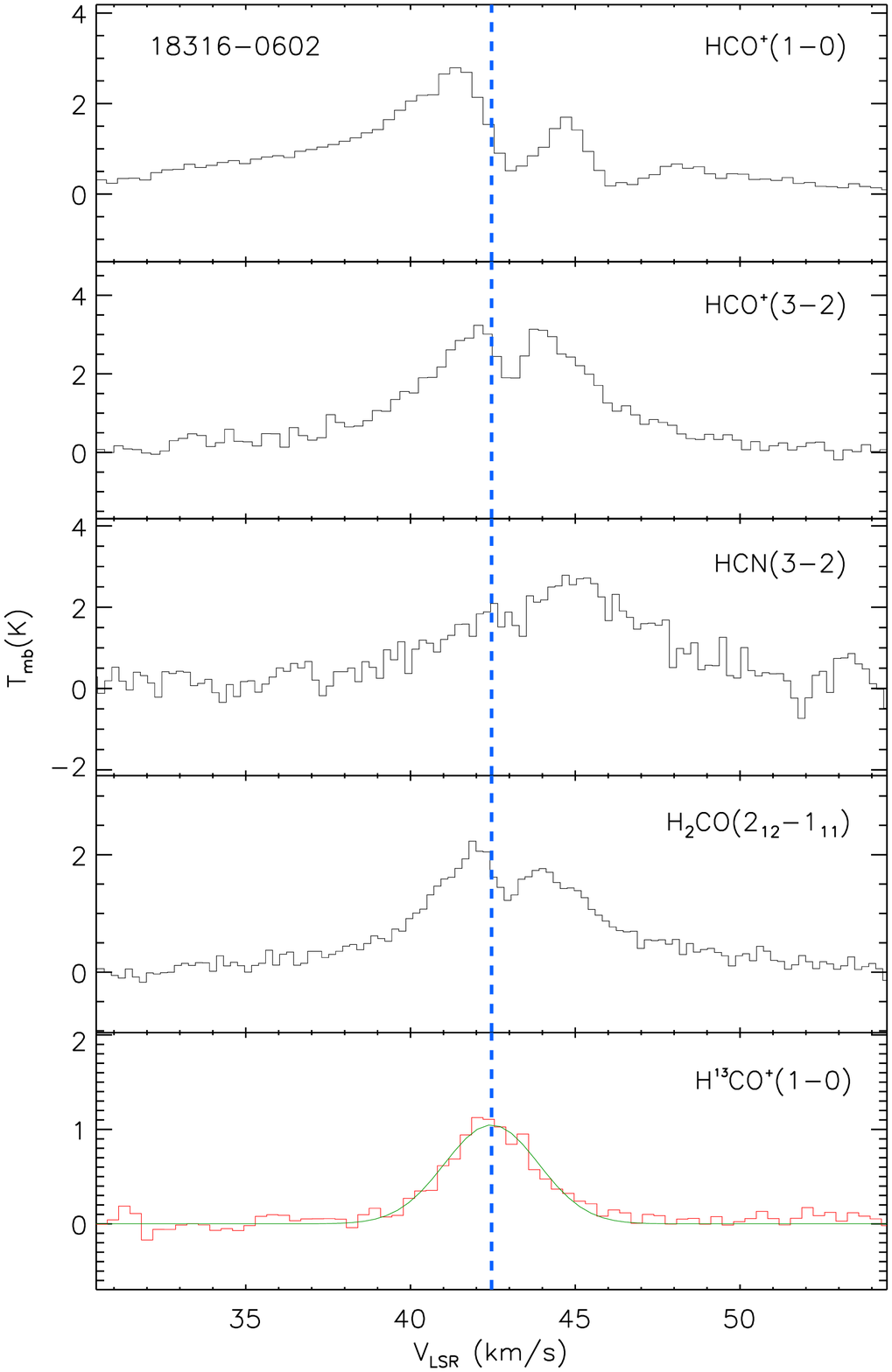} 
  \includegraphics[width=0.25\textwidth]{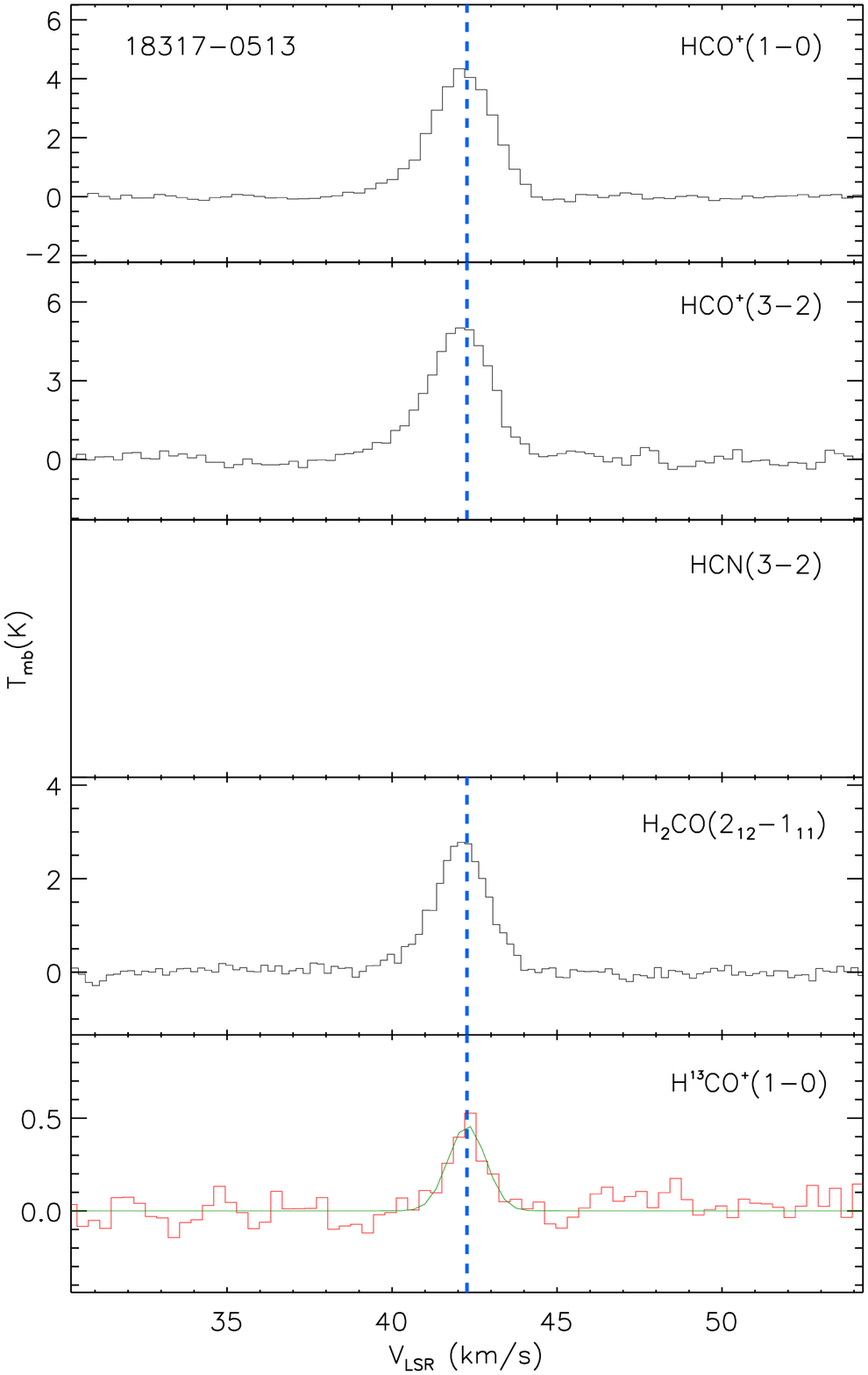}
  \includegraphics[width=0.25\textwidth]{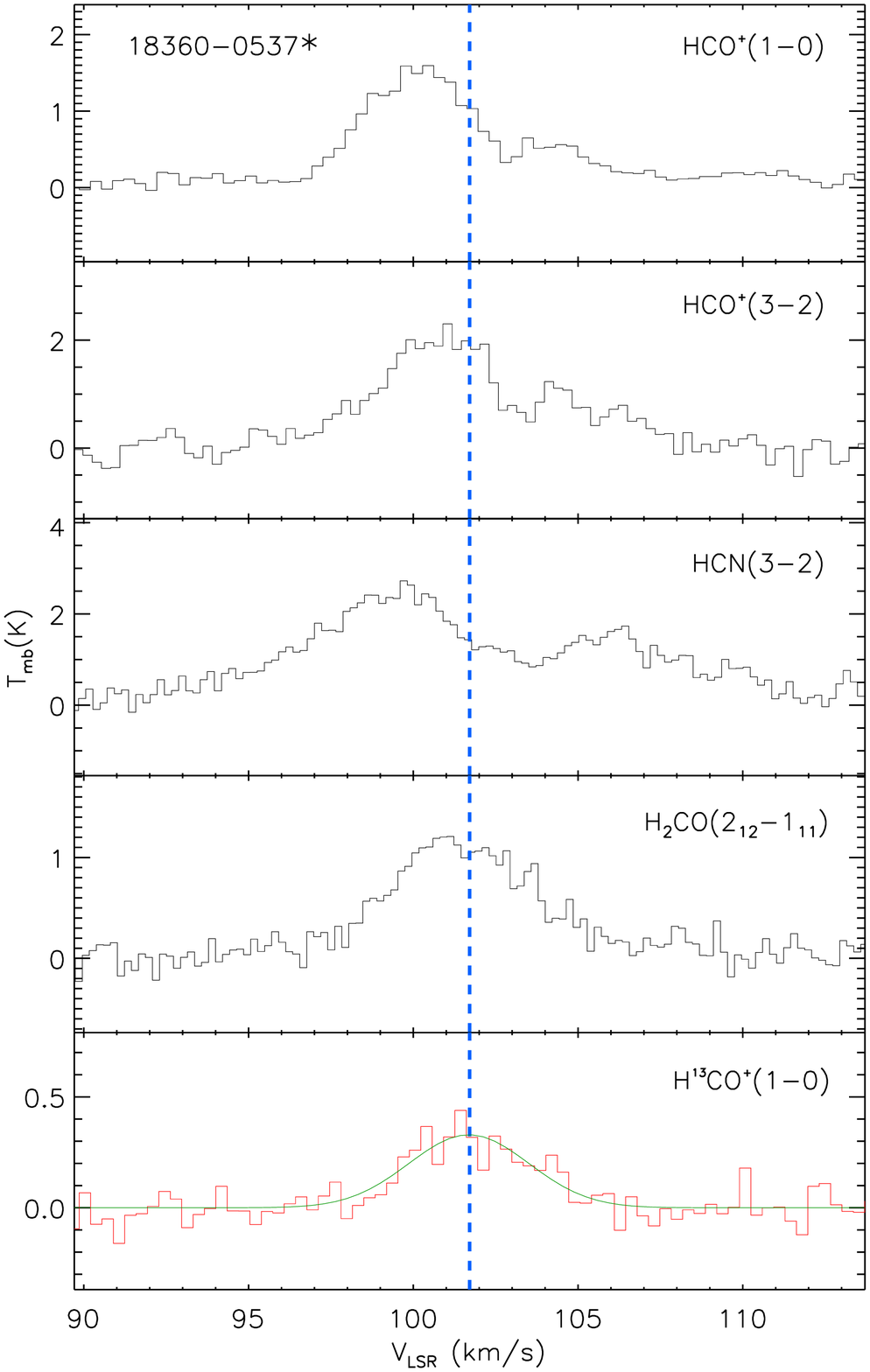}
  \includegraphics[width=0.25\textwidth]{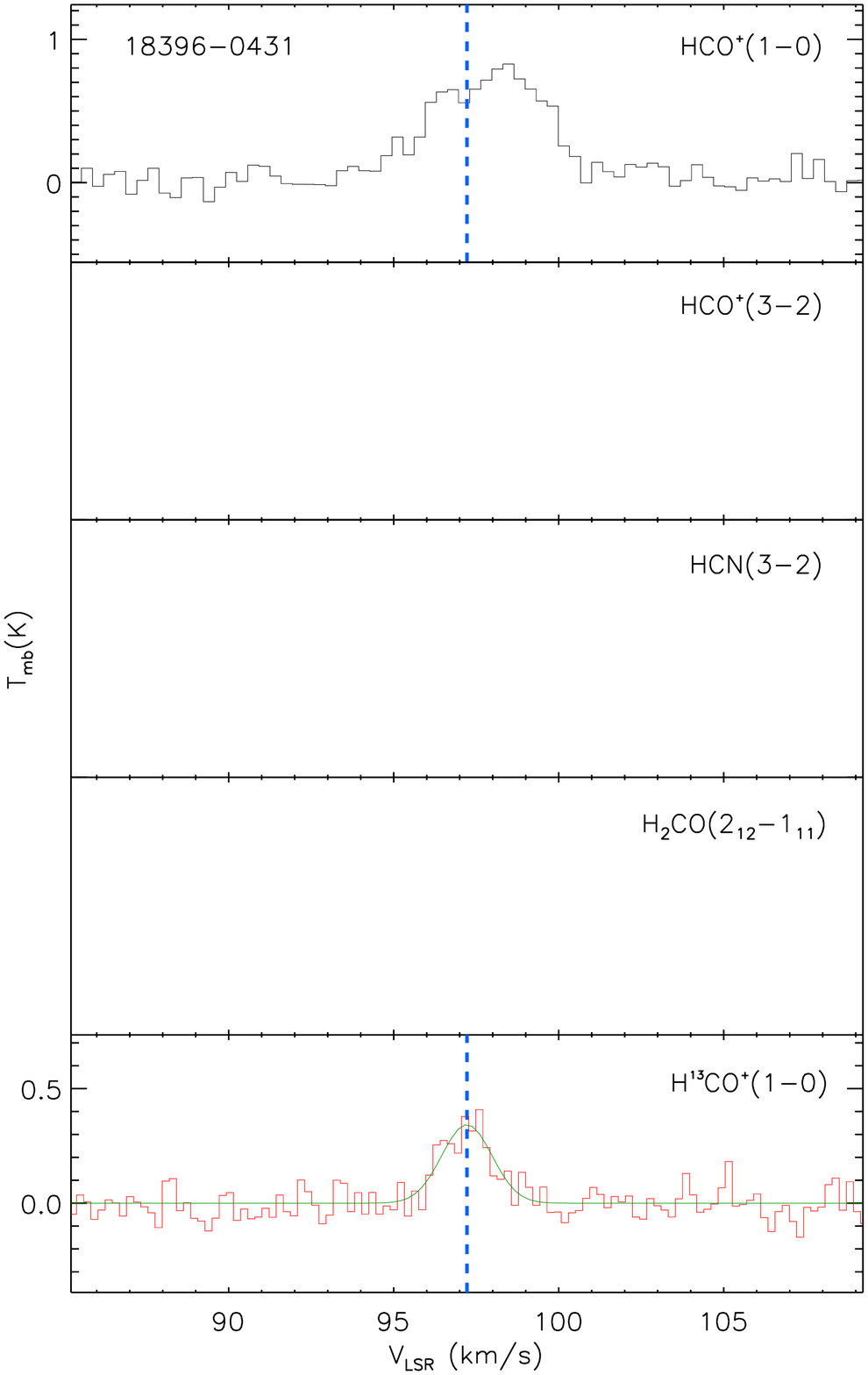} \\
  \includegraphics[width=0.25\textwidth]{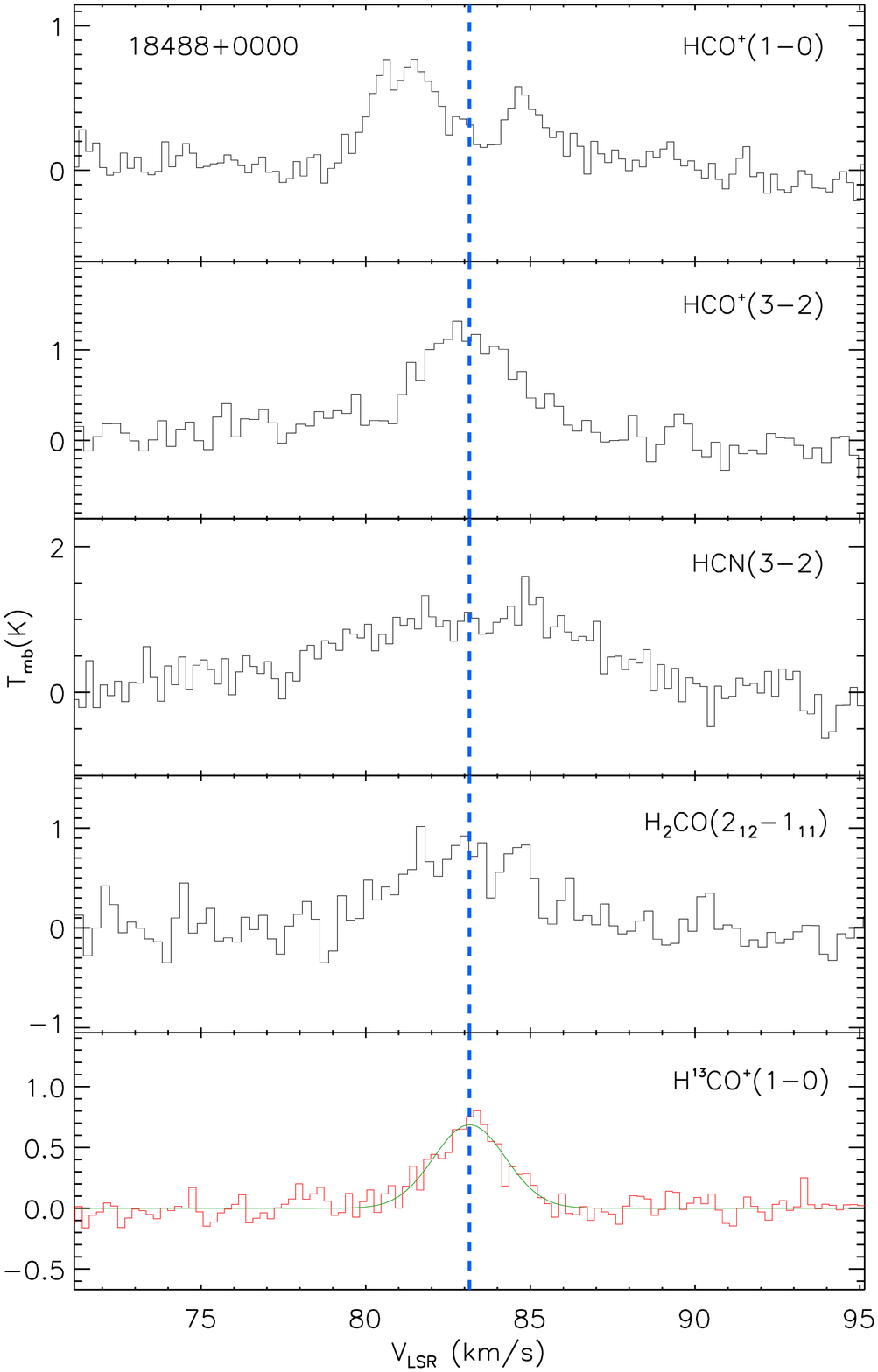}  
  \includegraphics[width=0.25\textwidth]{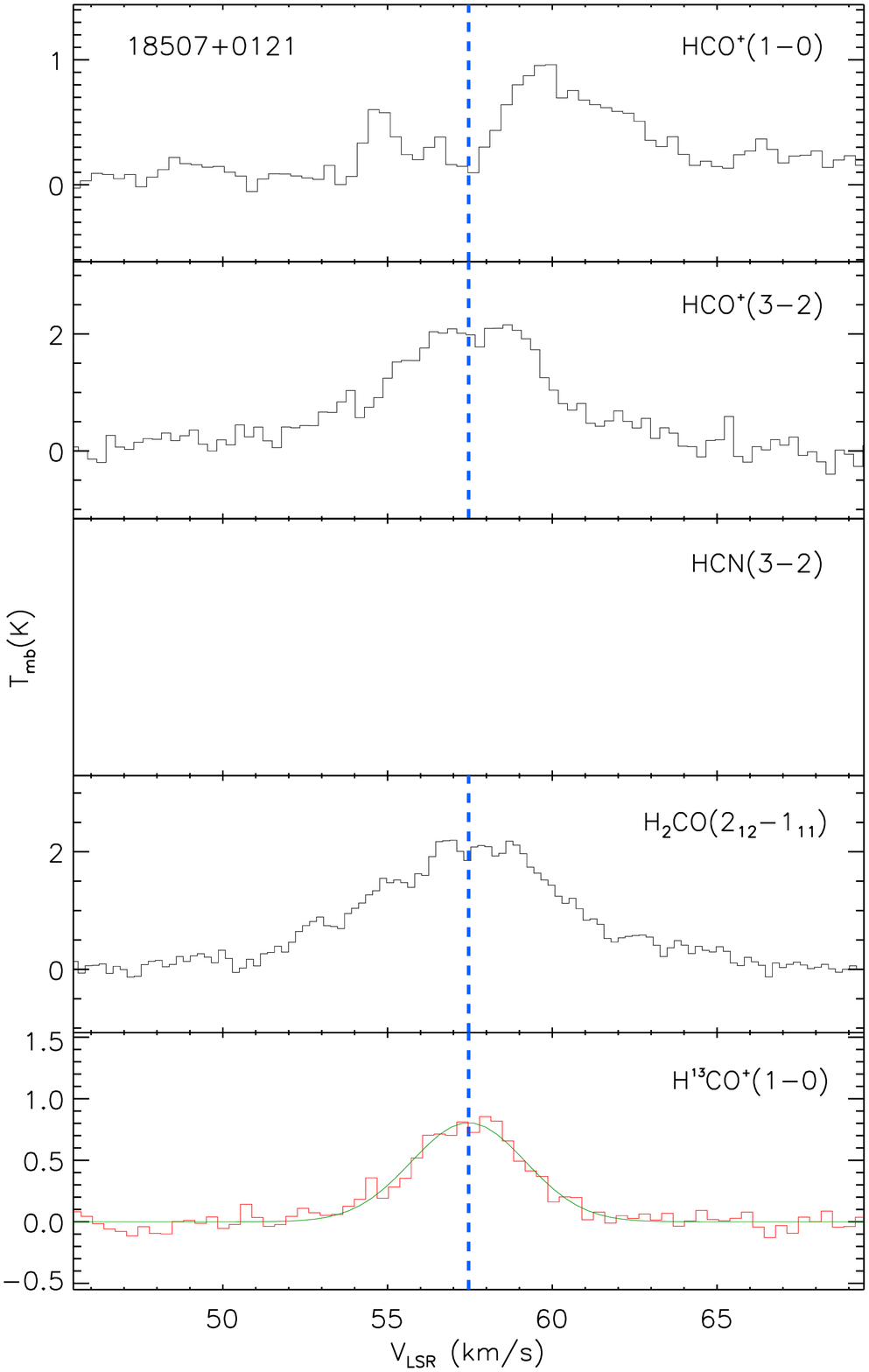} 
  \includegraphics[width=0.25\textwidth]{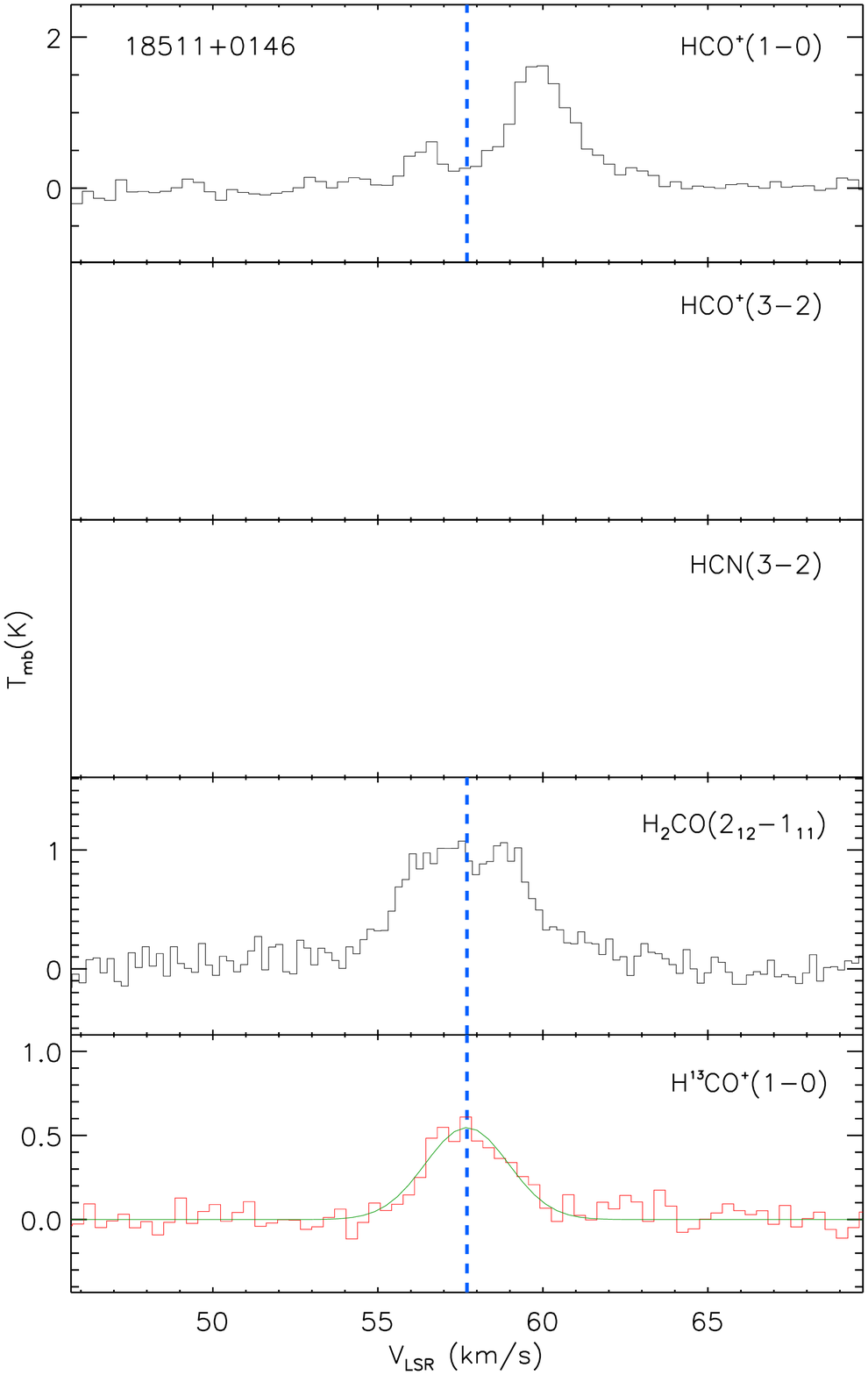} 
  \includegraphics[width=0.25\textwidth]{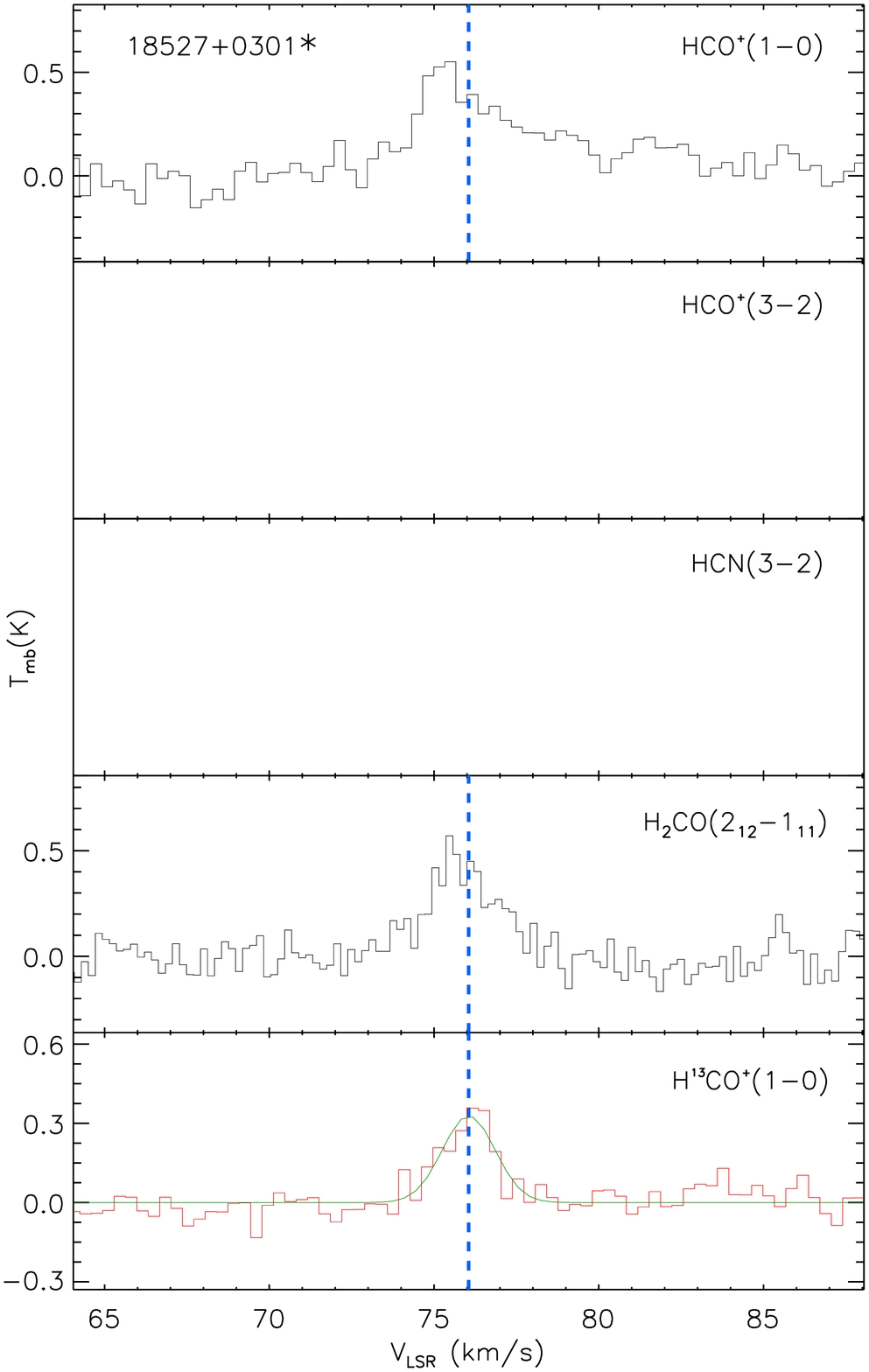} \\
  \includegraphics[width=0.25\textwidth]{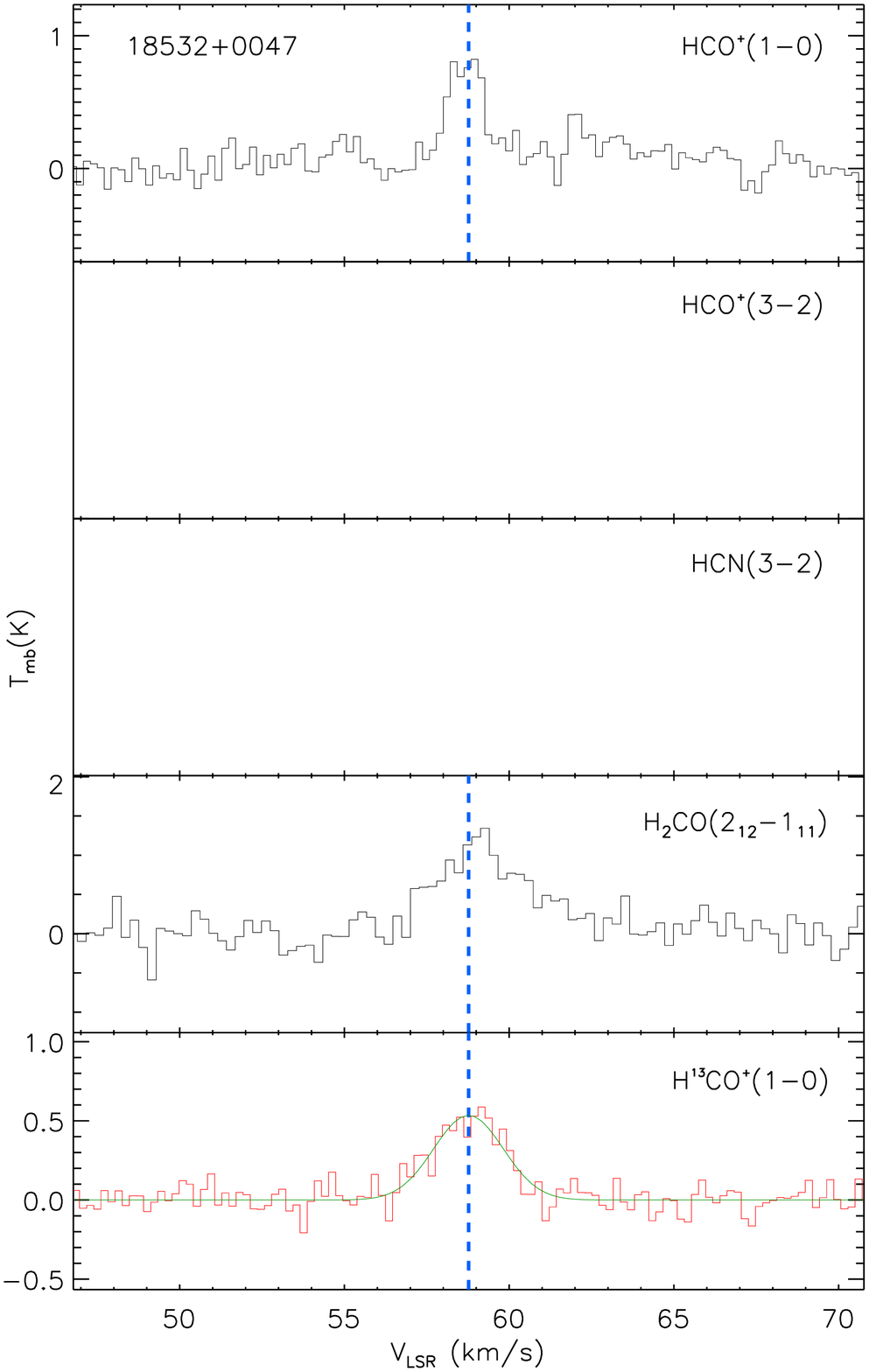} 
  \includegraphics[width=0.25\textwidth]{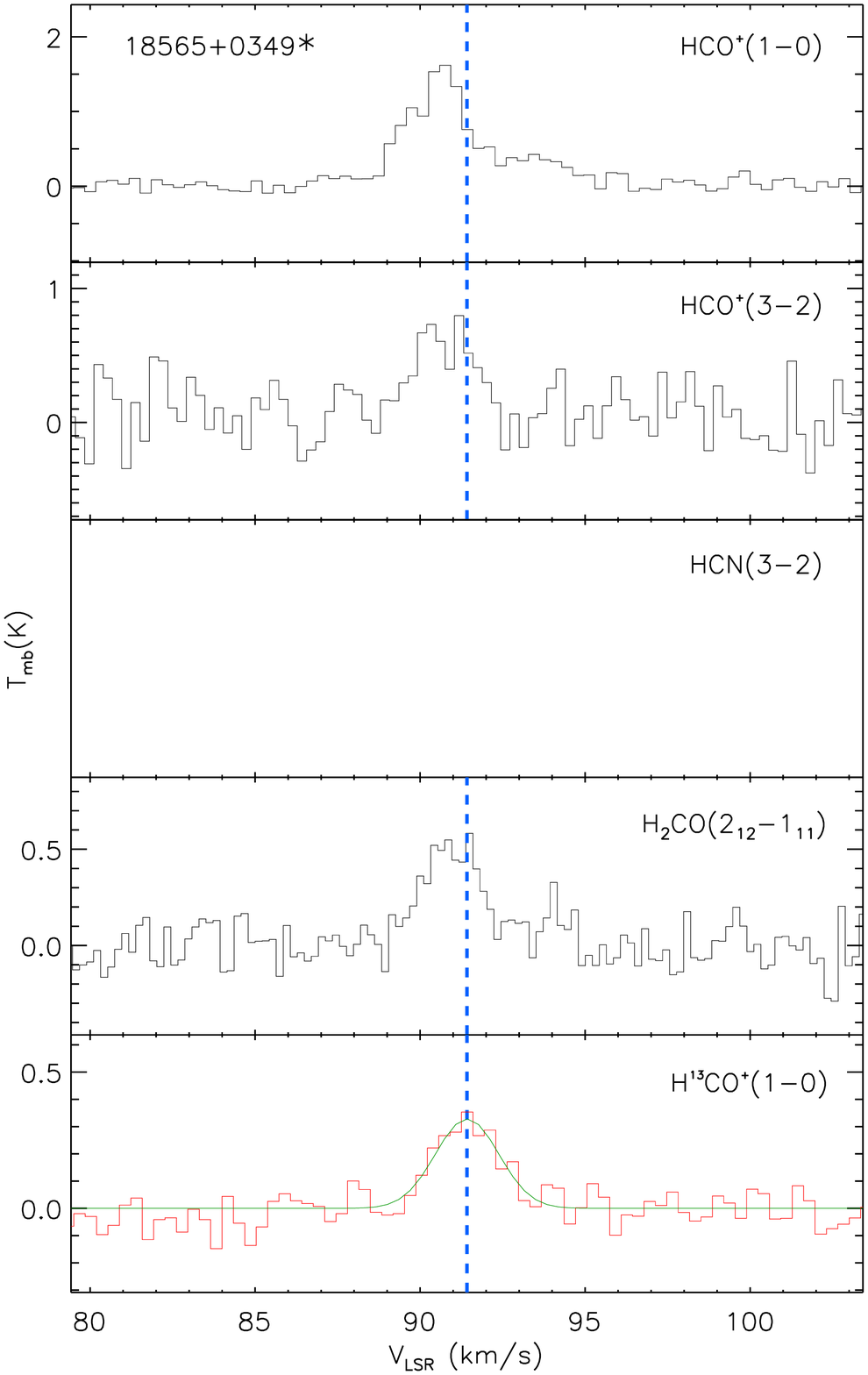} 
  \includegraphics[width=0.25\textwidth]{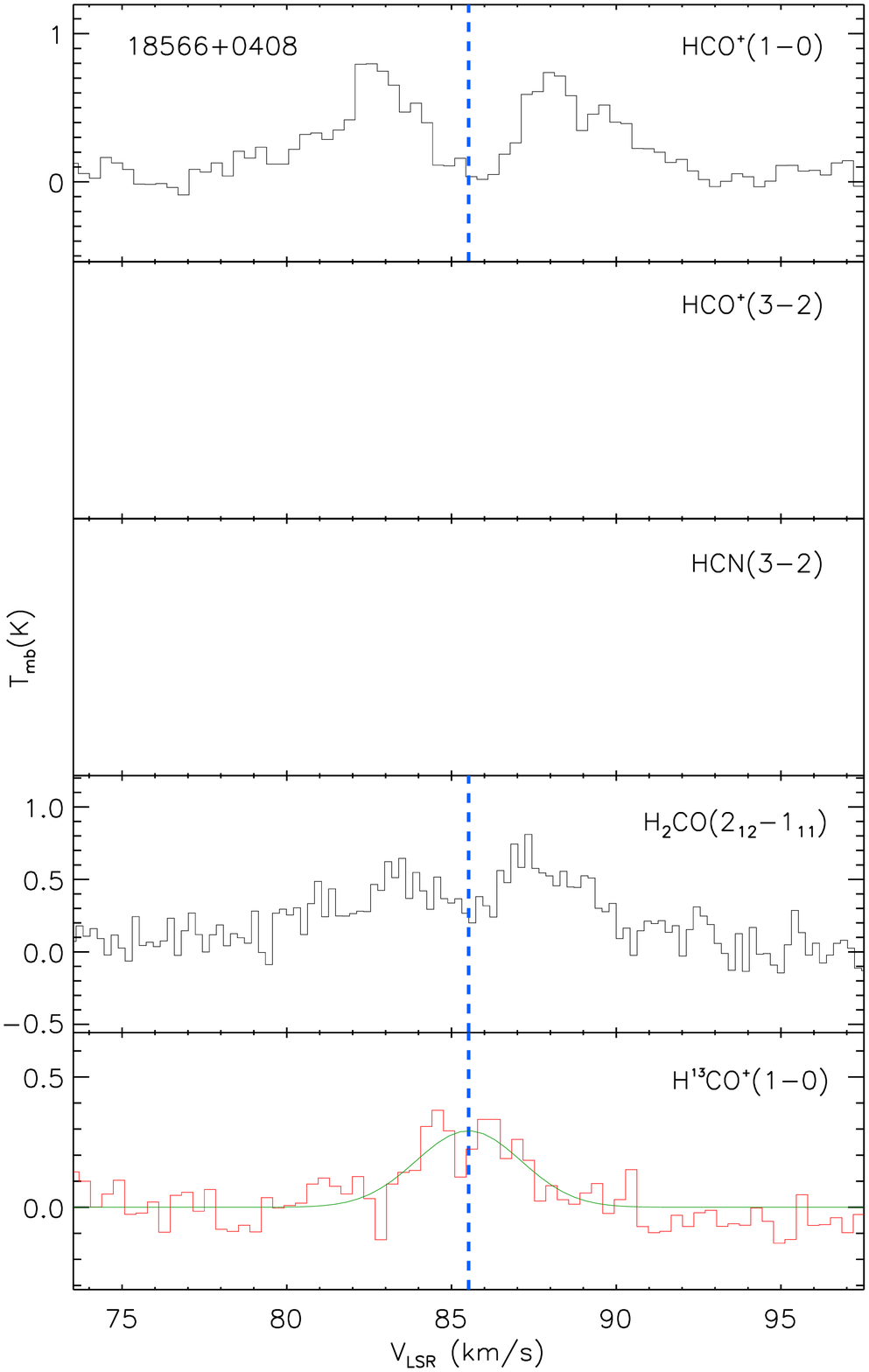} 
  \includegraphics[width=0.25\textwidth]{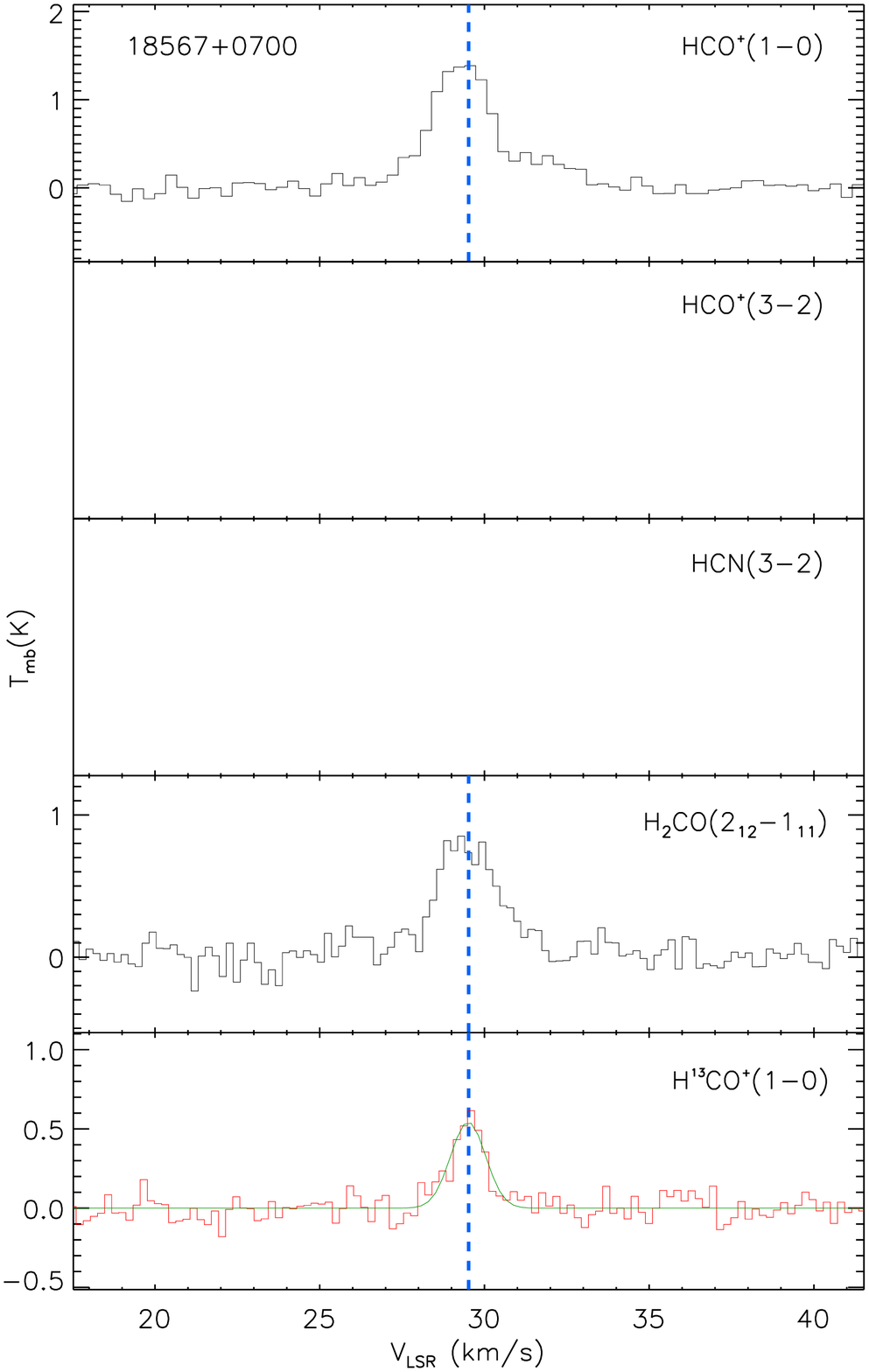} 
\end{tabbing}
\center{\textbf{Figure A1.} continued.}
\label{fA1}
\end{figure*}

\begin{figure*}
\begin{tabbing}
  \includegraphics[width=0.25\textwidth]{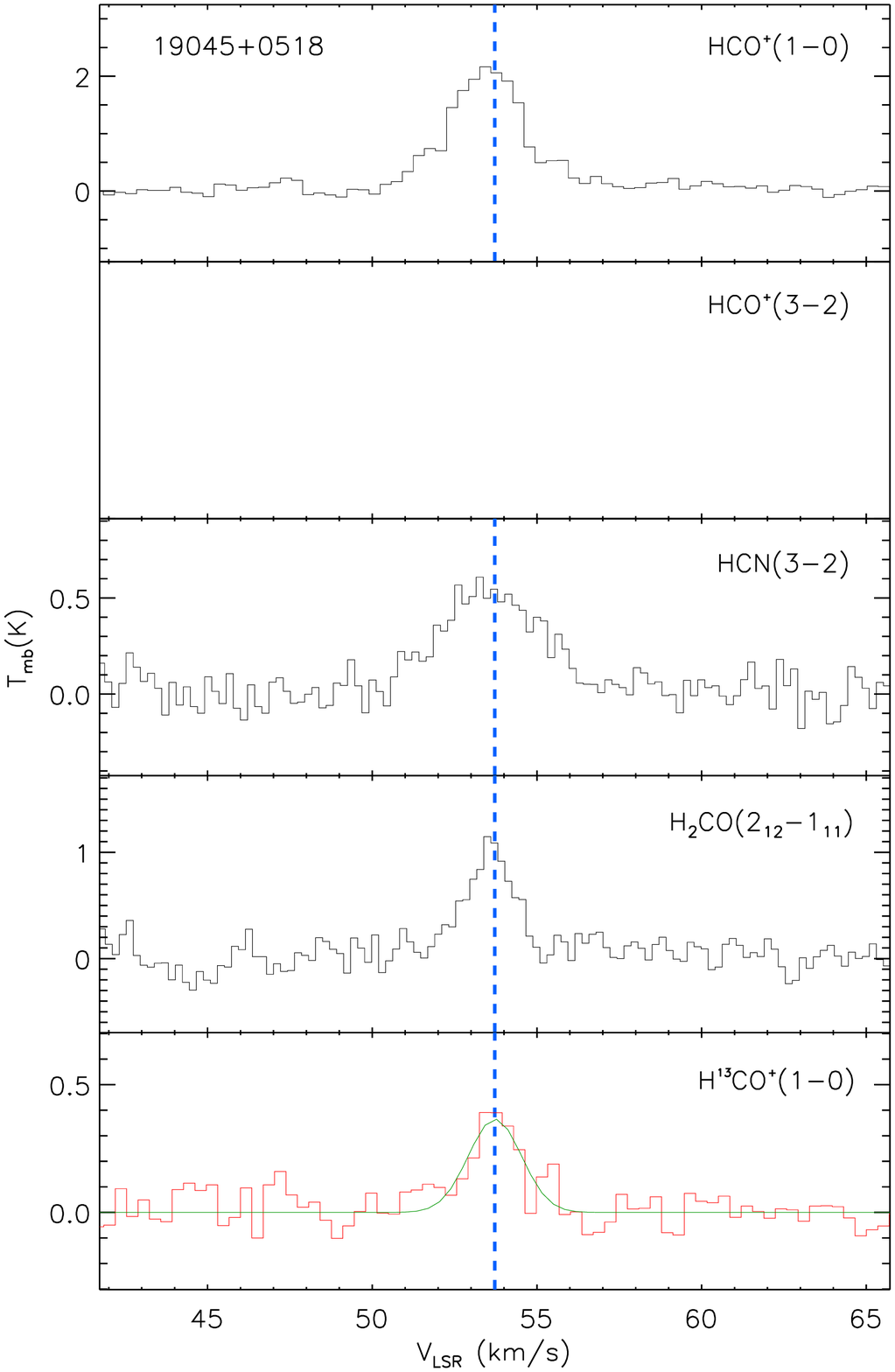} 
  \includegraphics[width=0.25\textwidth]{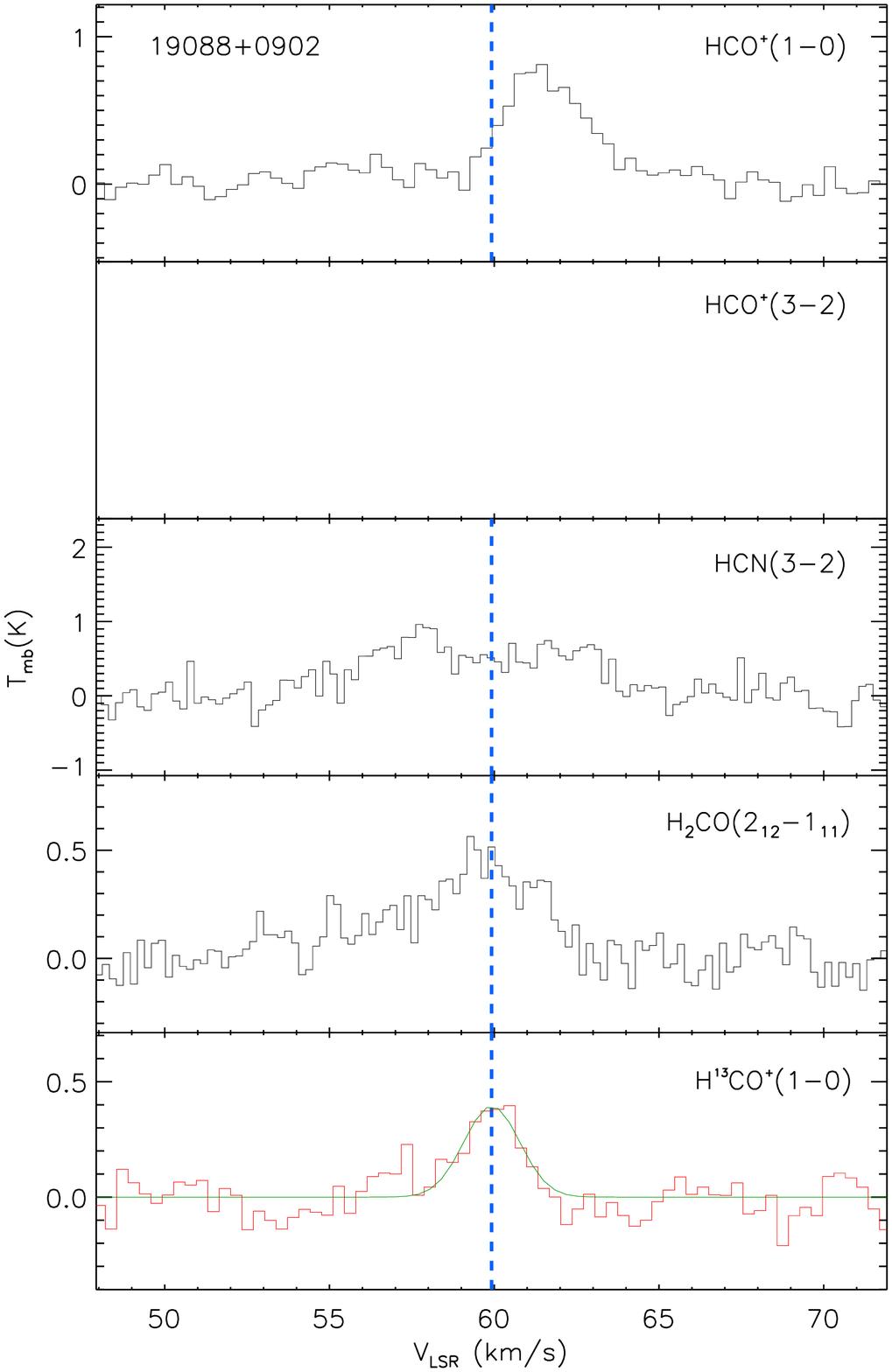} 
  \includegraphics[width=0.25\textwidth]{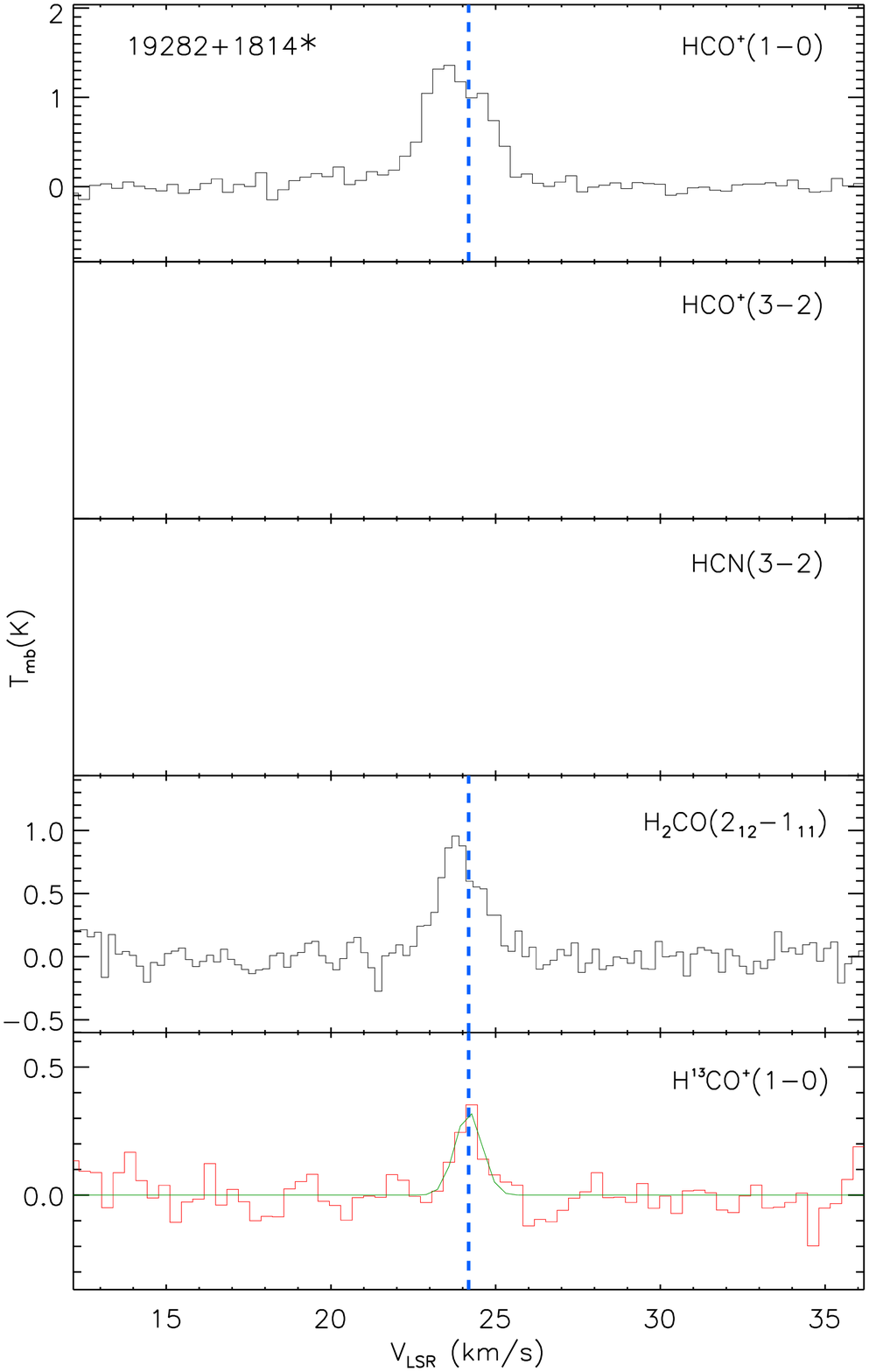}
  \includegraphics[width=0.25\textwidth]{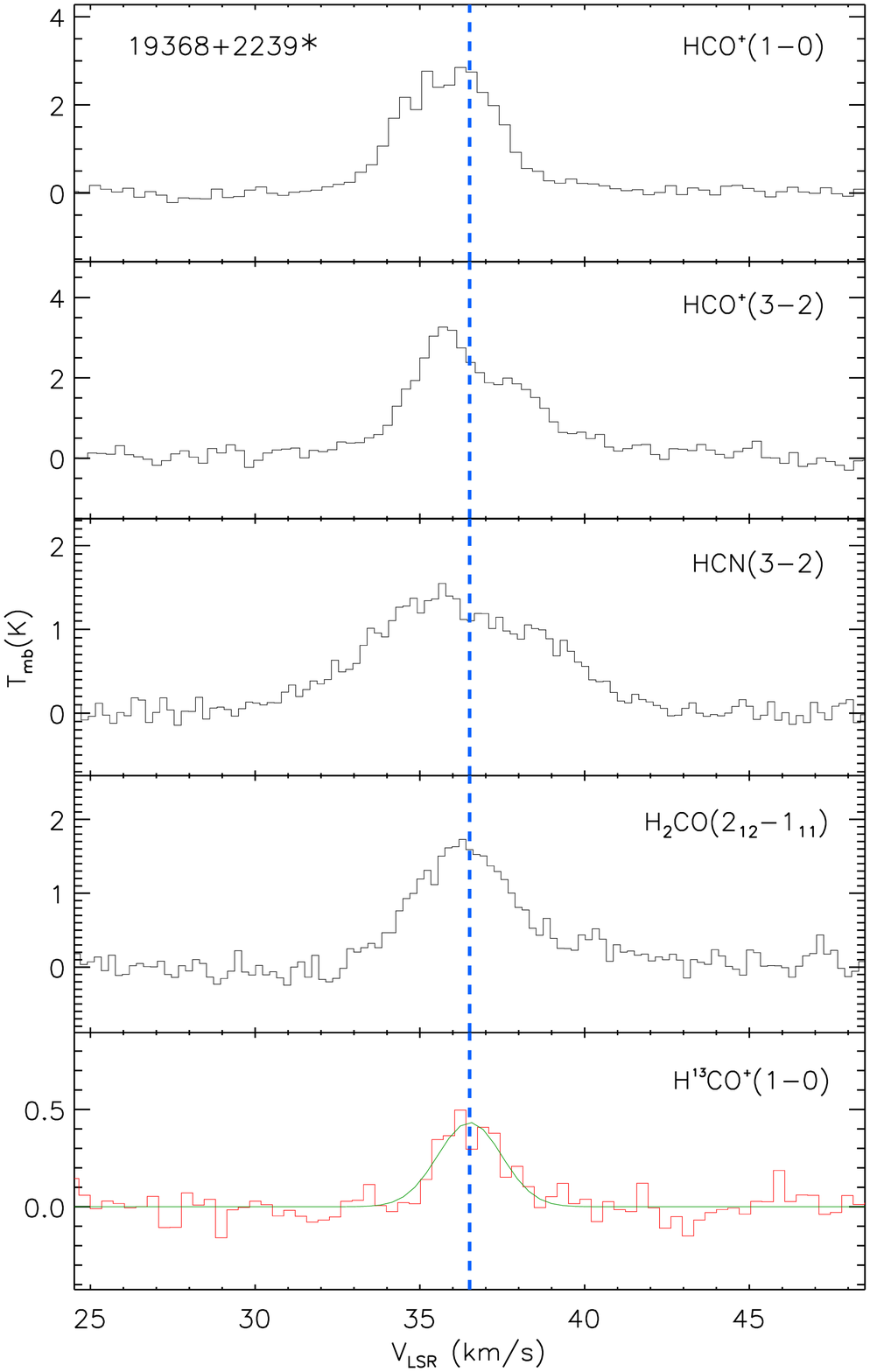} \\
  \includegraphics[width=0.25\textwidth]{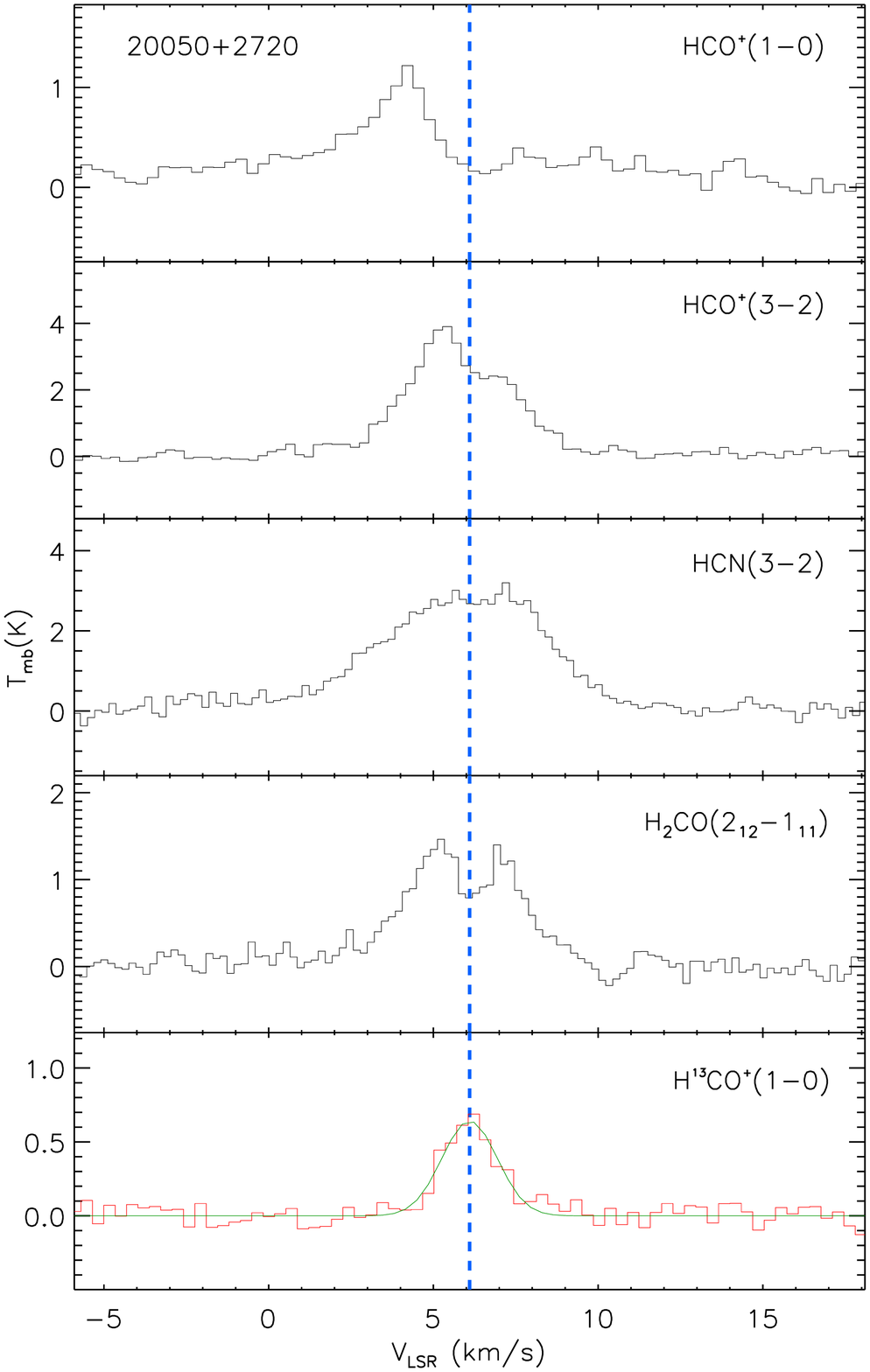} 
  \includegraphics[width=0.25\textwidth]{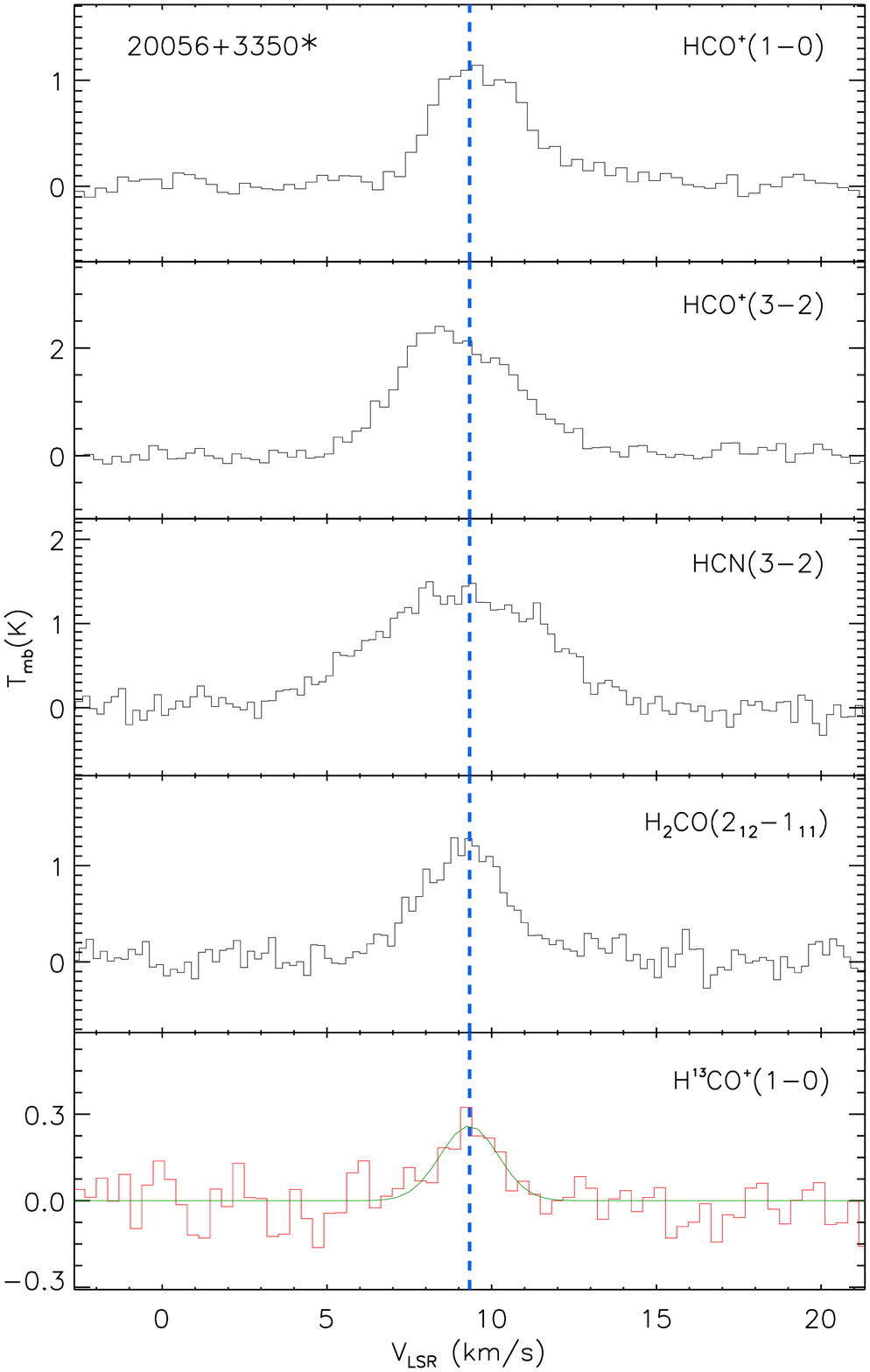} 
  \includegraphics[width=0.25\textwidth]{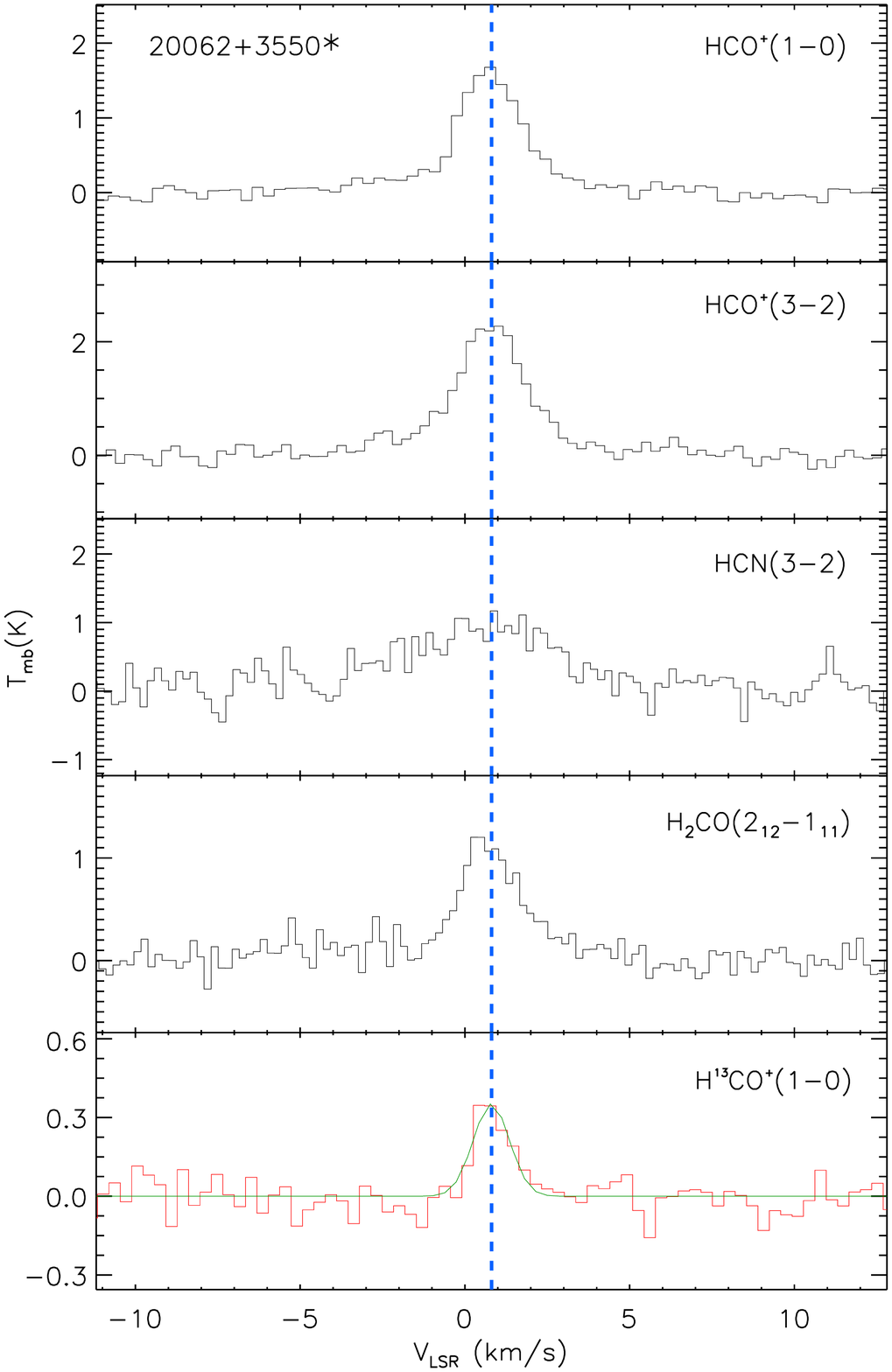}
  \includegraphics[width=0.25\textwidth]{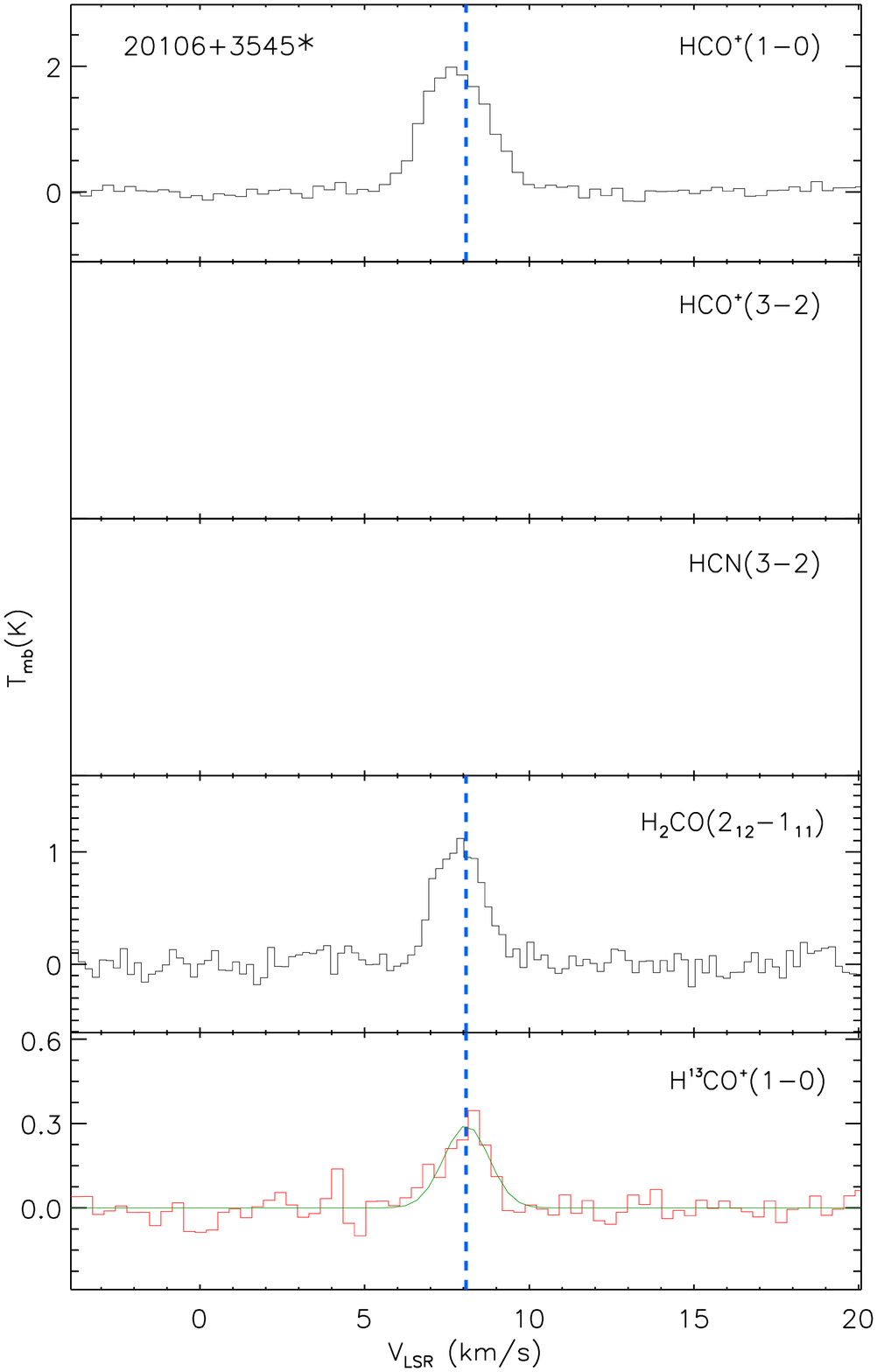} \\
  \includegraphics[width=0.25\textwidth]{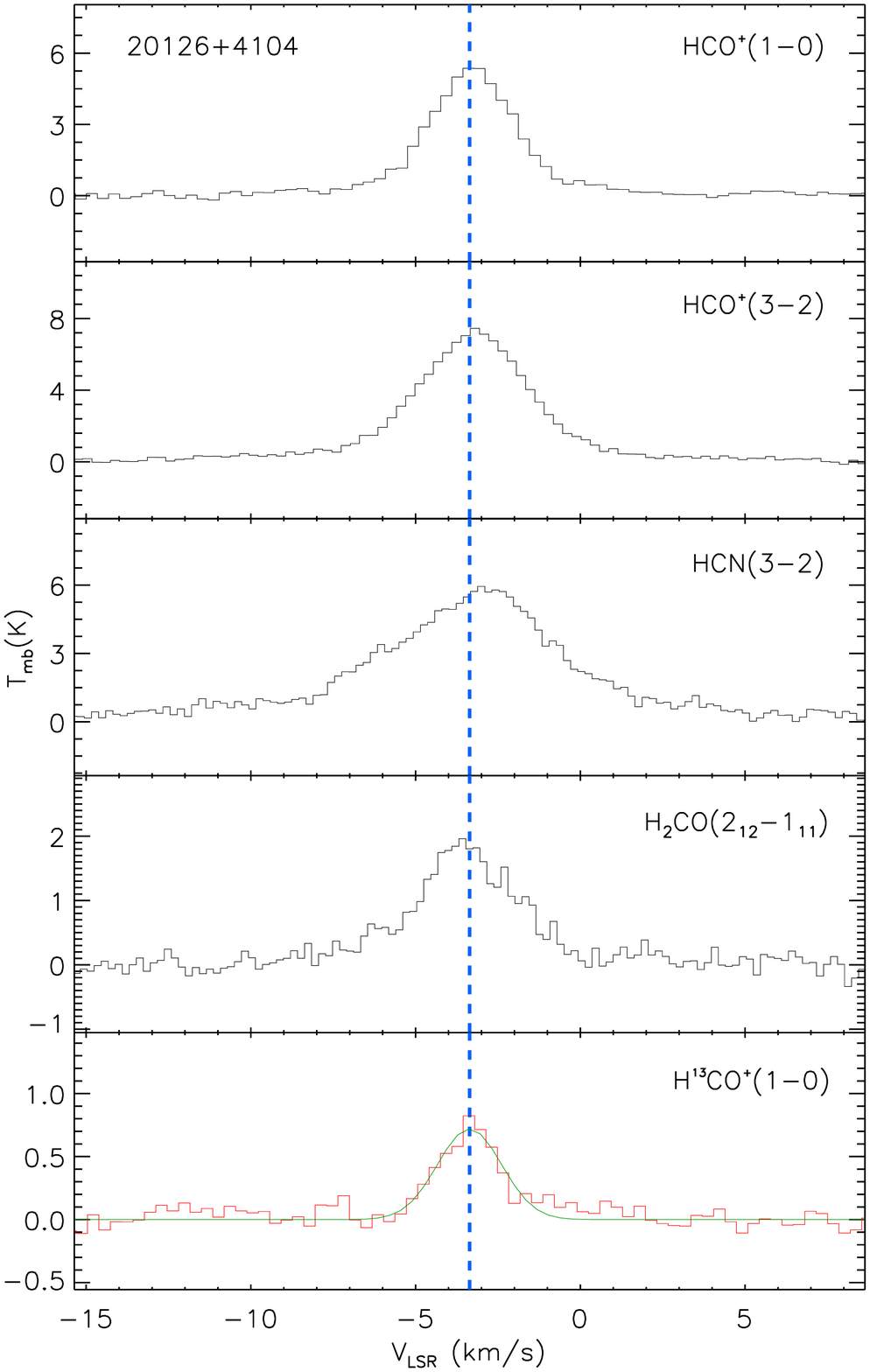} 
  \includegraphics[width=0.25\textwidth]{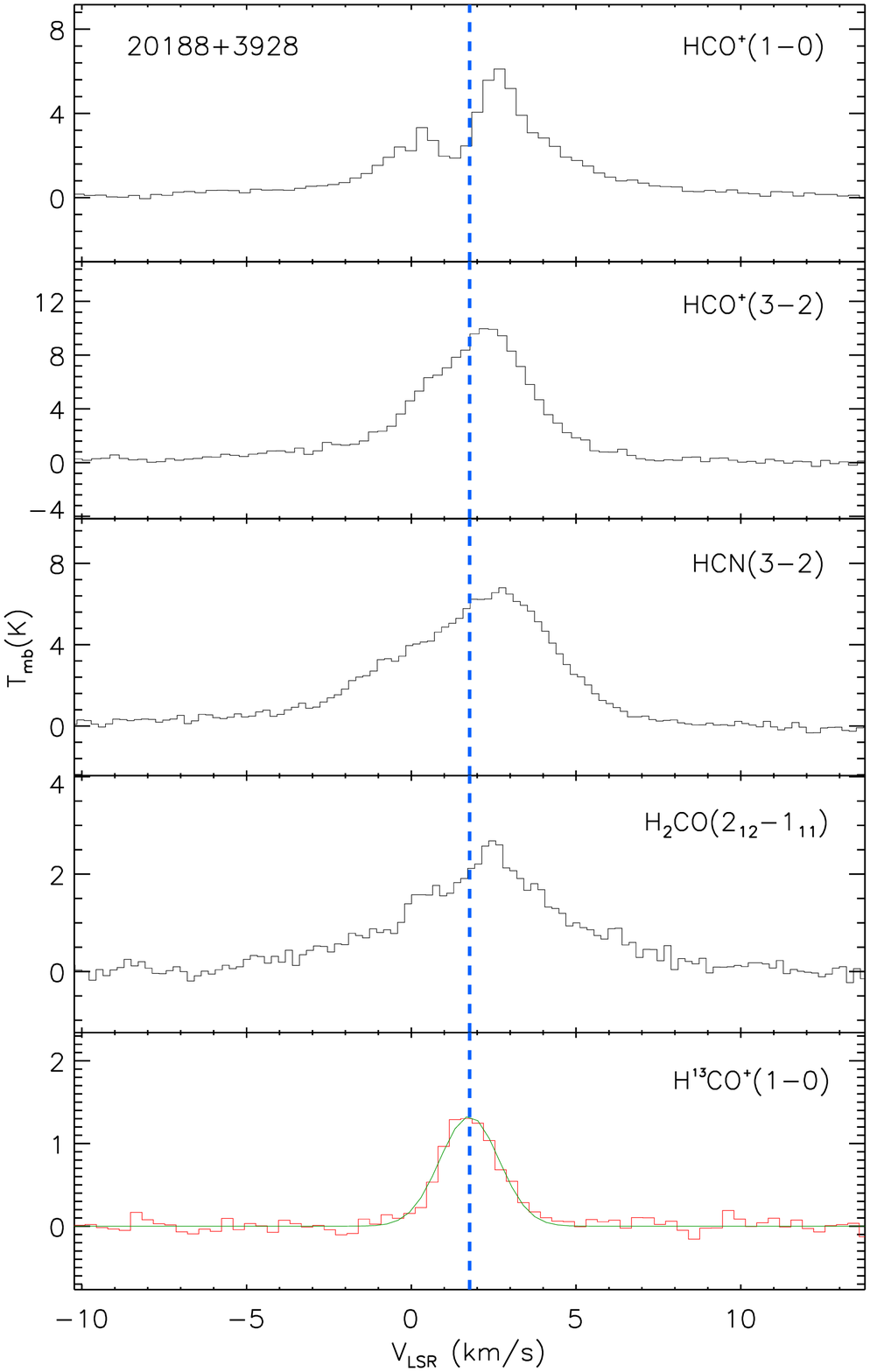}
  \includegraphics[width=0.25\textwidth]{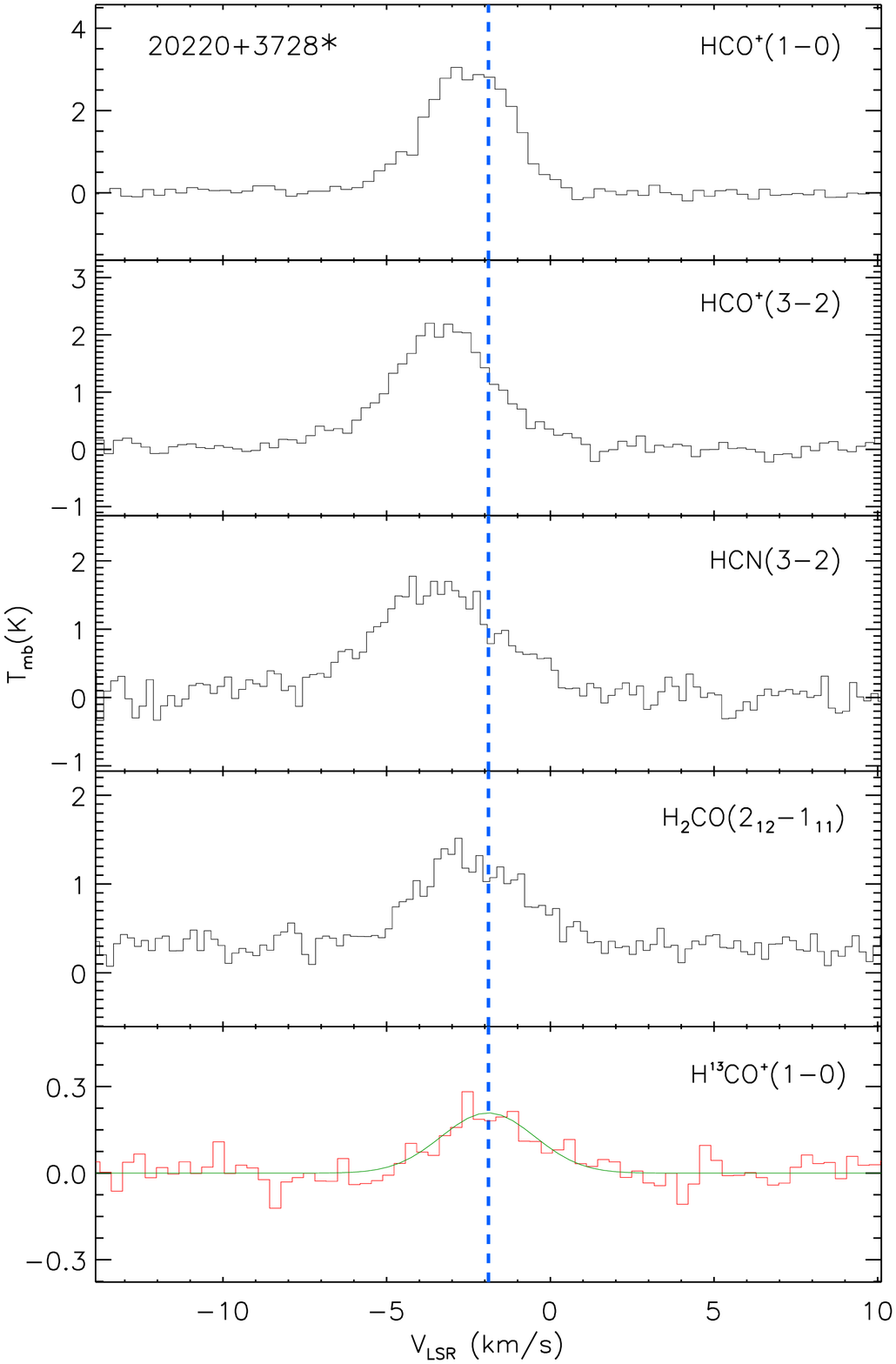}
  \includegraphics[width=0.25\textwidth]{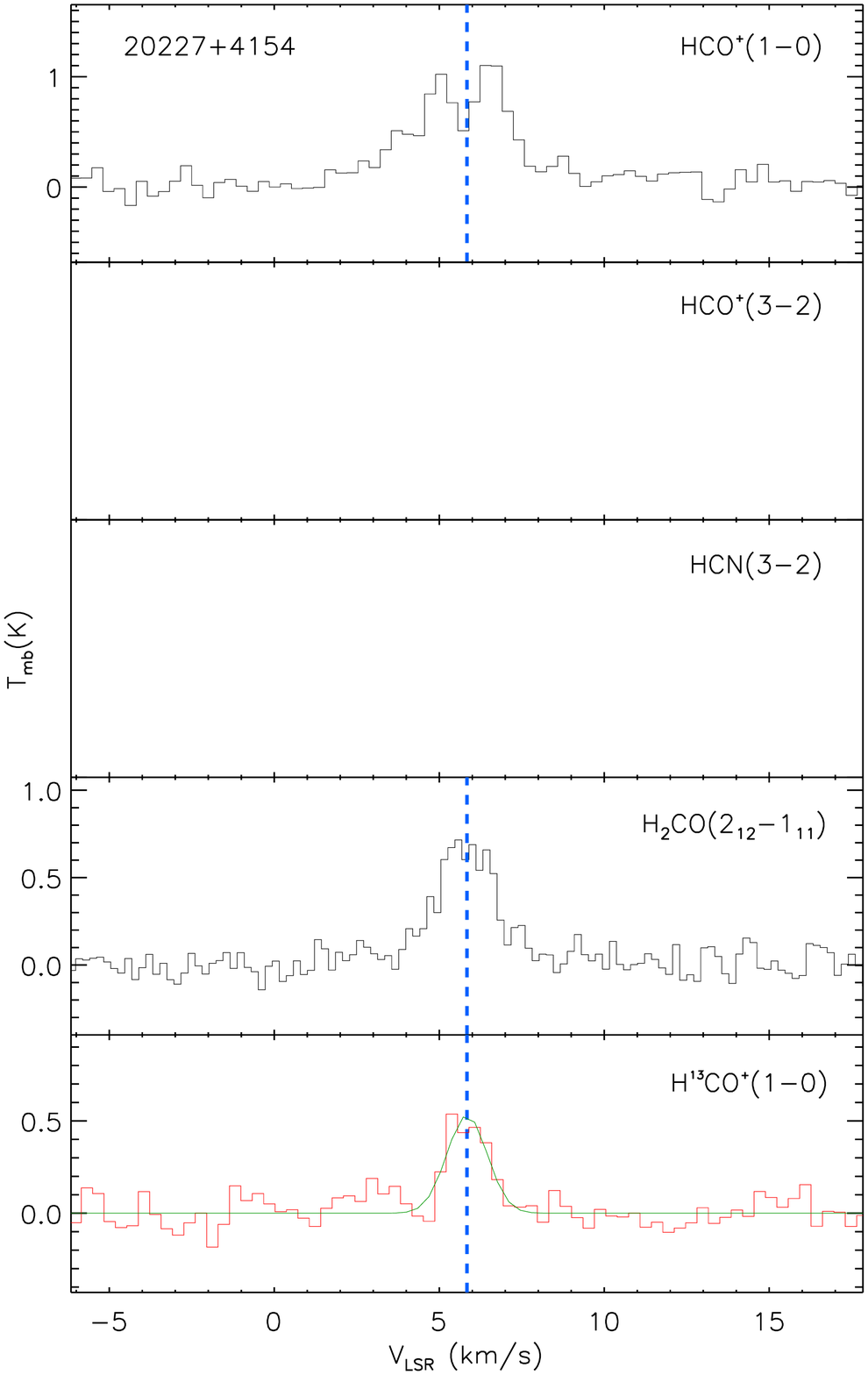}
\end{tabbing}
\center{\textbf{Figure A1.} continued.}
\label{fA1}
\end{figure*}

\begin{figure*}
\begin{tabbing}
  \includegraphics[width=0.25\textwidth]{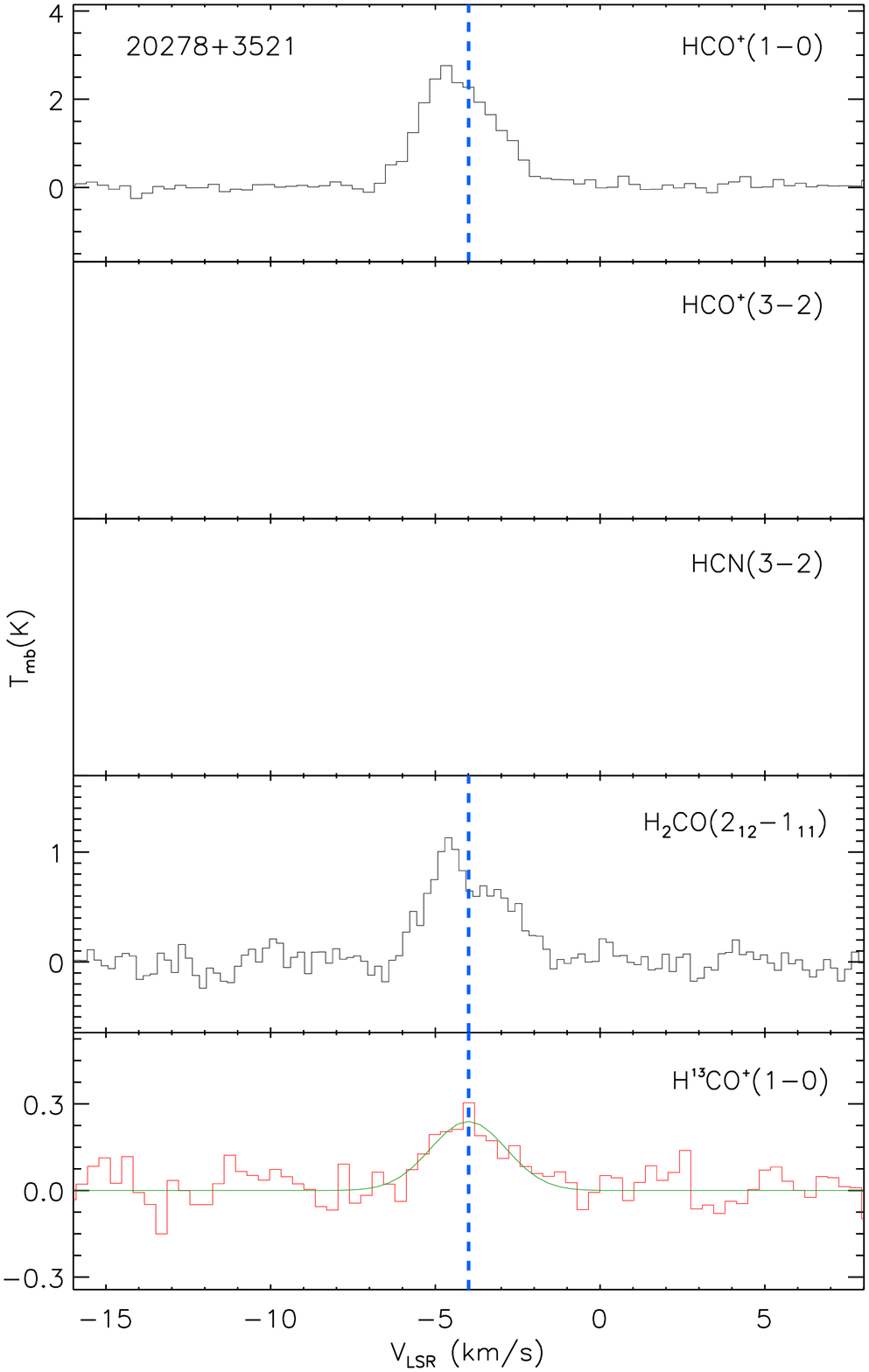}
  \includegraphics[width=0.25\textwidth]{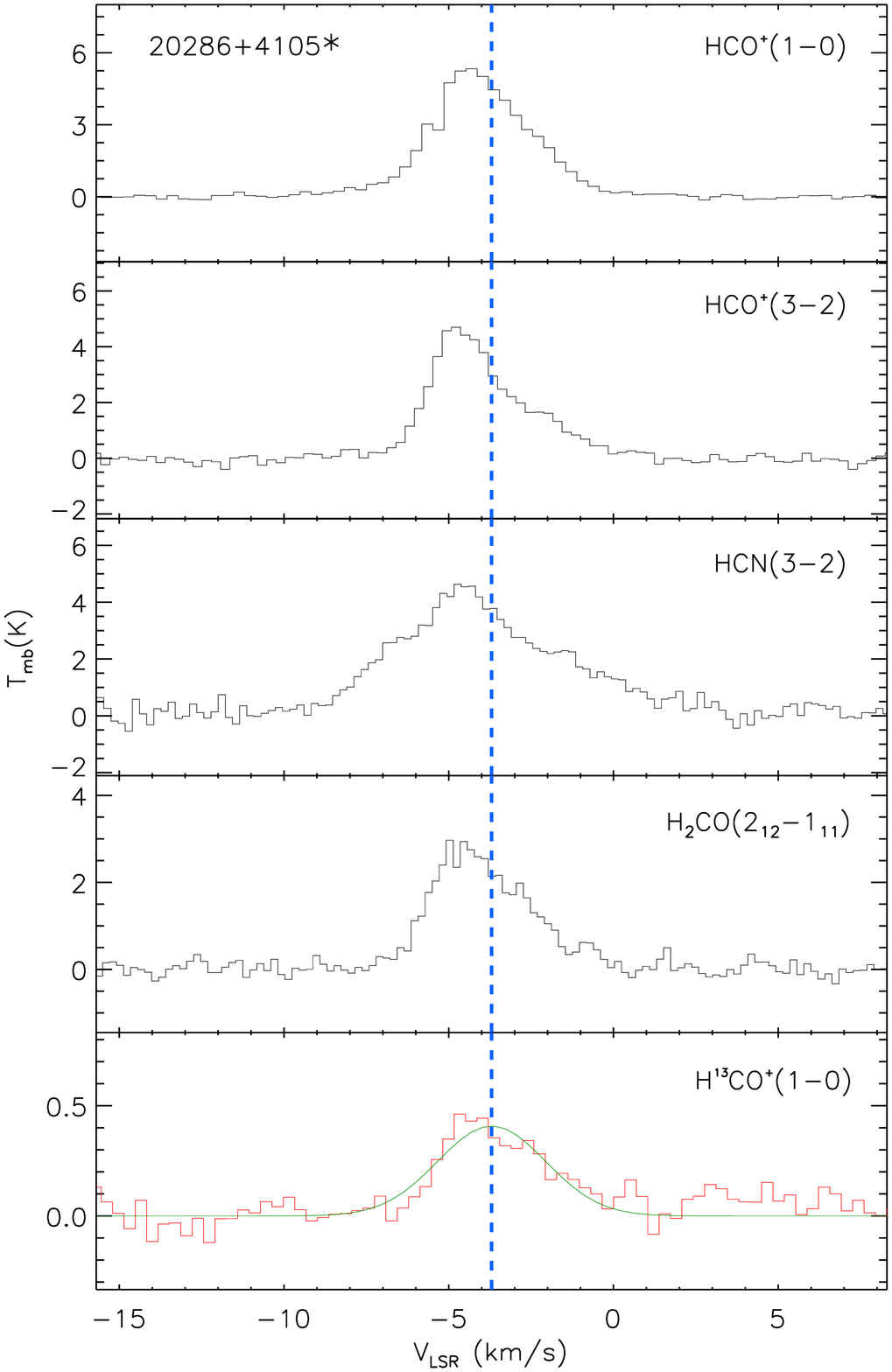}
  \includegraphics[width=0.25\textwidth]{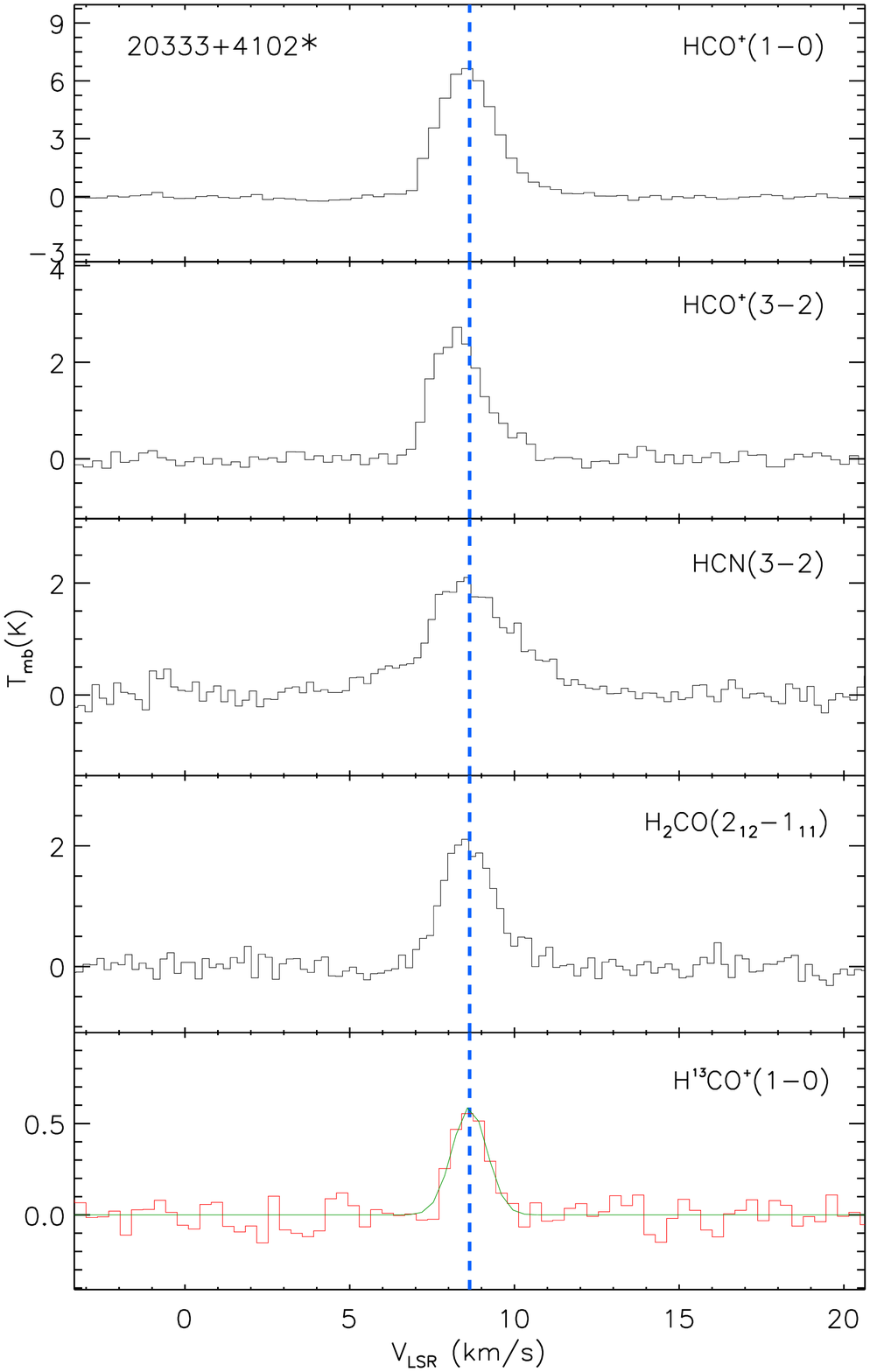}
  \includegraphics[width=0.25\textwidth]{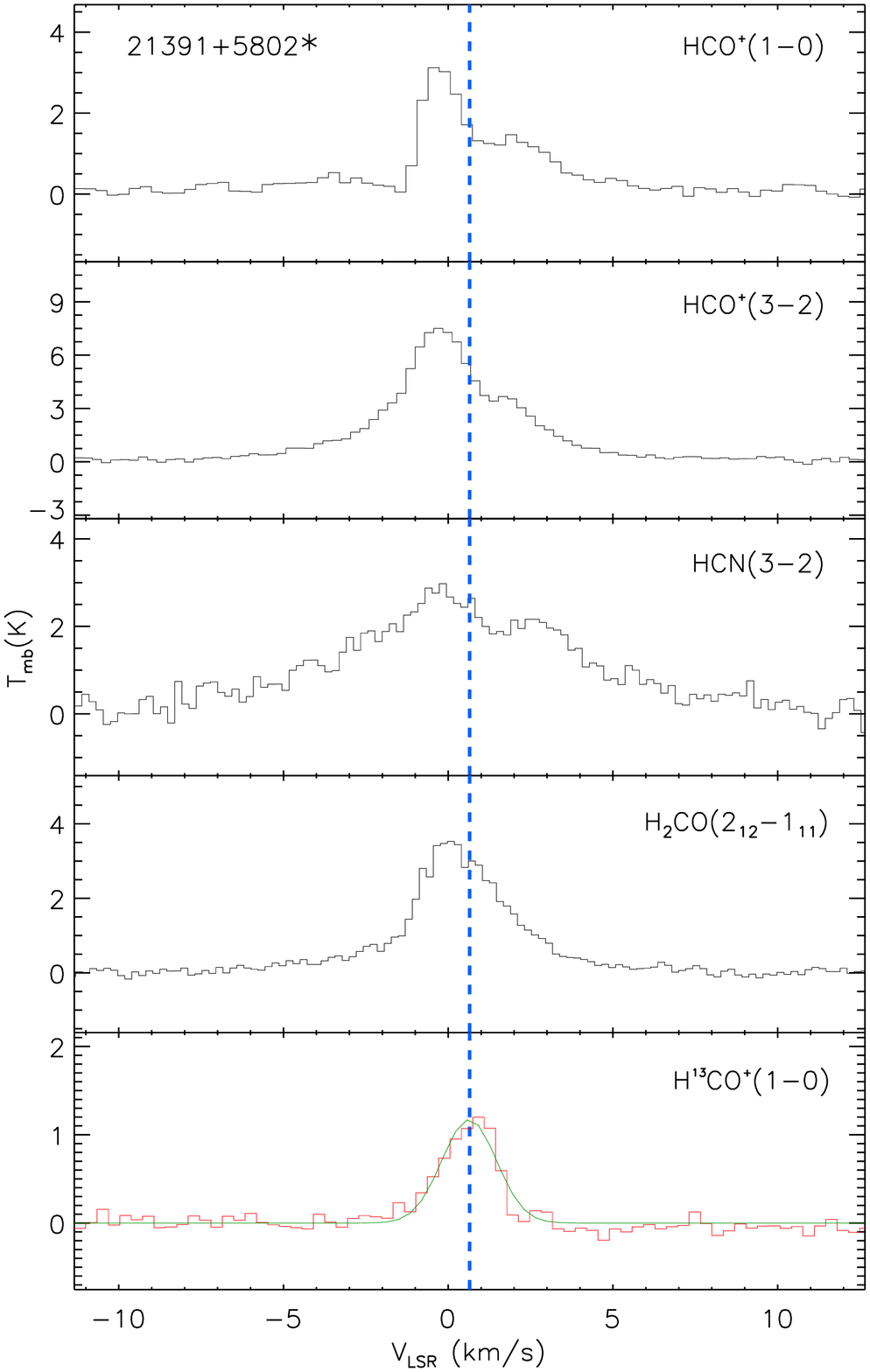} \\
  \includegraphics[width=0.25\textwidth]{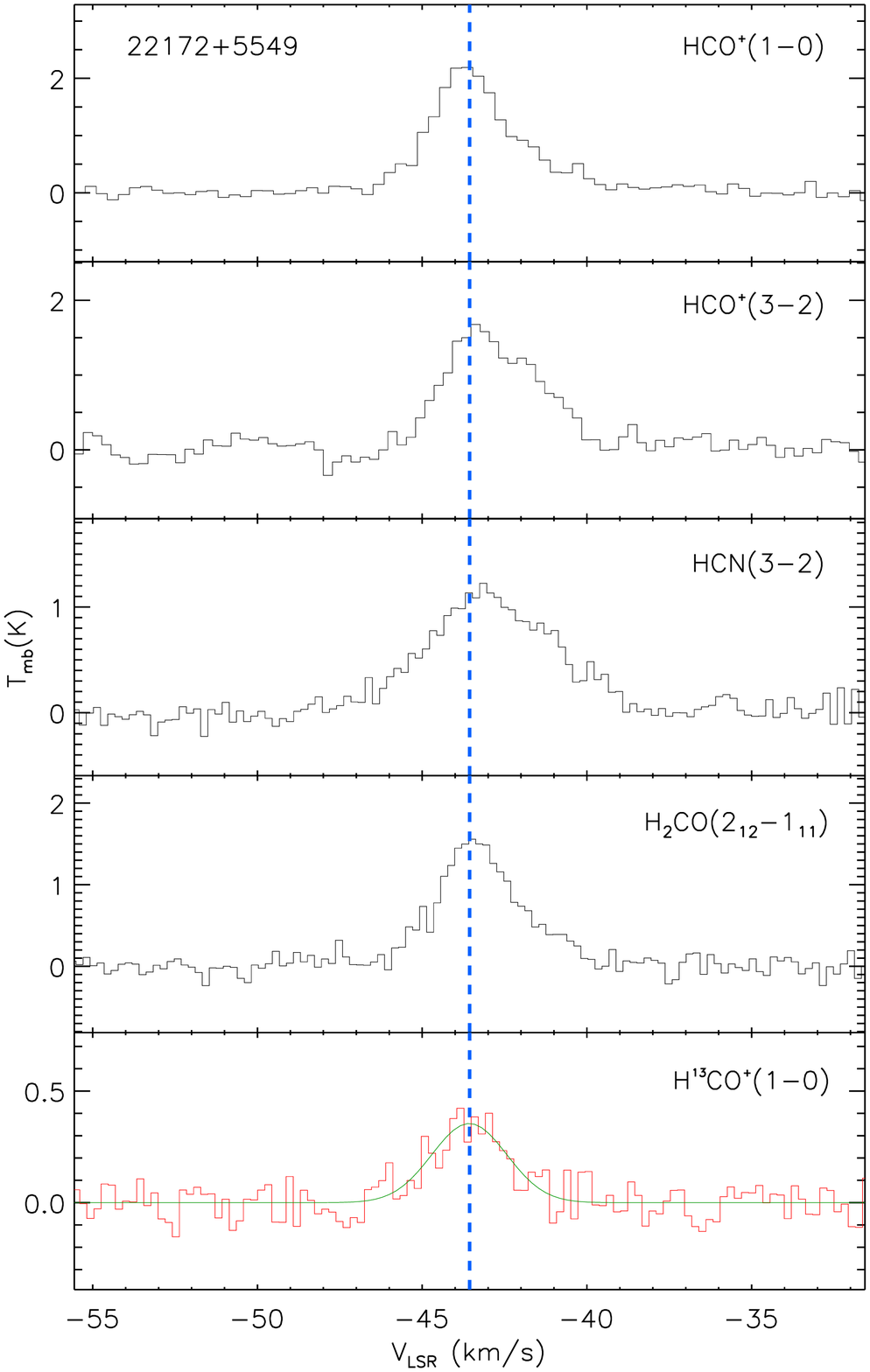} 
  \includegraphics[width=0.25\textwidth]{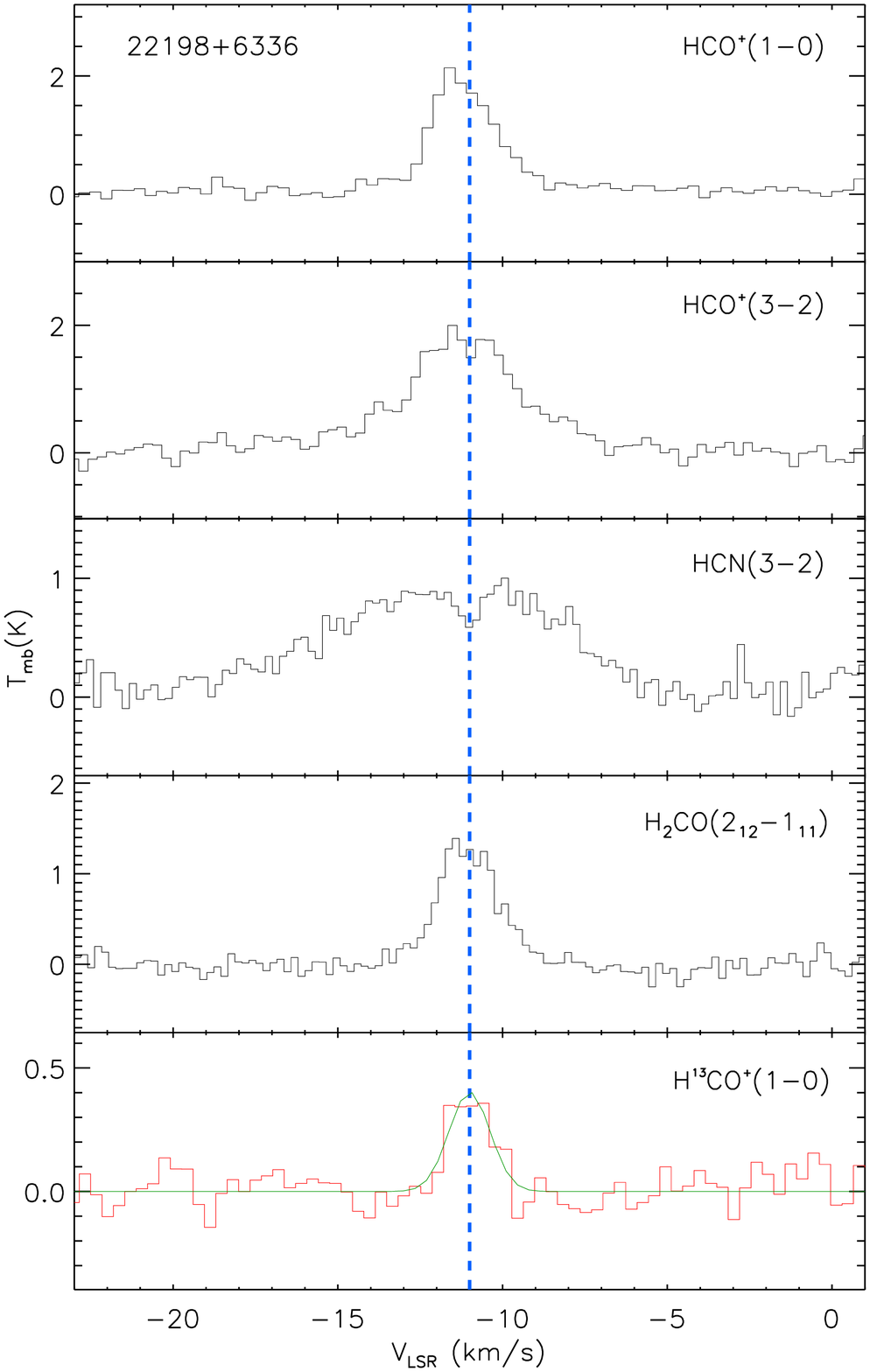} 
  \includegraphics[width=0.25\textwidth]{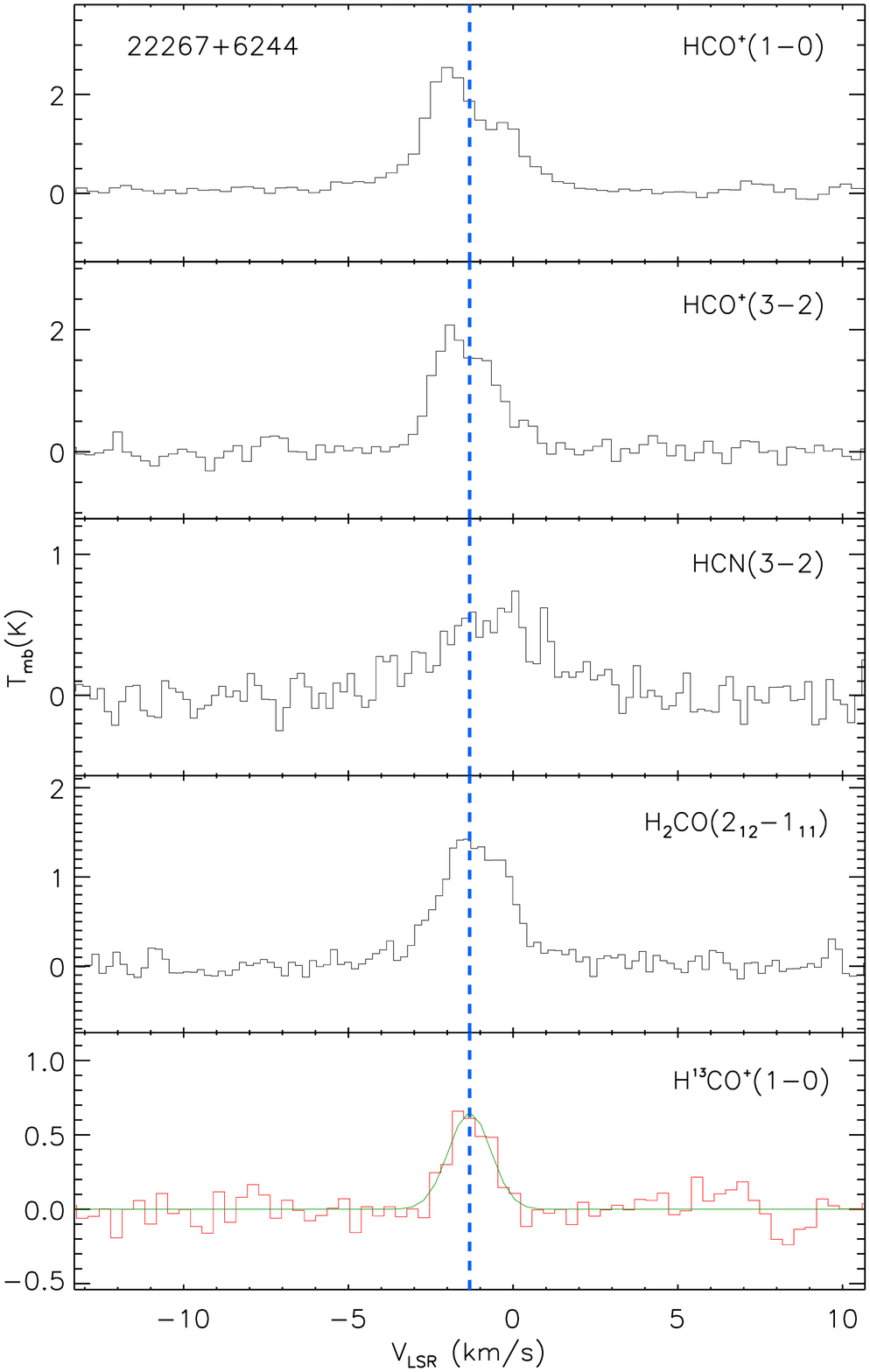}
  \includegraphics[width=0.25\textwidth]{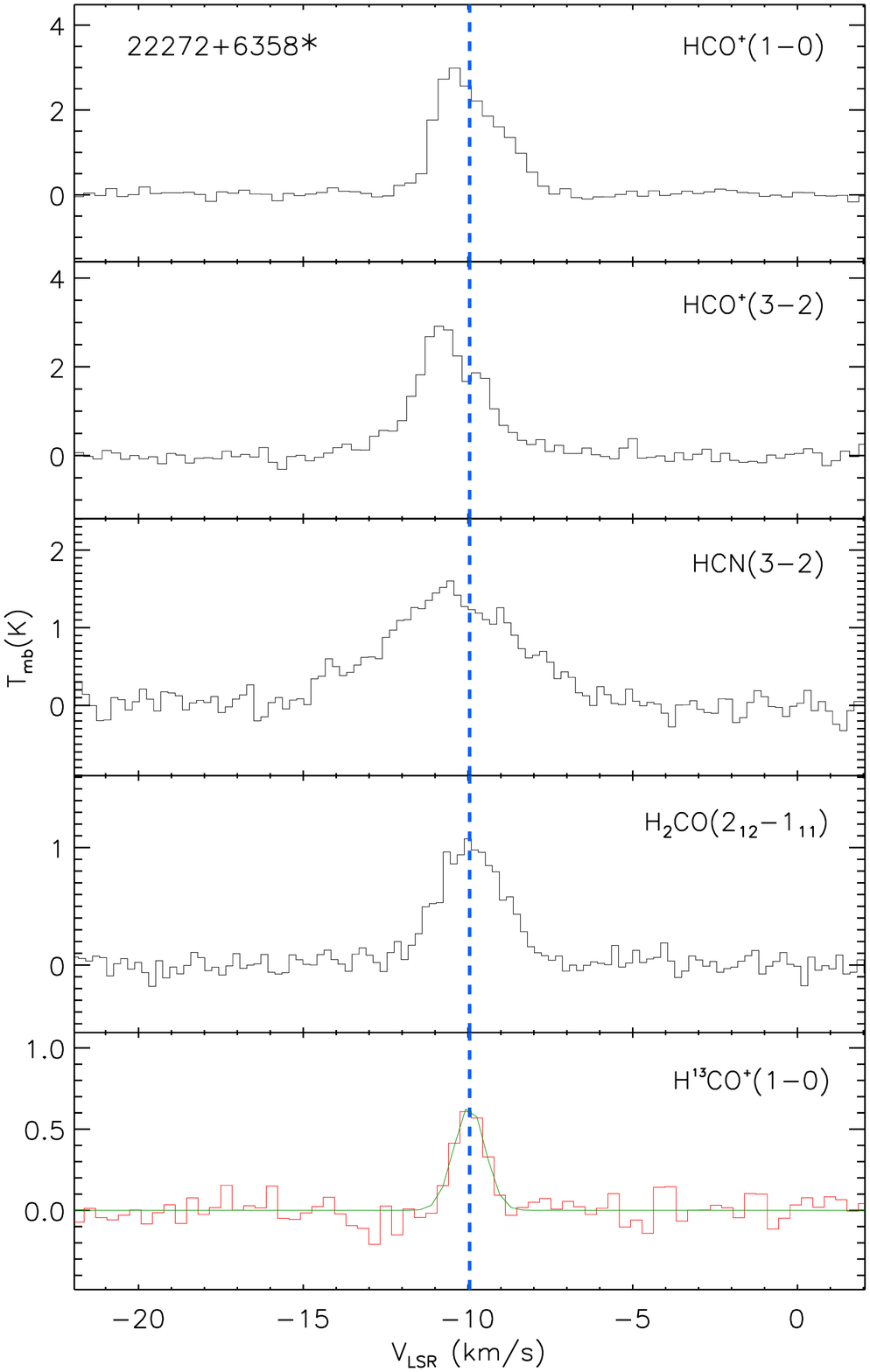} \\
  \includegraphics[width=0.25\textwidth]{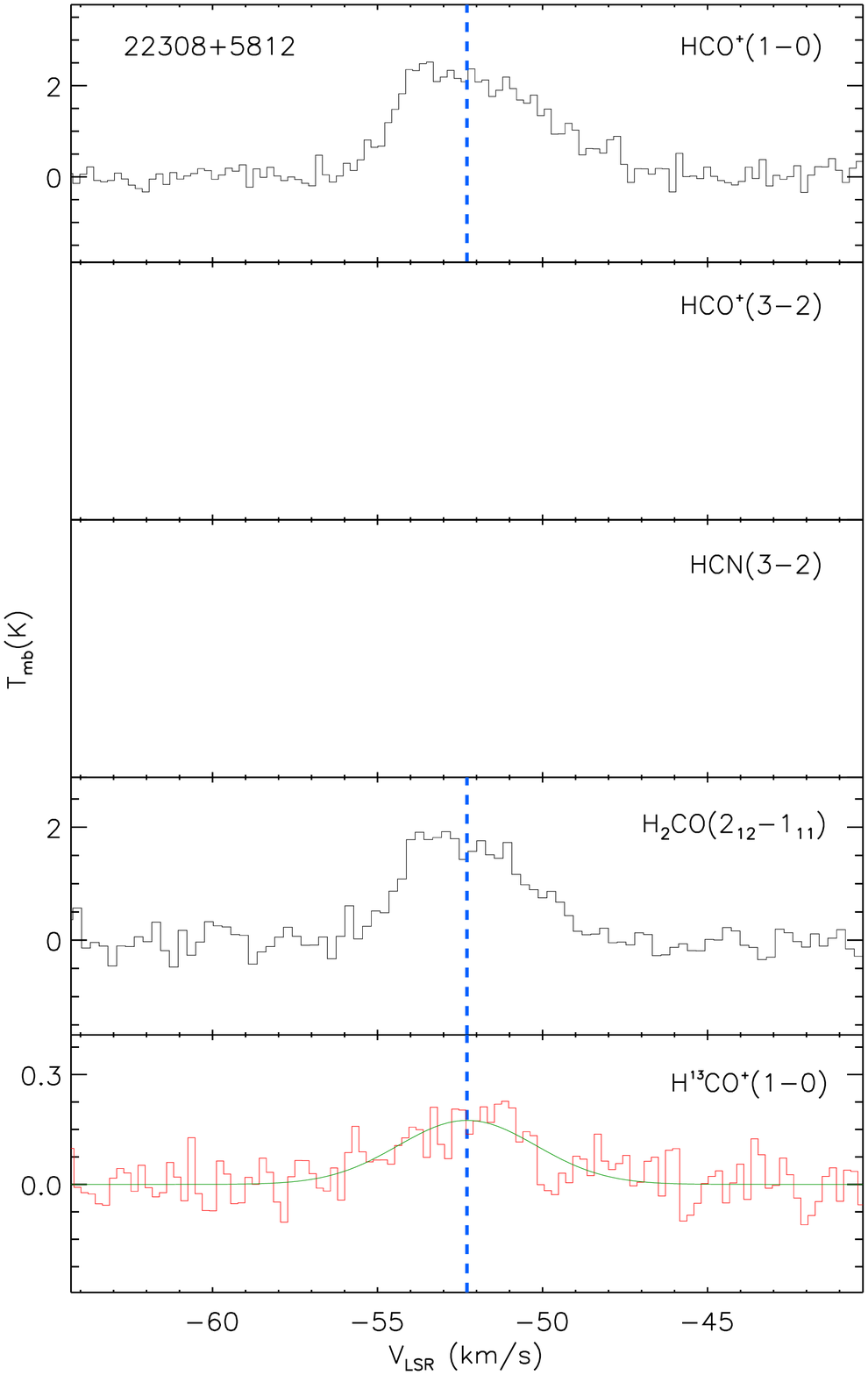} 
  \includegraphics[width=0.25\textwidth]{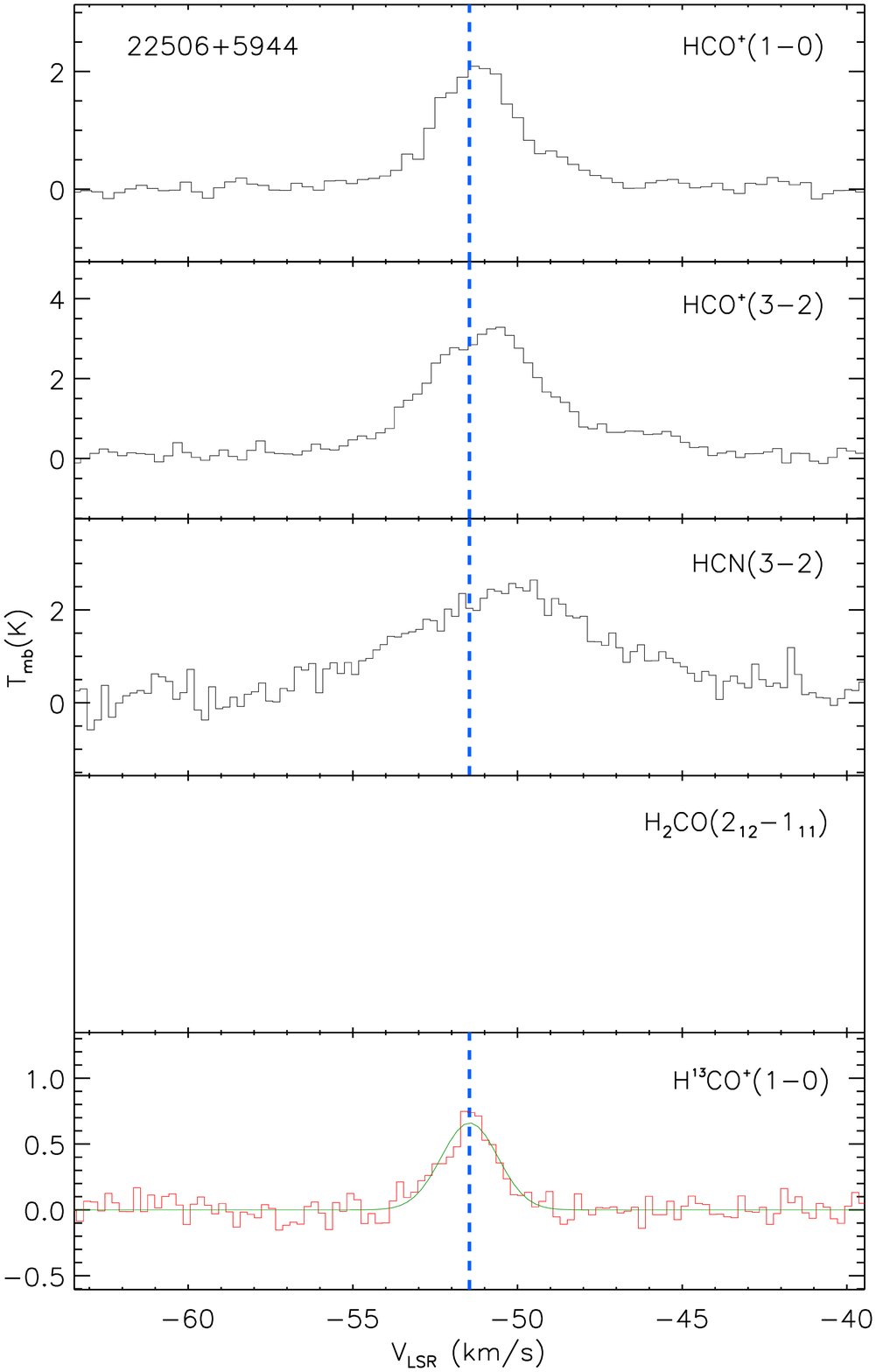}
  \includegraphics[width=0.25\textwidth]{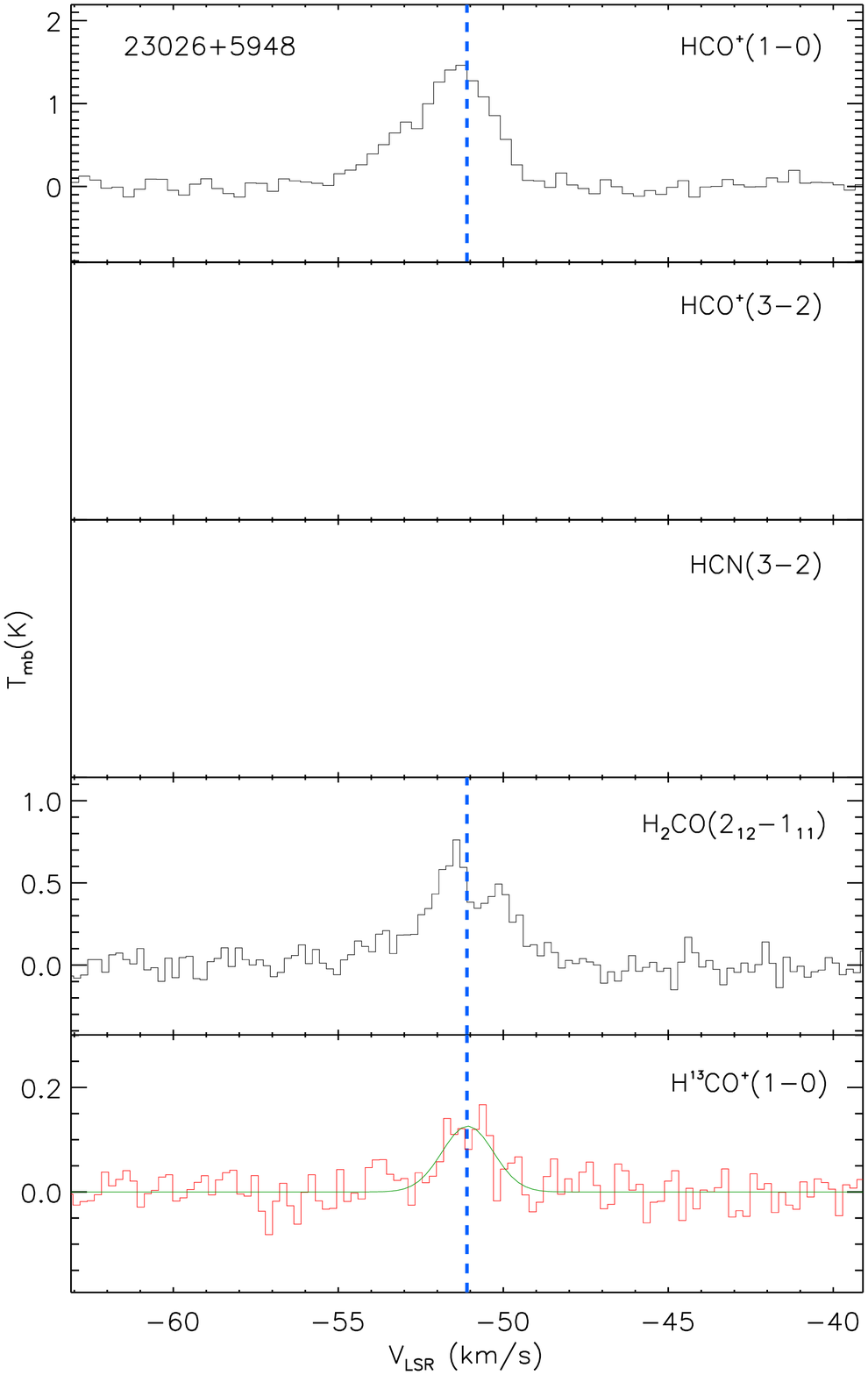}
  \includegraphics[width=0.25\textwidth]{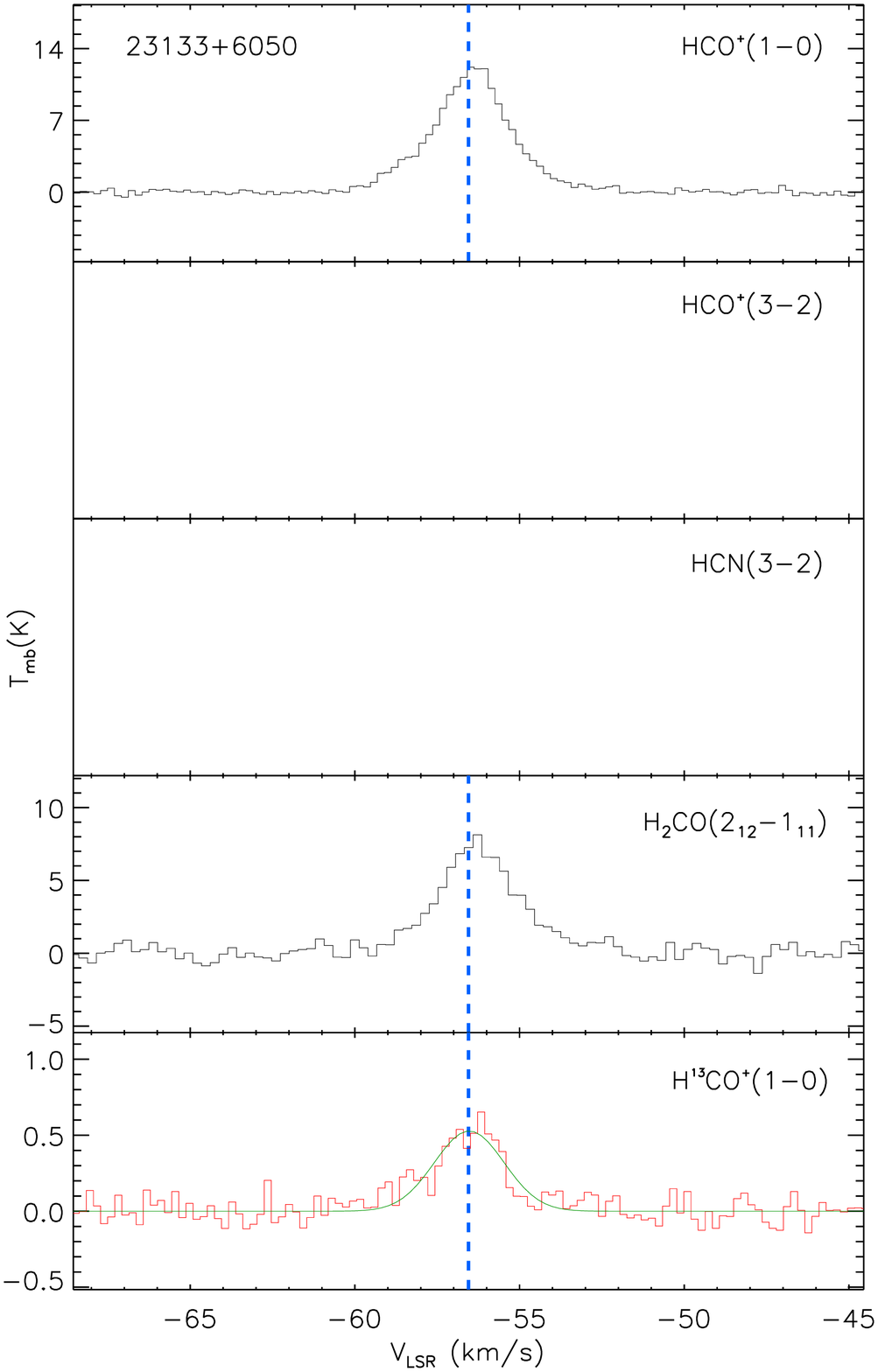}
\end{tabbing}
\center{\textbf{Figure A1.} continued.}
\label{fA1}
\end{figure*}

\begin{figure*}
\begin{tabbing}
  \includegraphics[width=0.25\textwidth]{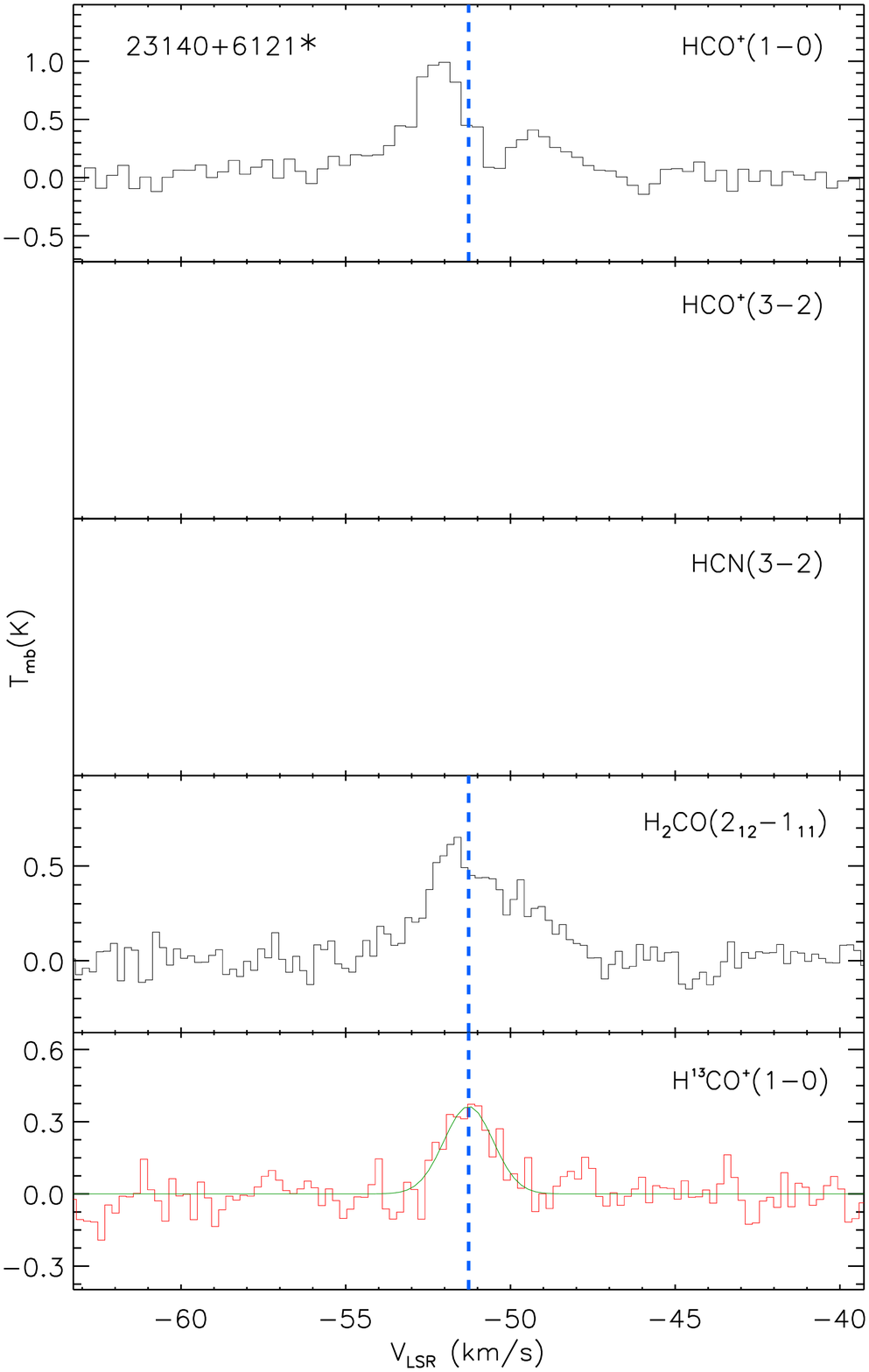}
  \includegraphics[width=0.25\textwidth]{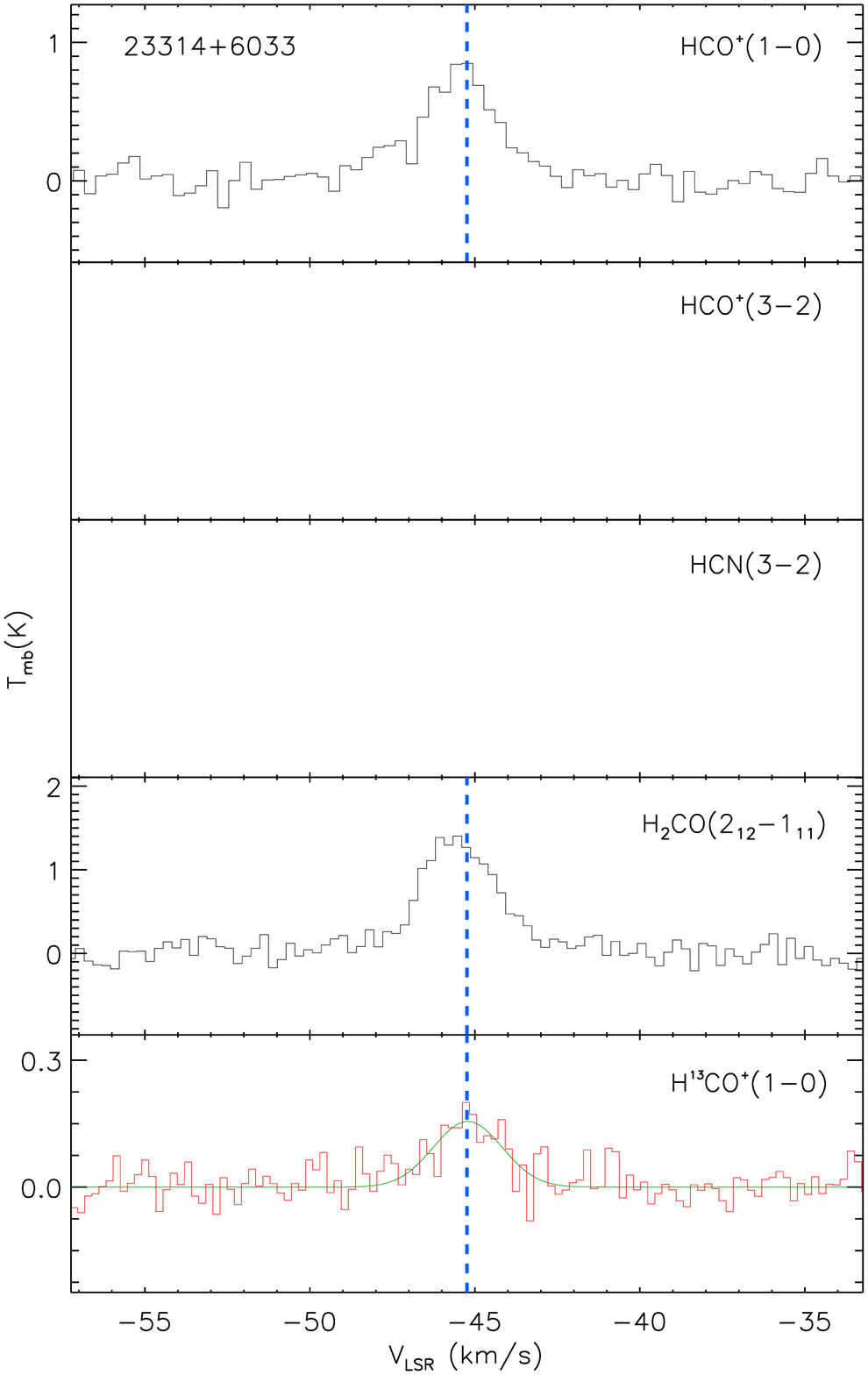}
  \includegraphics[width=0.25\textwidth]{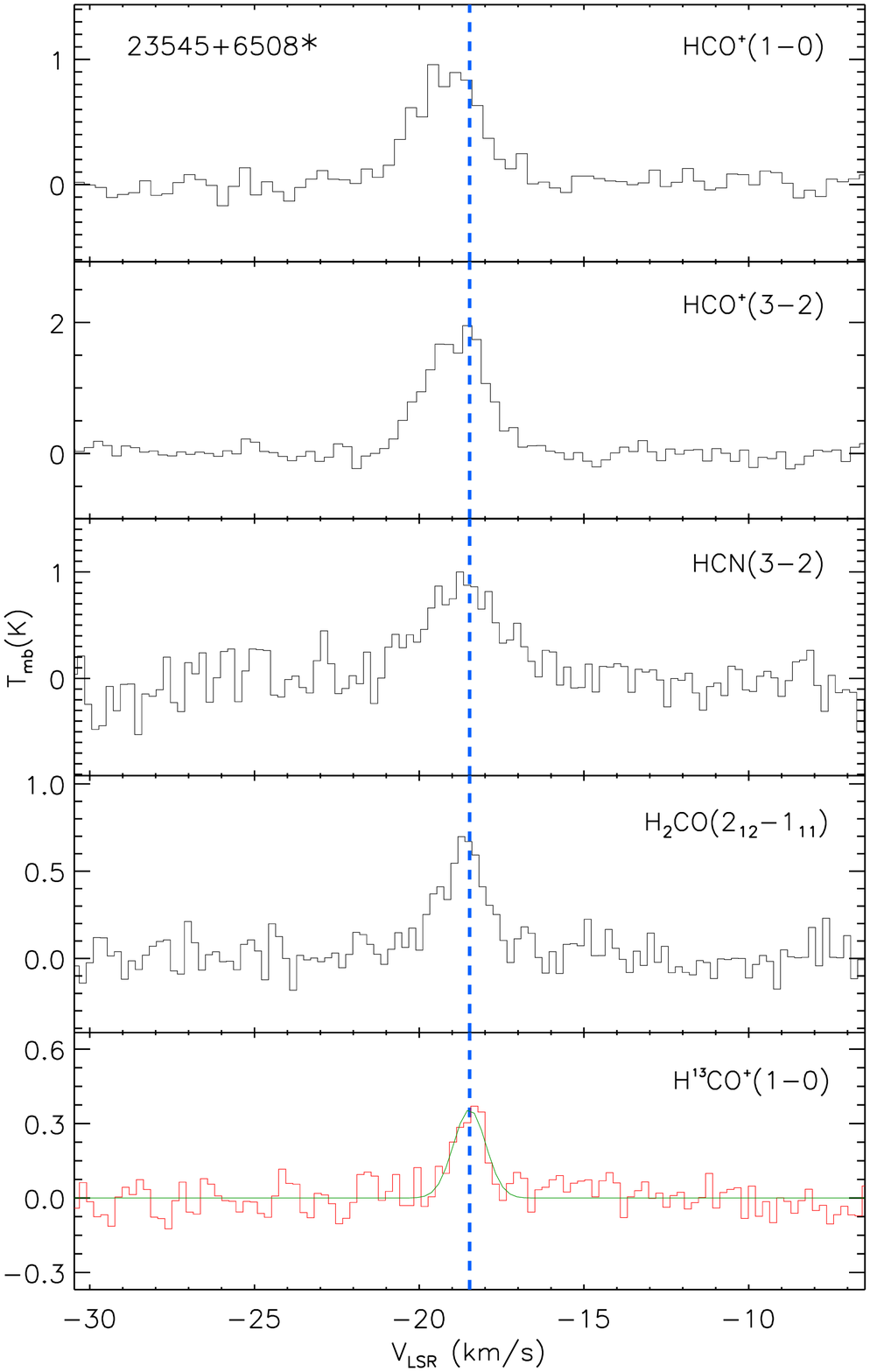}
\end{tabbing}
\center{\textbf{Figure A1.} continued.}
\label{fA1}
\end{figure*}




\end{document}